\documentclass[12pt,reqno]{amsart}
\usepackage{graphicx,graphics,amsmath,grffile}
\usepackage{longtable,booktabs}
\usepackage{url,subfig}

\usepackage[normalem]{ulem}
\usepackage{t1enc}              %% Ekezetes betuk begepelese:
\usepackage[utf8]{inputenc}    %% unicode

\def\re#1{(\ref{#1})}   %% Note: AMSTeX's  \eqref  also does  (\ref{#1})
\def\dnum{741} %number of days for long term data collection

\begin{document}
	
	\title[MGGL: Run-1]{Long term measurements from the M\'atra Gravitational and Geophysical Laboratory}
	\markright{Long term measurements in MGGL}

	\author[Ván et al.]{P. V\'an$^{1,5A}$, G.G. Barnaf\"oldi$^1$, T. Bulik$^{7}$, T. Biró$^2$, S. Czellár$^3$, M. Cieślar$^{9}$, Cs. Czanik$^2$, E. Dávid$^{1}$,  E. Debreceni$^1$, M. Denys$^7$, M. Dobróka$^{15}$, E. Fenyvesi$^3$, D. Gondek-Rosińska$^7$,  Z. Gráczer$^{2}$, G. Hamar$^1$, G. Huba$^{1}$, B. Kacskovics$^1$, Á. Kis$^{2}$., I. Kovács$^2$, R. Kovács$^{1,5A}$, I. Lemperger$^{2}$, P. Lévai$^{1}$, S. Lökös$^{12,13}$, J. Mlynarczyk$^{14}$, J. Molnár$^3$,  N. Singh$^7$, A. Novák$^{2}$, L. Ol\'ah$^{1,11}$, T. Starecki$^{6}$, M. Suchenek$^{6}$, G. Sur\'anyi$^{10}$, S. Szalai$^{2}$, M. C. Tringali$^7$, D. Varga$^{1}$, M. Vasúth$^{1}$, B. V\'as\'arhelyi$^{5B}$	V. Wesztergom$^{2}$,  Z. Wéber$^{2}$, Z. Zimborás$^1$, L. Somlai$^1$}

	\address{$^1$MTA Wigner Research Centre for Physics, Institute of Particle and Nuclear Physics, 1121 Budapest, Konkoly Thege Miklós út 29-33. \\
	$^2$MTA Research Centre for Astronomy and Earth Sciences, Geodetic and Geophysical Institute, H-9400, Sopron, Csatkai E. u. 6-8.\\
 	$^3$MTA Institute for Nuclear Research, Hungary, 4026 Debrecen, Bem tér 18/c \\
	$^{5A}$Budapest University of Technology and Economics, Faculty of Mechanical Engineering,  Department of Energy Engineering, Budapest, Hungary \\
	$^{5B}$Budapest University of Technology and Economics, Department of Engineering Geology and Geotechnics, Budapest, Hungary\\
	$^{6}$Institute of Electronic Systems, Warsaw University of Technology, Nowowiejska 15/19, 00-665 Warsaw, Poland\\
	$^{7}$Astronomical Observatory, University of Warsaw, Aleje Ujazdowskie 4, 00-478 Warsaw, Poland\\
    $^{8}$Instituto de Astronomía, Universidad Nacional Autónoma de México, Apartado Postal 877, Ensenada, Baja California, 22800 México\\
	$^9$ Nicolaus Copernicus Astronomical Center, Bartycka 18, 00-718 Warsaw, Poland,\\
 	$^{10}$MTA-ELTE Geological, Geophysical and Space Science Research Group, 1/A P\'azm\'any P. s., H-1117, Budapest, Hungary\\
	$^{11}$Earthquake Research Institute, The University of Tokyo, 1-1-1 Yayoi, Bunkyo-ku, Tokyo, Japan 113-0032\\
	$^{12}$Eszterh\'azy Károly University, H-3200 Gyöngyös, Mátrai út 36, Hungary\\
	$^{13}$Eötvös University, H-1111 Budapest, Pázmány Péter sétány 1/A, Hungary\\
	$^14$Department of Electronics, AGH University of Science and Technology, Krakow, Poland\\
    $^{15}$University of Miskolc, Hungary, H-3515 Miskolc-Egyetemváros
}

\date{\today}

\begin{abstract}
		Summary of the long term data taking, related to one of the proposed next generation ground-based gravitational detector's location is presented here. Results of seismic and infrasound noise,  electromagnetic attenuation and cosmic muon radiation measurements are reported in the underground Matra Gravitational and Geophysical Laboratory near Gy\"ongy\"osoroszi, Hungary. The collected seismic data of more than two years is evaluated from the point of view of the Einstein Telescope, a proposed third generation underground gravitational wave observatory. Applying our results for the site selection will significantly improve the signal to nose ratio of the multi-messenger astrophysics era, especially at the low frequency regime.
\end{abstract}
\maketitle

\section{Introduction}
% Lévai, P., Vasúth M., Barnaföldi G., Huba G. D\'avid E., Kov\'acs R., V\'an P., Zimborás Z. $^{1}$
% $^1$HAS, Wigner Research Centre for Physics, Institute of Particle and Nuclear Physics, \\
%1121 Budapest, Konkoly Thege Miklós út 29-33.
	
Preparation for the next generation ground-based gravitational detectors requires careful pre-analysis at the proposed locations. As for previous gravitational wave detectors, identifying noise sources and constructional risks is a key task. In case of underground installation and improved low frequency operation this must be based on long-term measurements to identify noise types and its origins. Parallel and combined measurements applying the most novel techniques for environmental tests is also beneficiary for the design of the proposed facility and for high-accuracy measurements with it.

The main advantage of underground operation for gravitational observatories is the improved sensitivity at the low frequency regime, 1-10Hz. The observation of gravitational waves with earth-based detectors and the birth of multimessenger astronomy in the last years \cite{LIGVIR16a,LIGVIR16a2,LIGVIR17a,LIGVIR17a1,LIGVIR17a2,LIGVIR17a3} revived interest in the scientific and technological challenges due to the underground operation \cite{Kagra_hp,ET_LoI_hp}. The scientific value of the existing plans of building an all-in-one observation facility increased considerably. Its improved sensitivity, extended frequency range, together with capabilities of measuring polarization and direction offer an unprecedented discovery potential \cite{ETdes11r}.

In order to explore the technological and scientific background of underground operation the M\'atra Gravitational and Geophysical Laboratory (MGGL)  was established by MTA Wigner Research Centre for Physics in the Gy\"ongy\"osoroszi ore mine, M\'atra Mountains, Hungary. The Laboratory is located in the coordinates (399 MAMSL,  47$^\circ$ 52' 42.10178", 19$^\circ$ 51' 57.77392" OGPSH 2007 (ETRS89)),  %(399 mBf, 711232.27, 281949.94 EOV),
along a horizontal tunnel of the mine, 1280~m from the entrance and 88~m below surface, however, the related research activities are extended to other areas of the mine and the surroundings in the last years. The initial instrumentation and the first results were reported in Refs. \cite{BarEta17a}.

In this paper we report the results of continued research summarizing the geophysical characteristics of the M\'atra Mountains as an Einstein Telescope site candidate. In this respect the analysis of the collected seismic noise data of more than two years are probably the most informative. We report them together with geophysical data, rock characteristics, infrasound, electromagnetic and muon flux measurements. The paper is organized as follows. First we introduce the geological and seismological environment of M\'atra Mountains. A short report of the mechanical properties of the grey andesite rock of the M\'atra is part of the geophysical survey. Then the seismological data collected in the last two years are shown and analysed. Here both the data from broadband seismometers of the Wigner RCP and also the custom made seismometers of the Warsaw University are shown.  In the third section underground infrasound measurements are reported. Then the damping of electromagnetic waves in the andesitic rocks of M\'atra is estimated using underground and surface electromagnetic measurements. Finally the evaluation of the collected muon flux  data is shown, demonstrating the tomographic capabilities of the installed detector technology with rock density maps.

In the following sections we shortly introduce the instruments and elaborate the long term data, whenever it is possible. This  includes all seismic information that was collected in MGGL. In the following the complete data collection period will be referred as Run-1 when compared to the previous, shorter, Run-0 data of our previous report \cite{BarEta17a}.

\section{The geophysical environment of MGGL}

\begin{figure}
	\centering
	\includegraphics[width=0.8\textwidth]{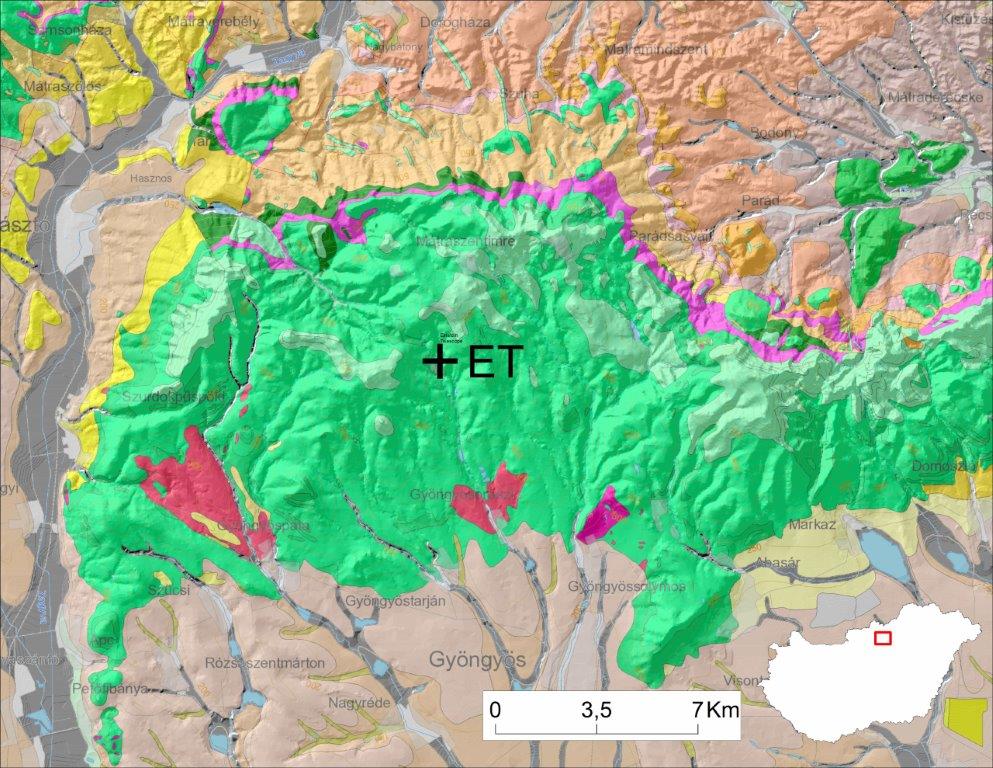}
	\caption{The surface lithology of the M\'atra area. The green part denotes various andesitic rocks.}\label{Matrand}
\end{figure}

The lithological composition of M\'atra mountain range is moderately blocky andesite, as it is shown in Fig. \ref{Matrand}, where the green areas denote various andesite types formed at the same geological era \cite{KovSza08a}. %The most important physical properties of a typical hard rock from Gy\"ongy\"osoroszi mine are given in Refs. \cite{BarEta17a,VanEta18k}.
In the following we characterise the seismic activity of the surrounding area and also the mechanical properties of the typical hard rock from Gy\"ongy\"osoroszi mine. In order to enhance comparison with other sites we also show quantitative measures by calculating the related ground displacement and seismic hazard of the M\'atra Mountains.

\subsection{Seismicity of the M\'atra Mountains and the surrounding areas}\label{seismicity-of-the-matra-mountains-and-the-surrounding-areas}

In general earthquakes are more important for the stability of the underground facility and small ones are not critical regarding the observations.

\begin{figure}
	\centering
	\includegraphics[width=\textwidth]{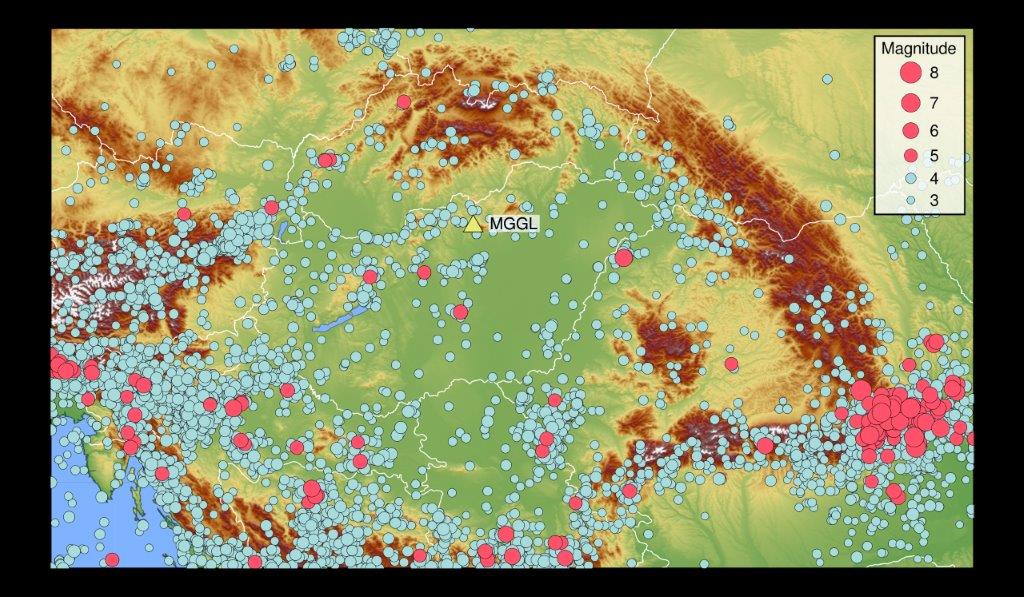}%png jobb
	\caption{Seismic activity in the Carpatho-Pannonian area. The triangle shows the location of the MGGL. The circles mark the epicentres of the known earthquakes occurred between the years 1800 and 2010. Blue circles: earthquakes with magnitudes 3.0$\leq$M\textless{}5.0, red circles: earthquakes with magnitudes M$\geq$5.0.}
\label{fig_Carp_seismo}\end{figure}

The seismicity of the Carpathian basin as a whole can be considered moderate. The known earthquakes of the area with magnitude larger than 3.0 are shown in Fig. \ref{fig_Carp_seismo}. According to the historical collected data on average one earthquake with magnitude $M \geq 5.0$ can be observed in the Carpatho-Pannonian region annualy. Also the seismic activity from neighbouring open pit mines can be identified by the ET1H seismological station deployed in the MGGL with the help of the stations of the Hungarian National Seismological Network \cite{GraEta18a}.

\begin{figure}
	\centering
	\includegraphics[width=\textwidth]{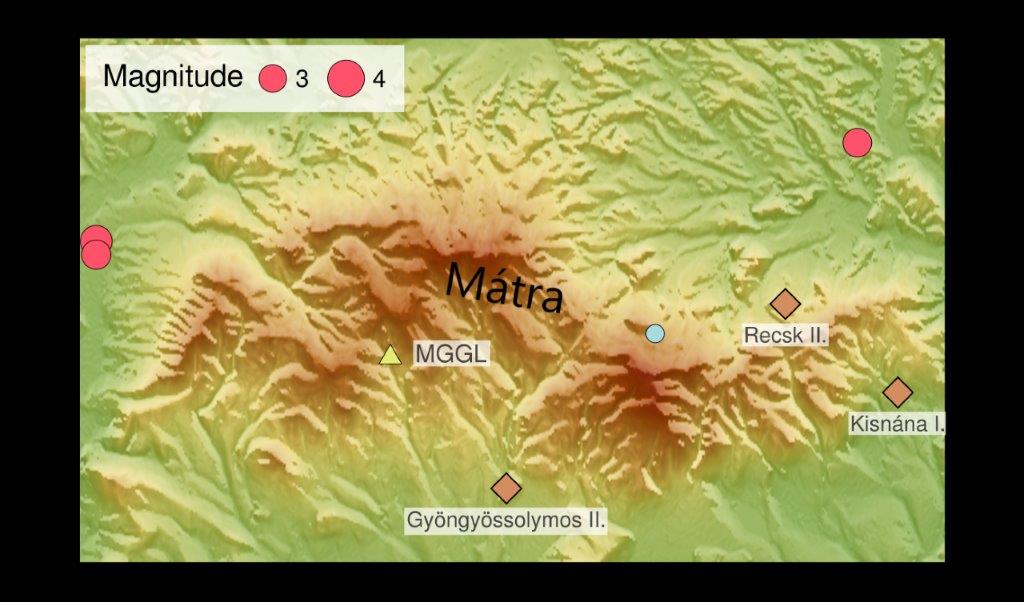}%png jobb
	\caption{The M\'atra area. The triangle shows the location of the MGGL, while the diamonds mark the position and name of the most active quarries. Red circles: known earthquakes in the area.}
	\label{fig_Matr_seismo}\end{figure}

The level of seismic activity in the M\'atra Mountains is quite low. Fig. \ref{fig_Matr_seismo} shows the epicentres of the  known  $M\geq 3$ earthquakes in the area (19.69-20.18E, 47.8-48.0N) based on the data of the MTA CSFK GGI Earthquake Catalogue \cite{HEqCat18} and Hungarian National Seismological Bulletins \cite{GraEta18r}. Only earthquakes with magnitude greater than 3.0 are shown as lower magnitude events can be misclassified quarry explosions. It can be seen that only three small earthquakes ($M \leq 3.5$) were ever observed in the area which occurred in 1879, 1895 and 1980.

\subsubsection{Ground displacements caused by seismic events}\label{ground-displacements-caused-by-seismic-events}%}

We have selected four events to characterise the ground displacement in the MGGL caused by different seismic events. For the computations of displacements 100 sps streams were used. The data before the instrument correction were detrended and filtered by a second order high pass Butterworth filter with a corner frequency of 1 Hz.

As the seismicity in the M\'atra area is mainly determined by the mining activities in the quarries, we computed the ground displacement for characteristic explosions carried out in the three most active mines. Table 1 shows the maximum displacement amplitudes for the three components in the case of the selected events. The magnitude of the displacement is similar for the mines Gyöngyössolymos and Kisn\'ana, while the maximum displacement for the explosion belonging to Kisn\'ana mine is an order of magnitude smaller.

As the recent Tenk earthquake (2013-04-22, M=4.8, epicentral distance=42 km) occurred relatively close to the MGGL it may be of interest. Unfortunately, at the time of the earthquake the ET1H station was not yet installed. However we can use the data of Piszkéstető station, which is located on the surface, 4.88 km from MGGL and was operational. The fourth row of Table 1. shows the displacements observed there. It can be seen that the Tenk earthquake produced around 20-30 times larger displacements than the strongest explosion. It must be noted that in the MGGL - due to the large subsurface depth - this value probably would have been smaller.

\begin{longtable}[]{@{}llrrr@{}}
	\caption{Observed maximum ground displacement values for frequencies larger then 1 Hz in the case of four characteristic events.}\tabularnewline
	\toprule
	Station & Event & Z {[}\(\mu m\){]} & N {[}\(\mu m\){]} & E
	{[}\(\mu m\){]}\tabularnewline
	\midrule
	\endfirsthead
	\toprule
	Station & Event & Z {[}\(\mu m\){]} & N {[}\(\mu m\){]} & E
	{[}\(\mu m\){]}\tabularnewline
	\midrule
	\endhead
	ET1H & 2016-05-10 10:47 (explosion; Gyöngyössolymos; dist= 7.4 km) &
	0.88 & 0.66 & 0.53\tabularnewline
	ET1H & 2017-05-30 10:43 (explosion; Kisn\'ana; dist=21.5 km) & 0.10 & 0.07
	& 0.08\tabularnewline
	ET1H & 2017-06-22 11:21 (explosion; Recsk; dist=16.8 km) & 0.34 & 0.78 &
	0.43\tabularnewline
	PSZ & 2013-04-22 22:28 (earthquake; M= 4.8; Tenk; dist=42.0 km) & 20.76
	& 22.57 & 15.99\tabularnewline
	\bottomrule
\end{longtable}

%\hypertarget{seismic-hazard-in-the-matra-mountains}{%
\subsubsection{Seismic hazard in the M\'atra Mountains}\label{seismic-hazard-in-the-matra-mountains}%}

The SHARE (Seismic Hazard Harmonization in Europe) was a large-scale collaborative project under the European Community's Seventh Framework Program \cite{GiaEta14a}. Its main objective was to construct a community-based seismic hazard model for the Euro-Mediterranean region. Its product, the 2013 Euro-Mediterranean Seismic Hazard Model (ESHM13) provides, among others, ground motion hazard maps for the peak ground acceleration (PGA) for different exceedance probabilities. The hazard values are referenced to a rock velocity (vs30) of 800 m/s.

For the M\'atra area we obtained the mean PGA values of the SHARE Mean Hazard Model from the online data resource \cite{GiaEta13r} for 73, 102, 475, 975 and 2475 years return periods (Table 2.). These correspond to 50\%, 39\%, 10\%, 5\% and 2\% exceedance in 50 years, respectively.
\begin{longtable}[]{@{}rr@{}}
	\caption{Average peak ground accelaration values in the M\'atra Mountains for different return periods according to the SHARE Mean Hazard Model}
	\tabularnewline\toprule
	Return period {[}\emph{years}{]} & PGA {[}\(m/s^2\){]}\tabularnewline
	\midrule
	\endfirsthead
	\toprule
	Return period {[}\emph{years}{]} & PGA {[}\(m/s^2\){]}\tabularnewline
	\midrule
	\endhead
	73 & 0.016\tabularnewline
	102 & 0.021\tabularnewline
	475 & 0.051\tabularnewline
	975 & 0.077\tabularnewline
	2475 & 0.123\tabularnewline
	\bottomrule
\end{longtable}

\subsection{Elastic and rheological properties of andesite from Gy\"ongy\"osoroszi}
%Here we report measuerements regarding rock rheology and interpret the differences between dynamic and static elasitc moduli of andesite from Matra.
%   V\'an P., Lökös S., Kov\'acs R., Poly\'ak Z.

In the following we give the static and dynamic elastic moduli of the andesitic rock of the Gy\"ongy\"osoroszi mine from time dependent laboratory experiments. We interpret these measurements in a rheological modelling framework and justify the obtained model parameters with elastic propagation speed measurements of the P and S waves. These investigations show that the hard rock of M\'atra cannot be considered ideal elastic in a characteristic frequency range. The influence of this property for Newtonian noise should be investigated.

\subsubsection{Measured elastic moduli and rheological parameters}

As it is well known, the static elastic moduli of rocks measured in laboratory are different from the dynamic ones determined from wave propagation speeds \cite{LamVut78b,Sav84a,Hee87p,MocPan03a,Ger07a,NajEta15a}. The difference is usually attributed to various heterogeneities, like microcracks, porosity and grain structure. The Kluitenberg-Verh\'as body of thermodynamic rheology, which is derived from non-equilibrium thermodynamics with internal variables, provides a simple modelling possibility and explanation \cite{AssEta15a,BarEta17a}. %The main advantage of this thermodynamic theory is the universal and uniform characterisation of both dispersive and dissipative phenomena.
If the material relaxation times of geometric effects are in the order of the operational domain of the low frequency part of ET, then rheological properties can influence the reliable detection of gravitational waves. Therefore we have performed laboratory measurements in order to determine the rheological properties of the gray andesite of M\'atra.

The investigated samples were cutted from blocks originated in construction works at the vicinity of MGGL. This is a middle gray, small grained, isotropic piroxen type andesitic rock, which is a differentiated upper miocen (tortona type), about 14.5 million years old and considered typical in this region.

The measurements were performed in the rock mechanics laboratory of Kőmérő Ltd.~ with the help of a hydraulic instrument (maximal compression is 150kN), HBM C6A (1MN) load sensor and a HBM Spider 8 \& CatmanEasy collected the data. The diameter and the lengh of the cylindrical sample was $37.99$~mm, and $78.33$~mm and its mass was $0.2213$~kg. The uploading speed was $0.7$~kN/s in every cycles. A hysteresis type measurement with increasing stress amplitudes was chosen to determine rheological parameters. In our calculations only the uploading parts of the last two cycles are considered.

\begin{figure}
\centering
\includegraphics[width=0.45\textwidth]{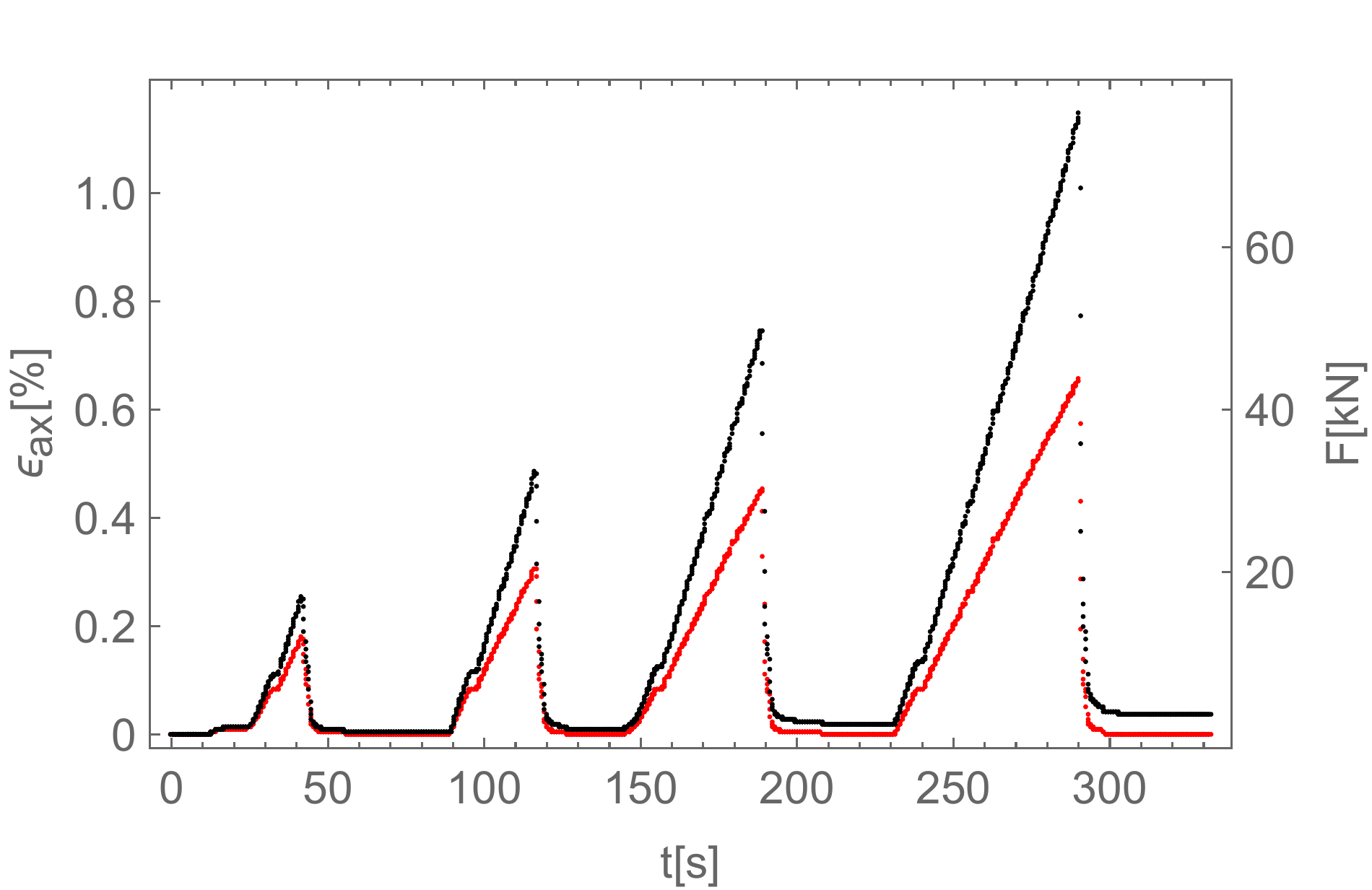}
\includegraphics[width=0.45\textwidth]{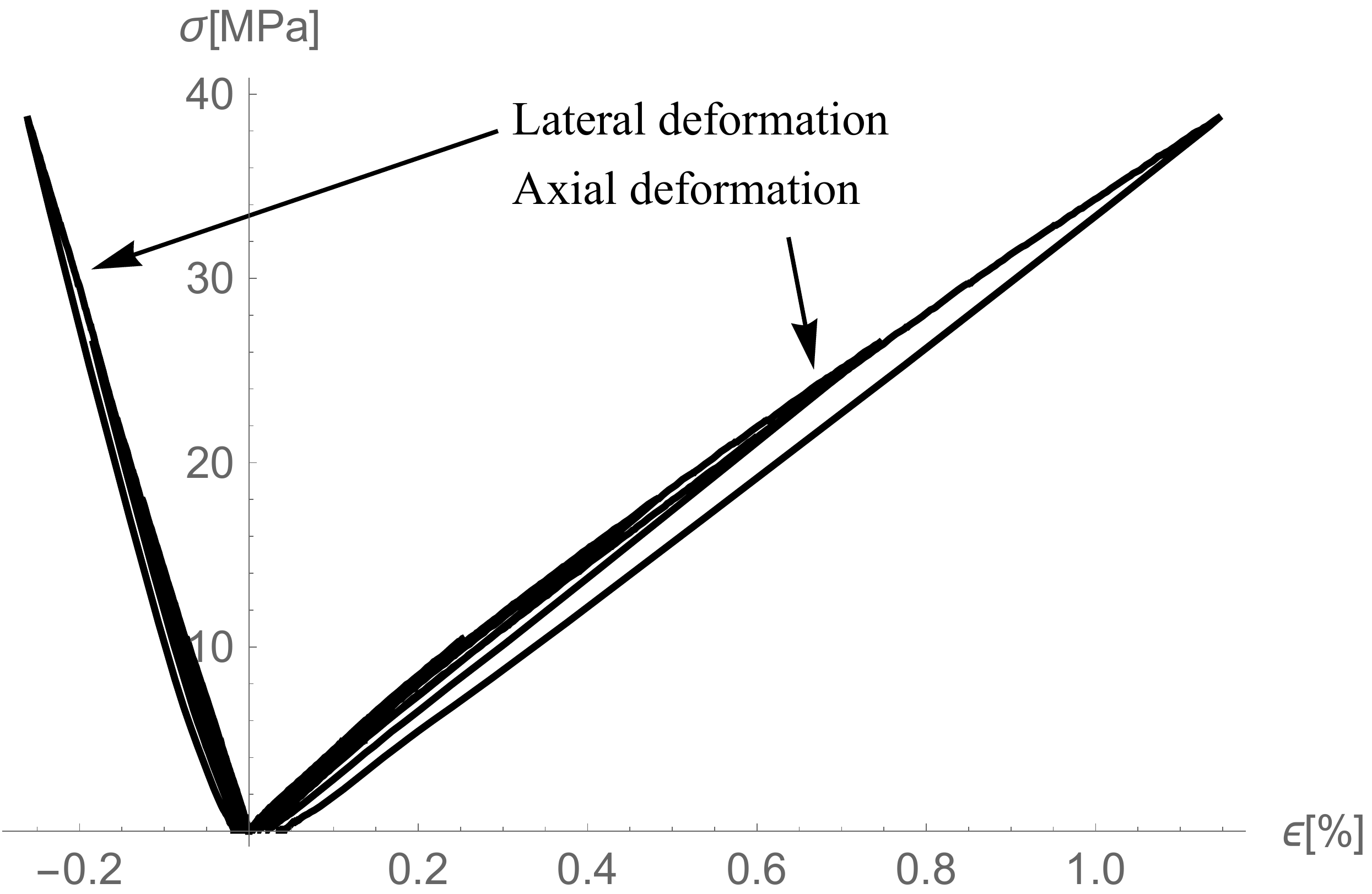}
\caption{\label{fig_fdtime} Left panel: Axial deformation (black) and loading force (red) with constant uploading speed as function of time. Right panel: Axial stress as function of the axial and lateral deformations (positive and negative horizontal axes, respectively) with hysteresis curves of rheological origin.}
\end{figure}

The force and the axial deformation are shown on the left side of Fig. \ref{fig_fdtime} as functions of time. The right side of Fig.~\ref{fig_fdtime} shows the axial stress as function of axial and lateral deformations at the positive and negative horizontal axes, respectively. The rheological hysteresis and also the apparent permanent deformation is visible at the end of creep periods. Therefore andesitic rocks from M\'atra show clear deviation from ideal elasticity with properly designed laboratory experiments.

\subsubsection{Kluitenberg-Verh\'as body and static-dynamic elasticity}

The isotropic Kluitenberg-Verh\'as body is given by the following relations between the stress $\sigma$ and strain $\epsilon$:
\begin{eqnarray}
\tau_d \dot \sigma_d +\sigma_d &=& E_{2d} \ddot \epsilon_d +  E_{1d} \dot \epsilon_d + E_{0d} \epsilon_d, \label{KVd}\\
\tau_s \dot \sigma_s +\sigma_s &=& E_{2s} \ddot \epsilon_s +  E_{1s} \dot \epsilon_s + E_{0s} \epsilon_s. \label{KVs}
\end{eqnarray}

Here the dot denotes time differentiation, the subscripts $d$ and $s$ refer to the deviatoric and spherical stresses  and deformations as well as to the related material parameters, like the $\tau_d$ and $\tau_s$ deviatoric and spherical relaxation times. In our case a cylindrical laboratory sample was prepared according to ISRM (International Society of Rock Mechanics) standards, therefore the uniaxial stress, $\sigma$, and the axial and lateral deformations $\epsilon_a$ and $\epsilon_l$ determine the spherical and deviatoric stresses and strains as $\sigma_d = 2\sigma/3$, $\sigma_s = \sigma/3$, $\epsilon_d = 2 (\epsilon_a-\epsilon_l)/3$ and $\epsilon_s = (\epsilon_a+2\epsilon_l)/3$.

The elastic moduli $E_{0d}=2G$ and $E_{0s}=3K$ are the well known static Lamé coefficients and $E_{1d}$ and $E_{1s}$ are the deviatoric and spherical viscoelastic material coefficients. If the $E_{2d}$ and $E_{2s}$ dynamic parameters are neglectable, then the Kluitenberg-Verh\'as body simplifies to the Poynting-Thomson-Zener body \cite{AssEta15a}, which is the standard model of creep and relaxation phenomena in rock mechanics (also with the names generalized Kelvin-Voigt and Hill-Maxwell body \cite{AydEta14a}). Then it is convenient to transform equations \re{KVd} and \re{KVs} into a hierarchical form:
\begin{eqnarray}
\tau_d \dot \sigma_d +\sigma_d -  E_{1d} \dot \epsilon_d - E_{0d} \epsilon_d &=&
	\tau_d \frac{\rm d}{{\rm d}t}\left(\sigma_d -  b_d E_{0d} \epsilon_d\right) +\sigma_d - E_{0d} \epsilon_d, \label{PTd}\\
\tau_s \dot \sigma_s +\sigma_s -  E_{1s} \dot \epsilon_s - E_{0s} \epsilon_s &=&
		\tau_s \frac{\rm d}{{\rm d}t}\left(\sigma_s -  b_s E_{0s} \epsilon_s\right) +\sigma_s - E_{0s} \epsilon_s. \label{PTs}
\end{eqnarray}
 Here the parameters $b_d = E_{1d}/(\tau_d E_{0d})$ and $b_s = E_{1s}/(\tau_s E_{0s})$ measure the deviation from the ideal elastic Hook body. If $b_d=b_s=1$, then the material is apparently completely elastic. The propagation speed of longitudinal and transversal waves determine the material parameters, $E_{d,dyn} = b_d E_{0d}$ and $E_{s,dyn}=b_s E_{0s} $, respectively. These are called dynamic Lamé coefficients.  Both for the static and dynamic cases the Young modulus and the Poisson coefficient are calculated from the Lamé coefficients as $Y = 3E_sE_d/(2E_s+E_d)$ and $\nu = (E_s-E_d)/(2E_s+E_d)$.

\subsubsection{Material parameters}

One can determine the static and dynamic elastic moduli both from the cyclic loading of the sample given above and the dynamic ones can be determined from direct laboratory measurements of the sound speeds. This direct laboratory measurement of the propagation speed of longitudinal and transversal waves gives the dynamic Young modulus $(38.6 \pm 1.1)$~GPa, and the dynamic Poisson coefficient $0.18 \pm 0.01$.

Furthermore, the time dependent data of the chosen rock sample was analysed using the differential equations of the rheological models. In particular we have assumed, that the measured deformation values are given and we have determined the best parameters from the differential equations \re{KVd} and \re{KVs} of the Kluitenberg-Verh\'as body to obtain the stress, both for the deviatoric and spherical components. According to these calculations the coefficents of the second derivatives of the deformation, the $E_{2d}$ and $E_{2s}$ coefficients, can be neglected and the Poynting-Thomson-Zener model, \re{PTd} and \re{PTs},  was applied. The obtained best fit parameters are shown in Table \ref{tab_reopar}.
\begin{table}
\centering
\begin{tabular}{c|cccc}
Cycle 3 up& $E_d = 2G$ [GPa] & $E_s = 3K$ [GPa] & $E_{Young}$ [GPa] & $\nu$ \tabularnewline
\hline
static  & $23\pm 2$ & $33.9\pm 0.5$ & $25\pm 2$ & $0.13\pm0.03$  \tabularnewline
\hline
dynamic & $32\pm 4$ & $73\pm 1$ & $39\pm 4$ & $0.23\pm0.03$ \tabularnewline
\hline
Cycle 4 up&  &  &  &  \tabularnewline
\hline
static & $17\pm 5$ & $50.1\pm 0.3$ &$ 22\pm 5$ & $0.28\pm 0.05$\tabularnewline
\hline
dynamic & $ 27\pm 5$ & $58\pm 3$ & $33\pm5$ & $0.22\pm 0.05$ \tabularnewline
\hline
\end{tabular}
\caption{\label{tab_reopar} The static and dynamic elastic coefficients according to rheological data evaluation of the uploading parts of the 3\textsuperscript{th} and 4\textsuperscript{th} cycles of the laboratory experiment. The dynamic Young modulus and dynamic Poisson ratio $\nu$ are consistent with the values obtained from propagation speeds of the P and S waves.}
\end{table}

\begin{table}
\centering
\begin{tabular}{c|cc}
    & $b_d$ & $b_s$ \tabularnewline
\hline
Cycle 3 & $1.4\pm 0.2$ & $2.15\pm 0.05$ \tabularnewline
\hline
Cycle 4 & $1.6\pm 0.6$ & $1.16\pm 0.06$  \tabularnewline

\end{tabular}
\caption{\label{tab_bpar} The rheological parameter, $b=\frac{E_1}{E_0\tau}$, characterising the deviation from the ideal elastic regime both in the deviatoric and spherical cases.}
\end{table}

\begin{figure}
\centering
\includegraphics[width=0.5\textwidth]{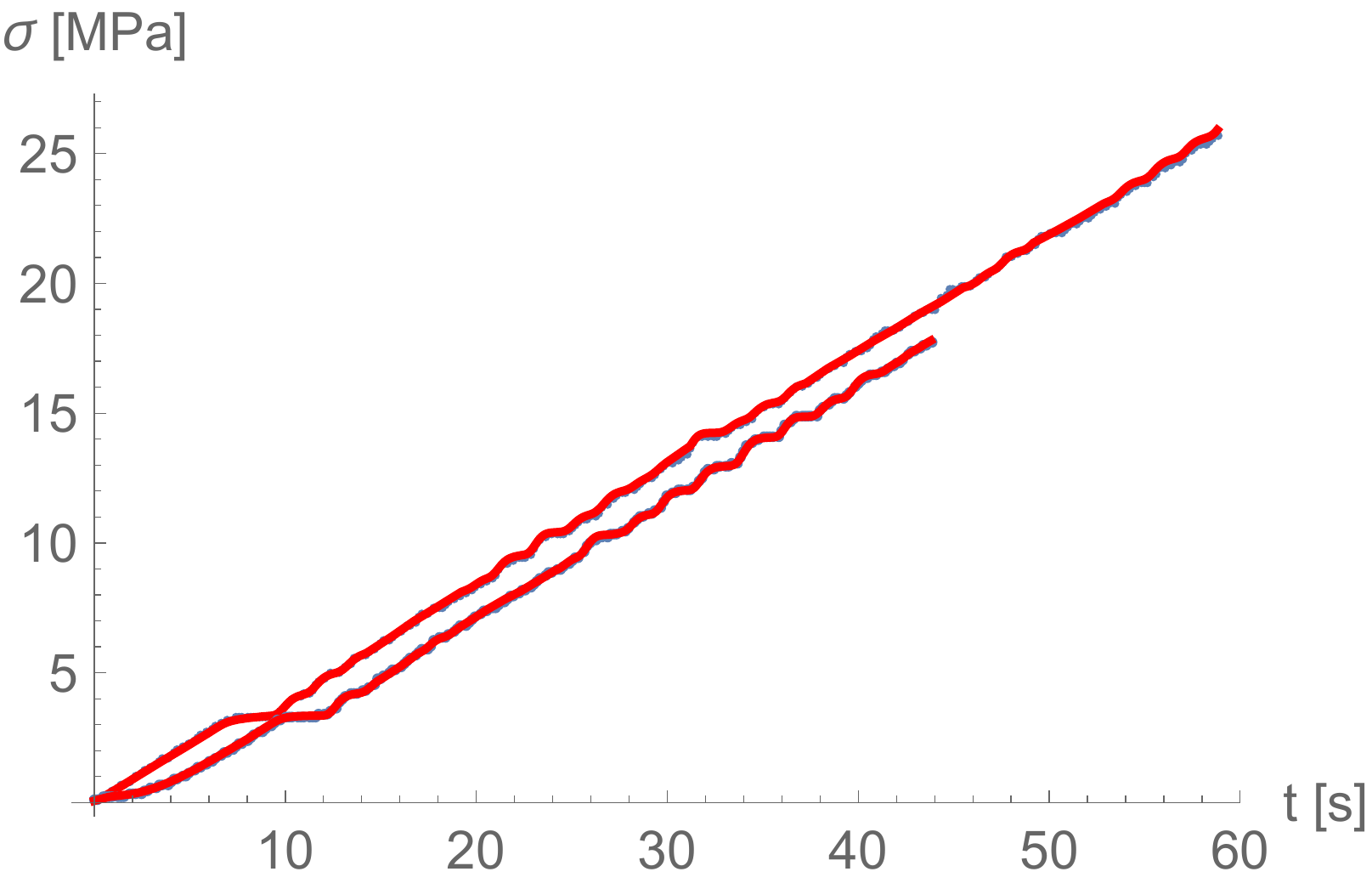}
\caption{\label{fig_devmes} The deviatoric stress as the function of time in the uploading part of the third and fourth loading cycles. The the blue data points are under the solid red fitted model solution curve.}
\end{figure}

Fig. \ref{fig_devmes} shows the measured stress values and also the stress determined from the rheological model with the parameters above for the uploading part of the third and fourth cycles in the deviatoric case. The deviation of the  two curves from the data is below 0.05 MPa in the whole domain.

According to our measurements the typical gray andesite of M\'atra is not ideally elastic and the deviation from elasticity is of rheological origin. The experimental parameters can be obtained both from the wave propagation speed measurements and from the cyclic loading experiments. These coefficients are consistent as it is shown by our preliminary calculations. The obtained transition regime, which is proportional to the inverse relaxation times, is 0.02-0.1Hz, which is below the low frequency sensitivity of ET. However, our experimental methodology prevents the detection of faster relaxation modes, therefore further investigations are necessary.

\section{Seismological measurements I.}\label{sec:seismo}

The ET site selection preparation measurements \cite{ETdes11r,Bek13t,BekEta15a} selected three best candidate sites for the Einstein Telescope in Europe. The selection is based on the low seismic noise level in the critical low frequency range 1-10Hz. The survey investigated several underground locations, each of them about for a week. In order to have a better estimate of the average seismic noise it is reasonable to collect at least two years of data and analyse the annual and seasonal changes in noise level originating from natural sources and human activity. The above mentioned survey was the main motivation in the establishment of MGGL. Due to the ongoing reclamation activity in the Gy\"ongy\"osoroszi mine the vicinity of the laboratory was also subject of maintenance works. Therefore in our case both internal and external human activities were contributed to the noise level.

Seismological data collection was performed by two Guralp CMG 3T low noise, broadband seismometers \cite{BarEta17a}, and also by the custom made seismic sensor developed in the Warsaw University. The Guralp instruments are sensitive to ground vibrations with flat velocity response in the frequency range 0.008-50 Hz. The Guralp seismometers were calibrated and cross calibrated, operated according to protocol of the seismometers in the Hungarian National Seismological Network. The self noise of the seismometers was below the low noise model from 0.02 Hz to 10 Hz \cite{Guralpmanual}.

The custom made Warsaw seismometer uses one vertical and two horizontal geophone sensors mounted firmly in a single aluminium block placed inside a metal housing along with a data acquisition system. The geophones (LGT-2.5 and LGT-2.5H) used as the sensors have the lower corner frequency of 2.5 Hz. The data acquisition system sampled the analog signal with the frequency of 125 Hz and the 32-bit resolution. The Warsaw seismometer was calibrated by comparing the data with a Trillium seismometer, and also with a Guralp CMG 3T during a data calibration session at the MTA Wigner Research Centre for Physics in January 2015.

\begin{figure}
	\centering
	\includegraphics[width=0.48\textwidth]{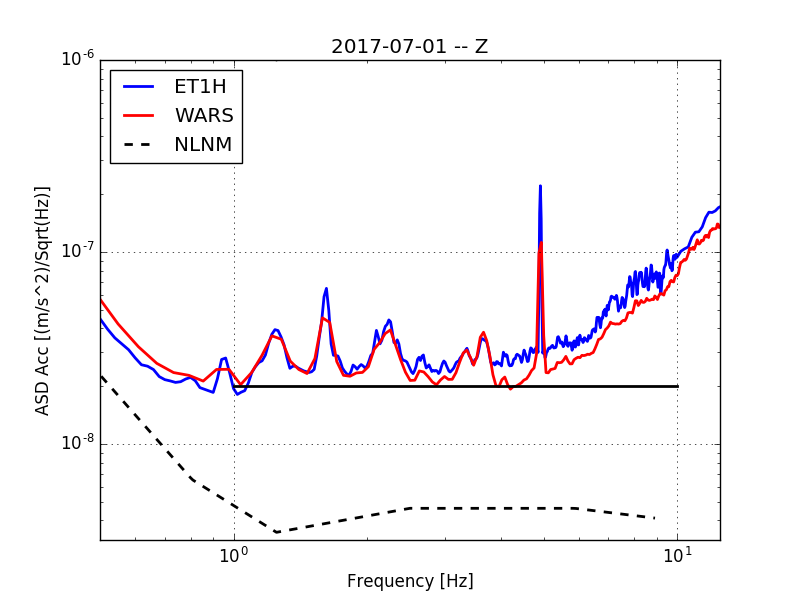}
	\includegraphics[width=0.48\textwidth]{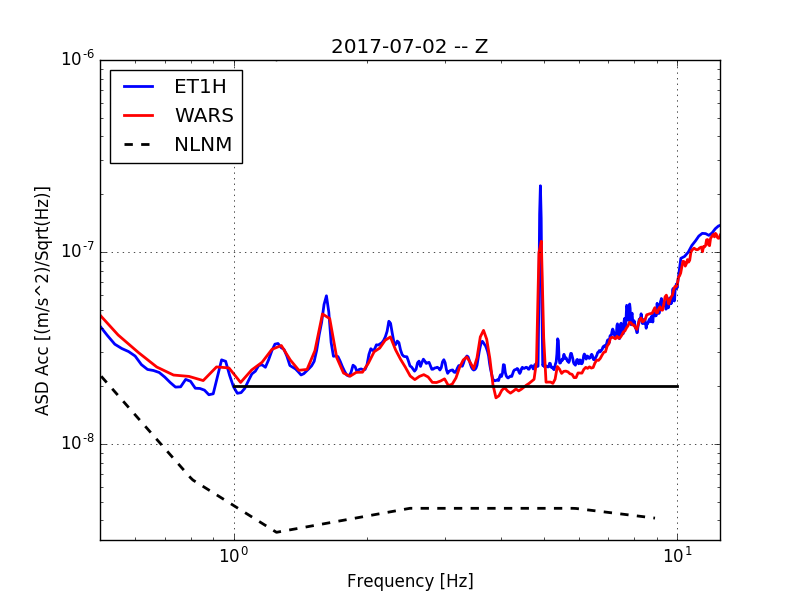}

	\includegraphics[width=0.5\textwidth]{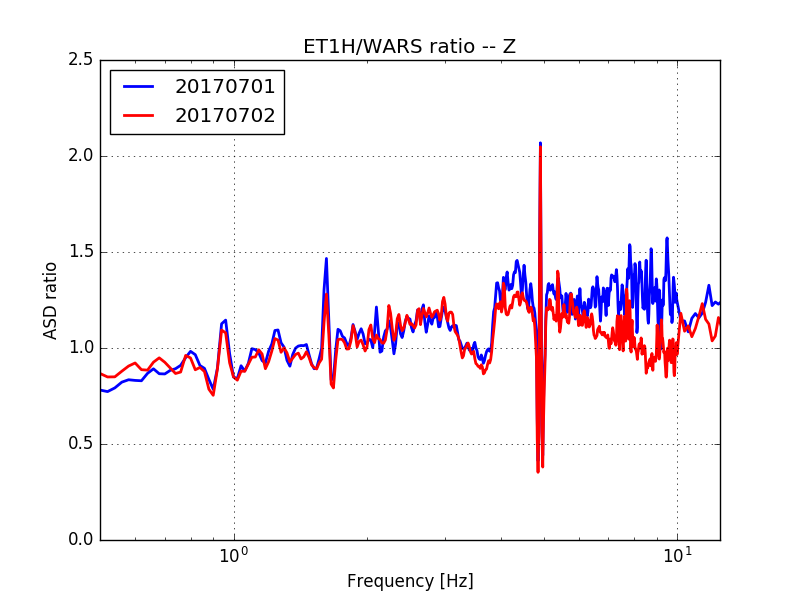}
	\caption{\label{fig-crosscal-seismo}The daily average ASDs of 2017-07-01 and 2017-07-02 from the data of ET1H and the Warsaw seismometer, Z direction. The last figure shows the ratio of ASDs for the two studied days. }
\end{figure}

One of the Guralp instruments (ET1H) and the Warsaw seismometer (WARS) were permanently installed in the MGGL. The seismometers were deployed on separated concrete piers which were connected to the bedrock. Between the piers and the seismometers a granite plate has been placed. The other Guralp instrument (hereafter GU02) was used in a measurement campaign in the first two weeks of June 2017 in a measurement cabin, constructed next to the main tunnel and prepared for seismometer installation. The mutual performance of the ET1H and WARS seismometers are demonstrated on Fig. \ref{fig-crosscal-seismo}, where the averages of the day 2017-07-01 and 2017-07-02 are calculated for each instruments and also the ratio of the averages at given frequency range. The agreement is similar to other customary chosen days during the measurement campaign. Therefore the cross-calibration of the instruments is satisfactory.

The data acquisition period of the two instruments slightly differs due to operational problems, however the amount of the collected data is similar. In the following first we analyse the ET1H data, then the WARS data. The methods and the elaboration are different, in order to show more aspects of the noise measurements.

\subsection{Run-1 data analysis of the ET1H seismometer}

The data collection period for ET1H was started on 2016-03-01 and in this paper we elaborate the data until 2018-06-28. From the point of view of ET related data the instrument was out of order in the periods 2016-09-09 --- 2016-09-18, 2016-10-13 --- 2016-10-16, 2016-10-19 --- 2016-11-16, 2017-12-16 --- 2018-01-31, 2018-03-14 --- 2018-04-04 and 2018-04-13 for ET1H. The Z direction data was compromised between 2016-09-22 ---2016-10-16, for 25 days. Therefore \dnum days of data was collected for the horizontal directions and 716 days in the vertical direction. In our analysis we followed the data processing method of Ref. \cite{BekEta15a}. The particular problems observed in long term data analysis suggested some methodological improvements, and additional characteristics. These were reported in Ref. \cite{SomEta18m} where the precise definitions and the justification of the methodology of data processing can be found. In this section some basic definitions are recalled for clarification.

The velocity Power Spectral Density (PSD) is defined as \cite{BekEta15a,SomEta18m}
\begin{equation}
	P^{(v)}=\frac{2}{f_{s}\cdot N\cdot W}\left|V_{k}\right|^{2},
\end{equation}
where $f_{s}$ is the sampling rate --- in our case it is $100$ Hz ---, $N$ is the length of the analysed data sample --- $ N\,=\,5000 $ ---, and $W=\frac{1}{N}\sum_{n=1}^{N}w[n]^{2}$ with the Nuttall window function $w[n]$. {We did not use the advantage of fast Fourier algorithm on the expense of increasing the lowest frequency value.} The coefficients $V_{k}=F(w[n]\cdot (v[n]-\langle v\rangle)$, represent the Fourier transform of the deviation of raw velocity data $v[n]$ from its average value $\langle v\rangle$. In our analysis PSD-s were calculated with 50 s data samples with $f_{s}=100$~Hz sampling rate. Before further processing, raw data were highpass-filtered with $f_{HP}=0.02$ Hz and the overlap is 3/4 due to the window function.
To characterize spectral properties the acceleration Amplitude Spectral Density (ASD),
\begin{equation}
A^{(a)}=A=\sqrt{P^{(v)}\cdot\omega^2}\label{eq:dASD}
\end{equation}
will be used and displacement $ rms $ will be applied as cumulative property. This is the square root of the integral of displacement PSD between two frequency values
\begin{equation}
rms = \sqrt{\frac{1}{T}\sum_{k=l}^{K}P_{k}^{(v)}/\omega^2},
\end{equation}
where $l$ is the low cut-off index --- in Beker's paper this is chosen to be $ 2$ Hz ---, the $ K $ is the high cut-off index --- usually the Nyquist frequency  --- and $T=\frac{N}{f_{s}}$. In our case $l$ is chosen to be $ 1$ Hz or $ 2$ Hz and $K$ is either the Nyquist-frequency or $ 10$ Hz, because we use $rms_{2Hz}$ of Beker and also calculate the $rms_{2-10Hz}$ and $rms_{1-10Hz}$ values. With these new measures we can drop the irrelevant frequency interval above $ 10$ Hz and consider the technologically already available $ 1-2$ Hz region   \cite{SomEta18m}. The commonly used comparative measure for the spectral properties of the sites is a particular value of the amplitude spectral density, the so called Black Forest line:
\begin{equation}
	A_{BF} = 2\cdot10^{-8}{\rm \frac{m/s^{2}}{Hz}} \label{eq:ETReqv}.
\end{equation}
It is worth to give the various rms values corresponding to the the reference spectral density represented by the Black Forest line:   $rms_{2Hz}^{BF} \approx 0.1\,nm$   $rms_{2-10Hz}^{BF} = 0.1\,nm$  and $ rms_{1-10Hz}^{BF} =0.29$ nm.

An other important aspects of the long term data evaluation are the use of percentiles and the choice of intermediate averaging periods. Previous studies applied the mode of the data for the representation of a characteristic mean value. However, the median, and also the other percentiles, are less sensitive for discretisation and the inevitable averaging. Moreover, naturally select the representative data without a necessity of filtering short large noise bursts. Therefore in the following the analysis of site properties is mostly based on the median of the data. However, we will demonstrate the most important differences in the spectral representation and also calculate the mode related $rms_{2Hz}$ of Beker for a clear comparison with the previous studies.

It is also worth to mention, that the basic Fourier length used at the Fourier transformation was 50 s for Guralp instruments and 128s for WARS. For the ET1H and GUO2 Guralp instruments we have introduced a short time averaging (STA) period of $ 300\,s $ and for the two-year data the use of daily averages was convenient. The daily periods were called intrinsic averaging (INA), because it considers the natural periodicity of the data. In other words in order to obtain the long term $ 10\textsuperscript{th}, 50\textsuperscript{th} $ and $ 90\textsuperscript{th} $ percentiles first the short therm averages (STA) were calculated and then particular chosen periods in each day (INA) enable the comparison of working hours, night time or whole day data. See \cite{SomEta18m} for more details about the chosen methodology.

\subsection{Long term seismic results}
A particular factor in our noise analysis is the ongoing mine reclamation activity in the Gy\"ongy\"osoroszi mine. As a result, the investigation and identification of various noise sources originating from external and internal human activity, machine noise, construction works, train noise, etc. proved to be difficult. In November 2016 a three-shift operation period has been started with increased industrial noise, present also during the nights. In order to compare the noise types of these kind of activities in our analysis we have defined three periods for each day for our study: (a) the whole day, (b) night period (20:00 - 2:00 UTC) and (c) working period (9:00 - 15:00 UTC). In the following we present a comparative analysis of long term data considering seasonal changes, external and internal human noise and also depth dependence for a shorter two weeks long measurement campaign performed by our second Guralp instrument, GUO2, at -404 m depth.

\subsubsection{Complete Run-1 results\label{subsec:Yearly-Whole-period-results}}

\begin{figure}
	\centering
	\includegraphics[width=.48\textwidth]{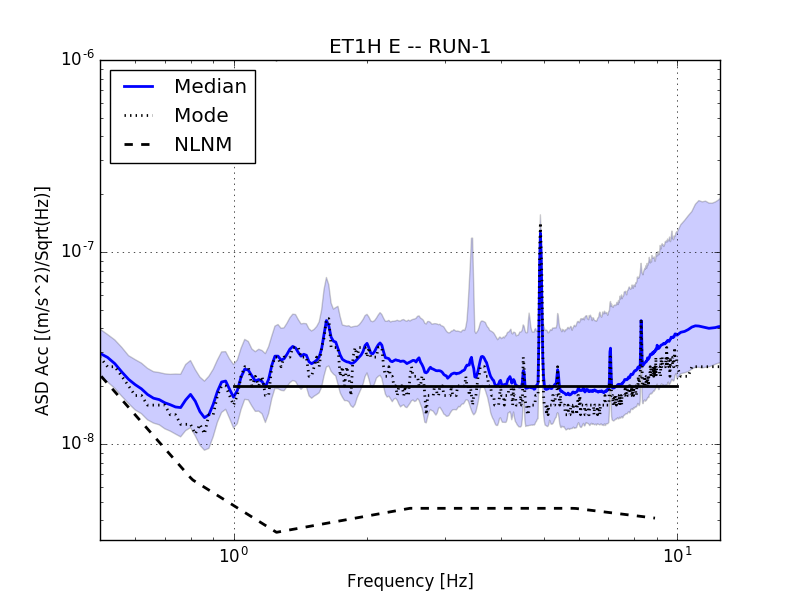}
	\includegraphics[width=.48\textwidth]{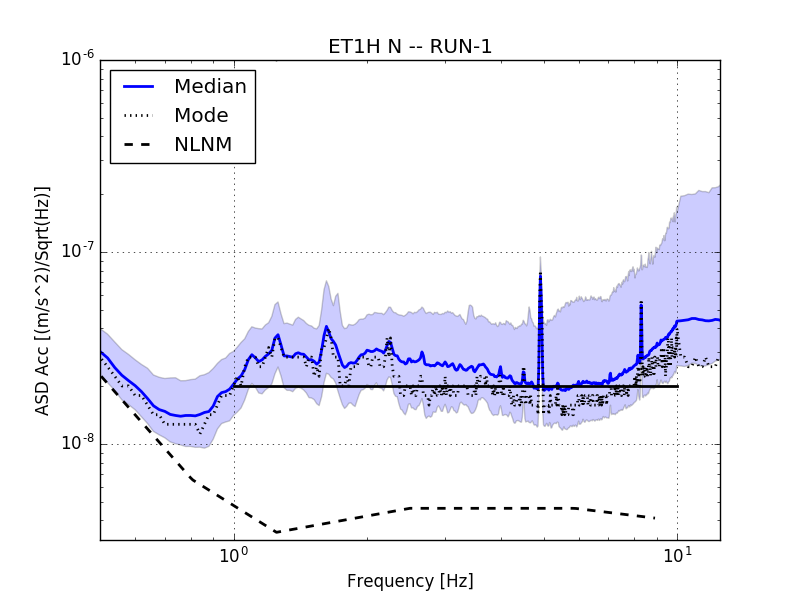}

	\includegraphics[width=.48\textwidth]{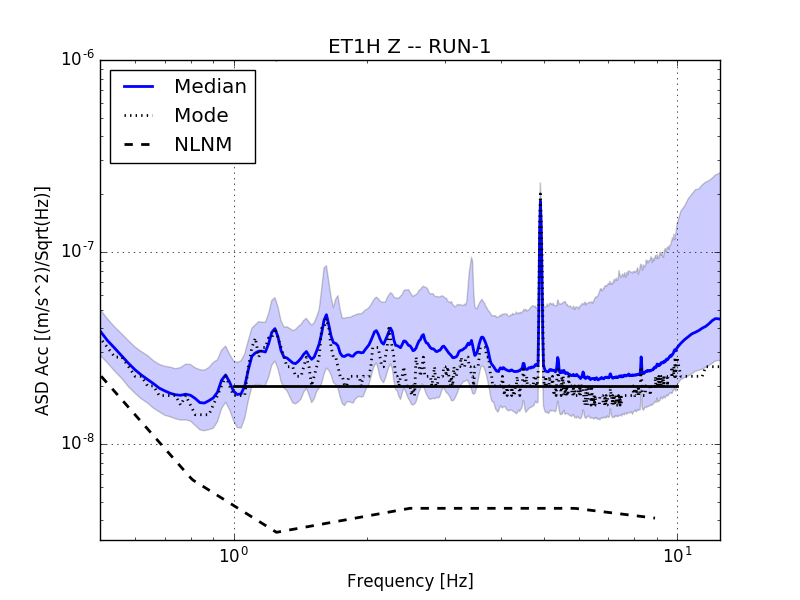}
	\caption{\label{fig:Whole-ASD} The acceleration ASD spectra of the East-West (ET1H E), North-Sud (ET1H N) and in the vertical directions (ET1H Z). The solid blue and the dotted black lines indicate the median and the mode of the the data, repectively. The borderlines of the blue colored area are the 10\textsuperscript{th} and 90\textsuperscript{th} percentiles. The Black Forest line is solid black and the dashed black lines show the New Low Noise Model of Peterson (NLNM) \cite{Pet93r}.}
\end{figure}

The acceleration ASD-s for the two-year observation period are shown in Fig. \ref{fig:Whole-ASD}. The borderlines of the blue colored area are the 10\textsuperscript{th} and 90\textsuperscript{th} percentiles of daily $300$ s data. The dotted black lines are the modes of the $1800$ s averages, according to the methodology in Ref. \cite{Bek13t}. The medians and modes are closer to the Black Forest level in horizontal directions. It is also remarkable that the mode underestimates the median, that is most of the days are noisier than the mode. at a given frequency. This is well represented in the corresponding $rms$ values given  in Tab. \ref{tab:Whole-rms}, too. Here the first line $ rms_{2Hz} $, calculated from the mode of the data. It is worth to recall that it was $ 0.12$ nm  in the previous short term measurements  of Beker in the same place. As it can be seen, the $ rms $ calculated from the median is much larger, about $ 20\% $ more. It is also worth to mention that the Black forest line reference value of $rms_{1-10Hz}$ is $0.29$ nm, and the $rms_{1-10Hz}$ normalized to this value is less than the $rms_{2-10Hz}$, therefore the site is less noisy at the lower frequencies, as it can be seen in Figs. \ref{fig:Whole-ASD}, as well.

%In our case the one-week and the two-year data have the same value.

\begin{table}
	\centering
	\begin{tabular}{|c|c|c|c|}
		\hline
		Whole day & East & North & vertical \tabularnewline
		\hline
		Mode $rms_{2Hz}$ &0.123 & 0.121 & 0.140\tabularnewline
		\hline
		Median $rms_{2-10Hz}$ & 0.144 & 0.147 & 0.176\tabularnewline
		\hline
		Median $rms_{1-10Hz}$ & 0.387 & 0.417 &0.436\tabularnewline
		\hline 	
	\end{tabular}
	\caption{\label{tab:Whole-rms}The various $rms$ values in nm of Figs. \ref{fig:Whole-ASD} are shown. The reference value for the Black Forest line is $0.29$ nm for $rms_{1-10Hz}$.}
\end{table}

\begin{figure}
	\centering
	\includegraphics[width=.48\textwidth]{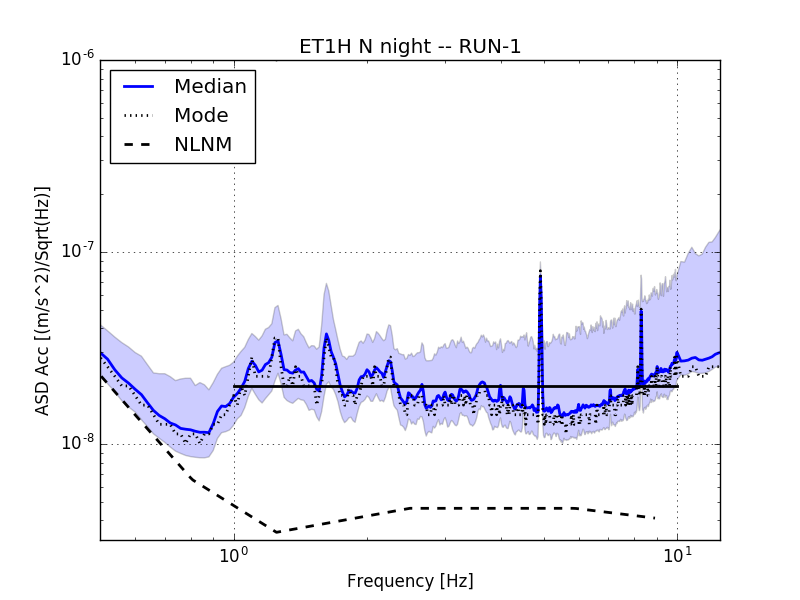}
	\includegraphics[width=.48\textwidth]{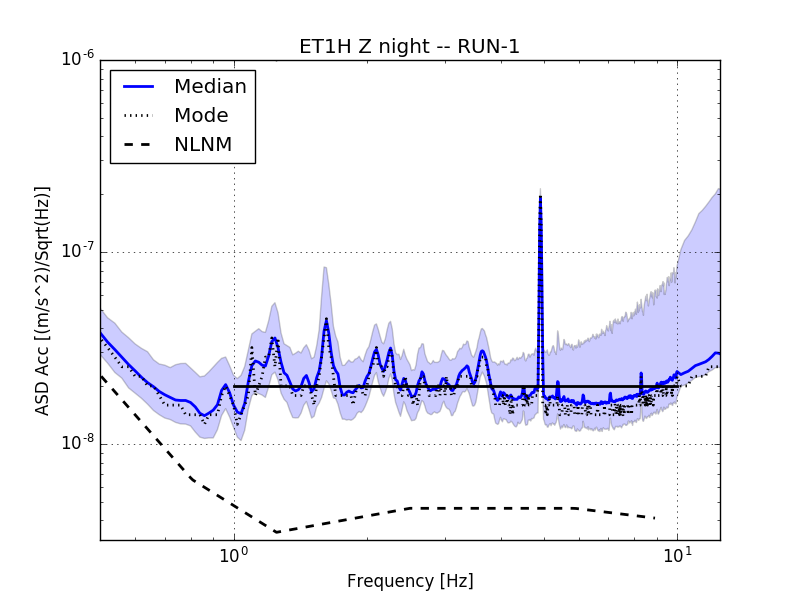}

	\includegraphics[width=.48\textwidth]{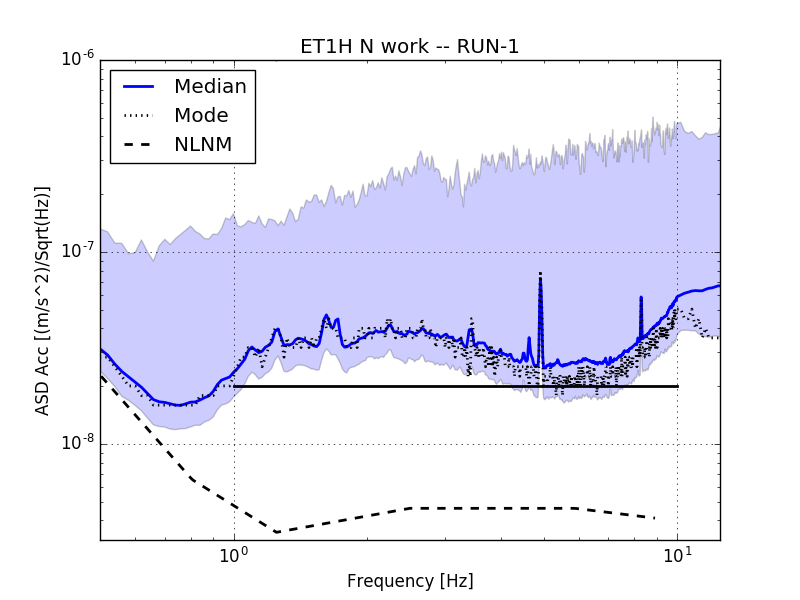}
	\includegraphics[width=.48\textwidth]{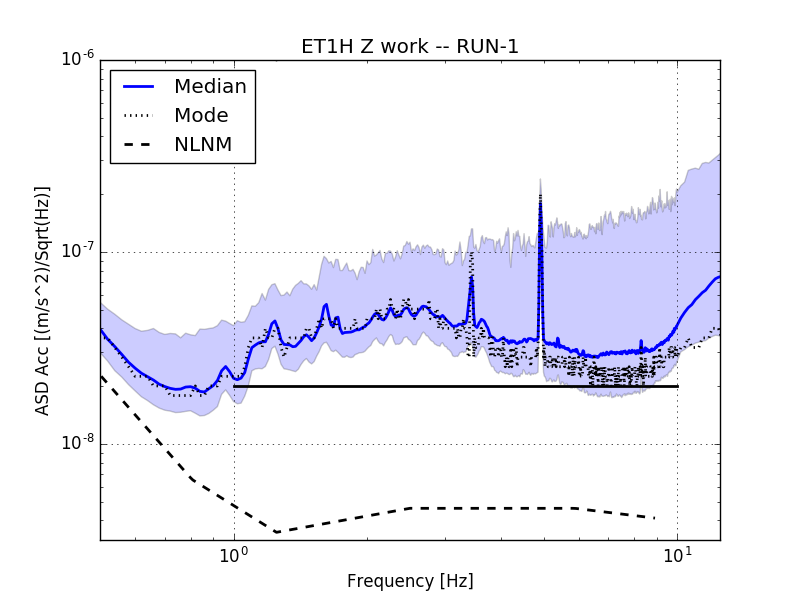}
	\caption{\label{fig:Whole-night-working-ASD}The acceleration ASD-s of the working and night periods of 741 days data for a horizontal and the vertical directions. The solid blue and the dotted black lines indicate the median and the mode of the the data, repectively. The borderlines of the blue colored area are the 10\textsuperscript{th} and 90\textsuperscript{th} percentiles. The Black Forest line is solid black and the dashed black lines are NLNM curves.}
\end{figure}

The role of the human and industrial noises is shown in Fig. \ref{fig:Whole-night-working-ASD}, where the spectral densities for the working and night periods is plotted in the North-South and the vertical directions. It is remarkable, that half of the frequency range is below the median of the data (blue line), the asymmetric relative position of the blue area at working period indicates the presence of short noisy periods between 9:00 - 15:00 UTC. We have to mention that from the end of 2016 the reclamation works were performed close to the MGGL in a three-shift schedule.

%For the 88 m deep station we can say that in the night period and for an average day the mode is almost everywhere lower than the Black Forest one, therefore it can be suitable for ET if we focus on the $ 2-10$ Hz interval \cite{BekEta15a}. Recall the three-shift working from the end of 2016, it should be much better if the reclamation will be finished.

In order to illustrate the cultural and industrial noises the frequency dependence of the ratio of the working and night periods are plotted in Fig. \ref{fig:Whole-night-working-ASD}. The cultural noises start at about $0.7$ Hz and reach their maxima  between $ 2-3$ Hz and $ 10-20$ Hz.
%\footnote{In Sec. \ref{subsec:2-weeks-results} we will show that GU02 station ($ -404\,m $) is much better in the $ 1-2$ Hz  interval than the ET1H.}
We may suppose that the night shifts cause less noises, the main works are done during the daytime. Then the observed noise level at the night periods can be considered as an upper limit for an operating underground GW detector facility, where the equipments are optimized for a low noise operation.
\begin{figure}
	\centering
	\includegraphics[width=0.8\textwidth]{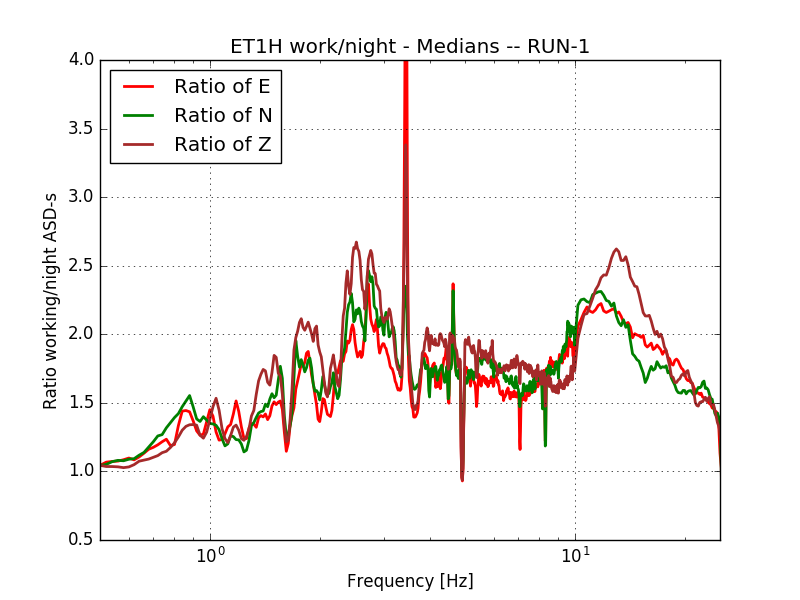}
	\caption{\label{fig:Whole-work-night-ratio} The frequency dependence of the ratios of the night and work period medians in the three directions. }
\end{figure}

The corresponding $ rms $ values are for the night and working periods are given  in Table (\ref{tab:Whole-night-rms}. We can see, that the $ rms $ of the noisier working periods can be two times higher than in the calm night ones.
\begin{table}
	\centering
	\begin{tabular}{|c|c|c|c|}
		\hline
		Night period & East & North & Z \tabularnewline
		\hline
		Mode $rms_{2Hz}$ &0.101 & 0.102 & 0.123\tabularnewline
		\hline
		Median $rms_{2-10Hz}$ & 0.114 & 0.115 & 0.128\tabularnewline
		\hline
		Median $rms_{1-10Hz}$ & 0.324 & 0.354 &0.353\tabularnewline
		\hline
	\end{tabular}
	\begin{tabular}{|c|c|c|c|}
		\hline
		Working period & East & North & Z \tabularnewline
		\hline
		Mode $rms_{2Hz}$ &0.196 & 0.189 & 0.240\tabularnewline
		\hline
		Median $rms_{2-10Hz}$ & 0.193 & 0.147 & 0.241\tabularnewline
		\hline
		Median $rms_{1-10Hz}$ & 0.456 & 0.490 &0.527\tabularnewline
		\hline
	\end{tabular}
	\caption{\label{tab:Whole-night-rms}The $rms$ values in nm for the night and working periods in Run-1, according to the panels of Fig. \ref{fig:Whole-night-working-ASD}. }
\end{table}

\begin{figure}
	\centering
	\includegraphics[width=0.48\textwidth]{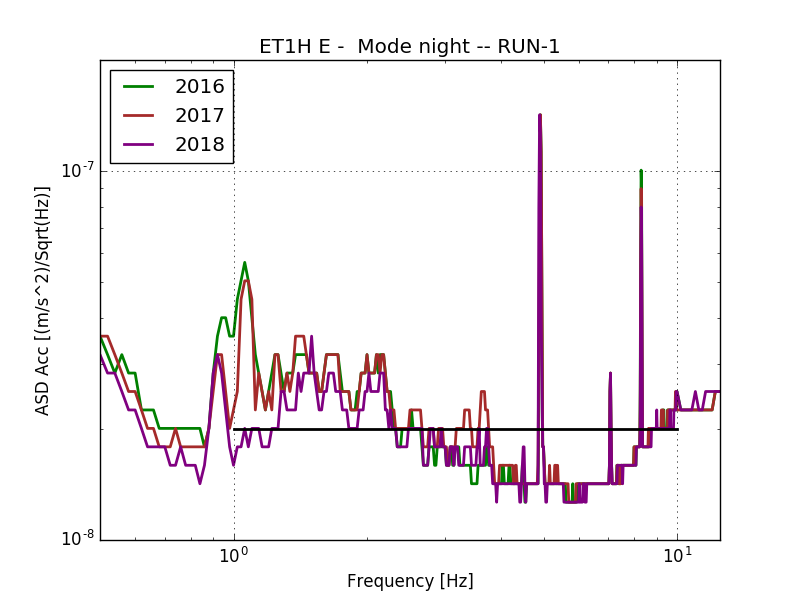}
	\includegraphics[width=0.48\textwidth]{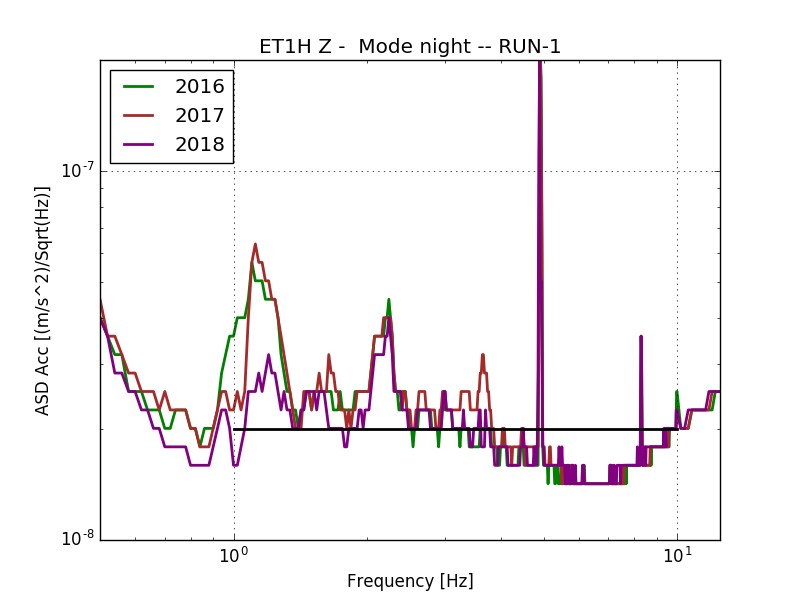}
	\includegraphics[width=0.48\textwidth]{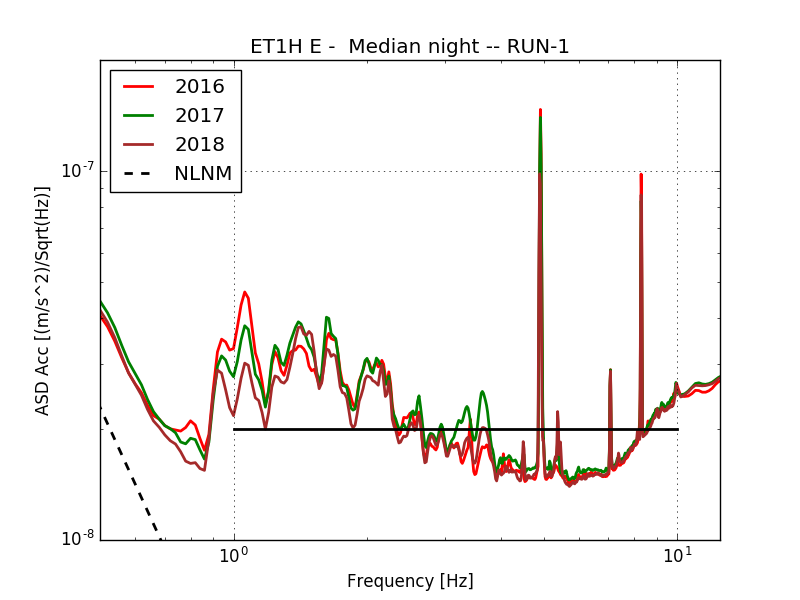}
	\includegraphics[width=0.48\textwidth]{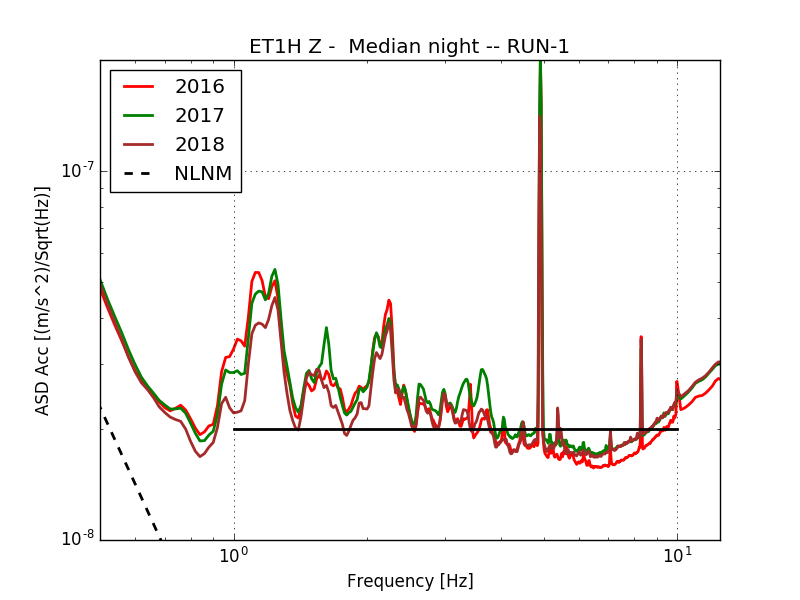}
	\includegraphics[width=0.48\textwidth]{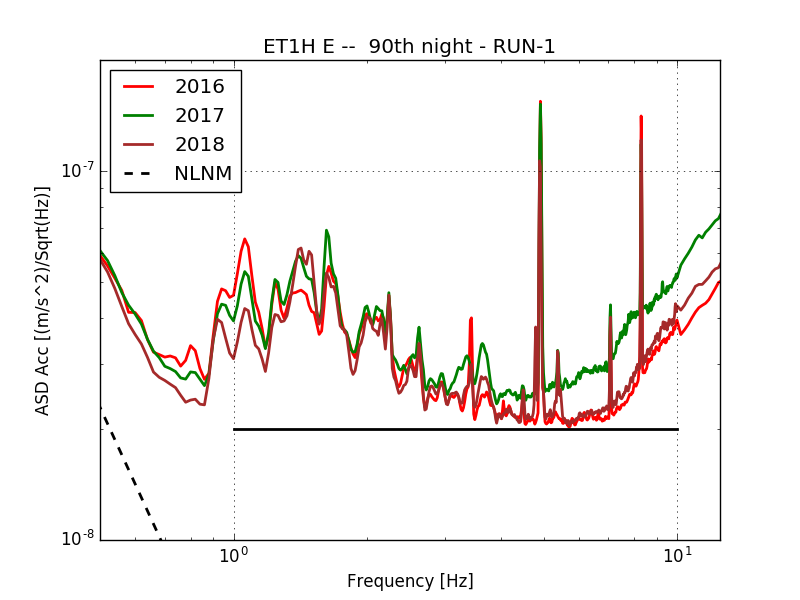}
	\includegraphics[width=0.48\textwidth]{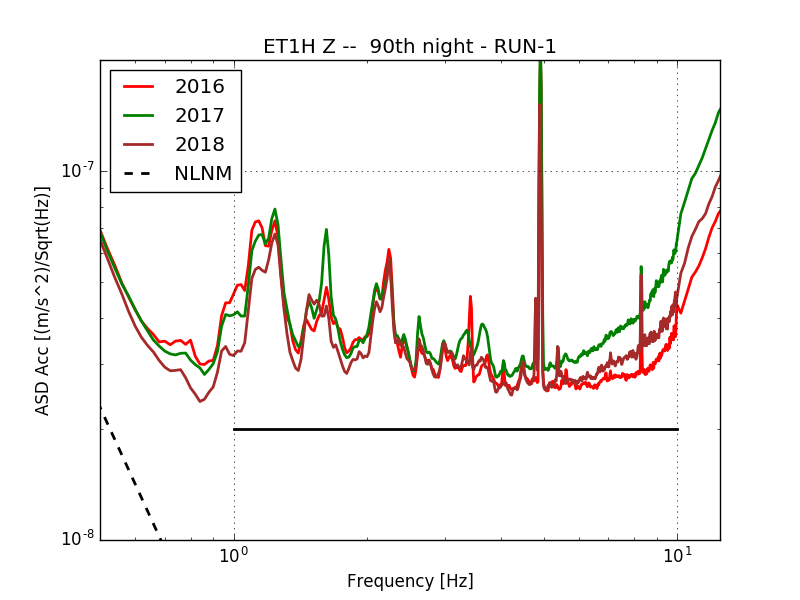}
	\caption{\label{fig:Year-night-ASD} The acceleration ASD-s for the years 2016, 2017 and 2018. The upper panels show the modes, the two figures in the middle row is for medians and the bottom graphs  are the $ 90\textsuperscript{th} $ percentiles for each year. The Black Forest line is solid black and the dashed black lines are NLNM curves.}
\end{figure}

%Let us recall, that the $rms_{2Hz}$ value of horizontal noise level in this depth was $0.12 nm$ in the evaluation of the former, short term measurements  \cite{Bek13t,BekEta15a,ETdes11r}.
The differences at longer periods in Run-1 were analysed by three different methods. First, the acceleration ASD-s of the night periods were plotted for each year in Fig. \ref{fig:Year-night-ASD}. %We will see in Fig. \ref{fig:seasonal-rms}, that  the differences in smaller frequencies are due to the difference of the data-taking periods in the different years, which ended in the mid-summer time.
The $ 90\textsuperscript{th} $ percentiles at higher frequencies show the increasing industrial noise at night due to the three-shift schedule of the mine works. Interestingly this is not apparent in the mode and in the median of the data, because the particular activity seemingly put up only short noisy periods. In Table \ref{tab:Yearly-night-rms-media} the $rms_{2-10Hz}$ values of the night periods of each year are shown, corresponding to the middle panels of Fig. \ref{fig:Year-night-ASD}. There is no significant annual variation during Run-1 of MGGL.

%As it can be seen from the modes and medians, only one difference shows up about $ 1$ Hz. This will be explained in the last point (Fig. (\ref{fig:seasonal-rms})) and it is  rather depends on the data-taking period, which ended in the mid-summer time. From the $ 90\textsuperscript{th} $ percentiles shows the start of the three-shift work period, increasing the human activities in the mine even for the night time. This may not be highlighted in the media/mode because, as we supposed earlier, there is only reduced work is in this period.

\begin{table}
\centering
	\begin{tabular}{|c|c|c|c|}
		\hline Night & 2016 & 2017 & 2018 \tabularnewline
		\hline E & 0.124 & 0.127 & 0.123 \tabularnewline
		\hline N & 0.122 & 0.123 & 0.128 \tabularnewline
		\hline Z & 0.149 & 0.151 & 0.137 \tabularnewline
		\hline
	\end{tabular}
	\caption{\label{tab:Yearly-night-rms-media} The median $ rms_{2-10Hz} $ values in nm of the night periods in the years of Run-1.}
\end{table}

\begin{figure}
	\centering
	\includegraphics[width=0.48\textwidth]{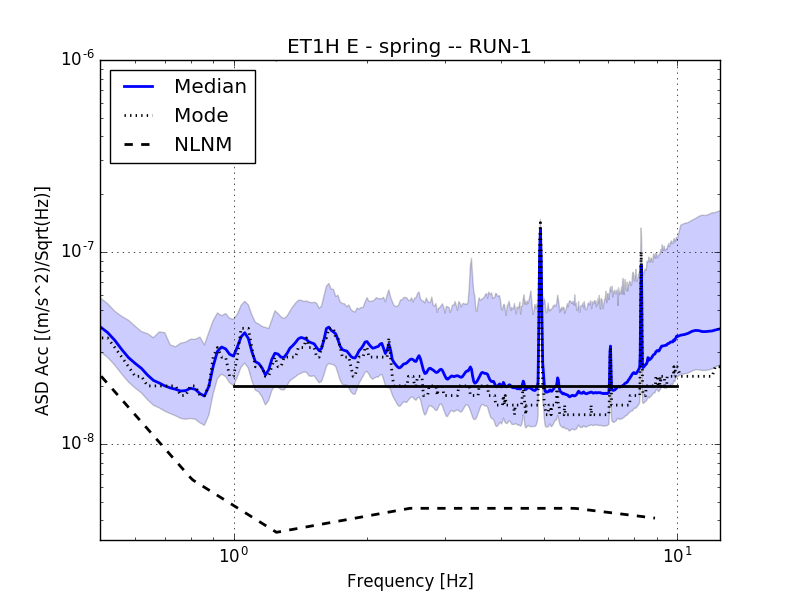}
	\includegraphics[width=0.48\textwidth]{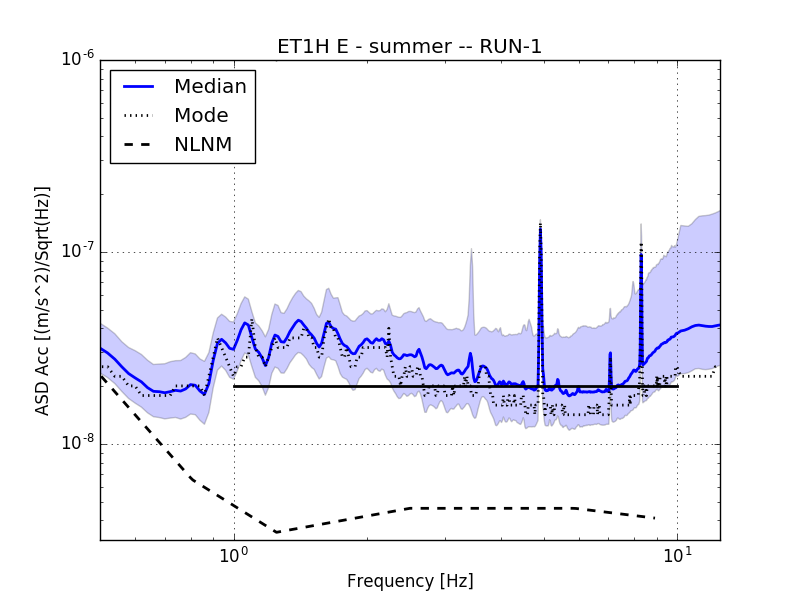}
	\includegraphics[width=0.48\textwidth]{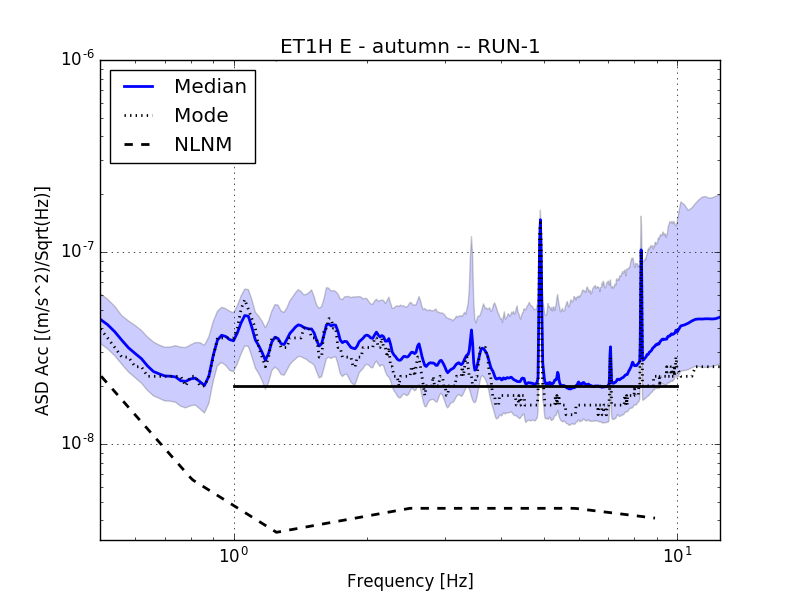}
	\includegraphics[width=0.48\textwidth]{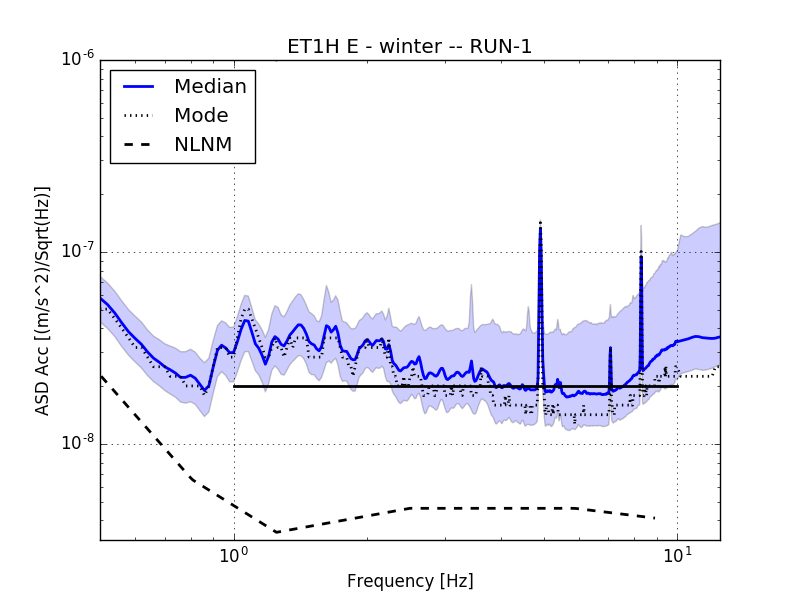}
	
	\caption{\label{fig:Seasonally-ASD} The acceleration ASD of the different seasons for a horizontal component calculated is shown here from whole-day periods. The solid blue and the dotted black lines indicate the median and the mode of the data, respectively. The borderlines of the blue colored area are the 10\textsuperscript{th} and 90\textsuperscript{th} percentiles. The Black Forest line is solid black and the dashed black lines are NLNM curves.}
\end{figure}

The seasonal averages are plotted in Fig. \ref{fig:Seasonally-ASD} including the $ rms $ values in Table \ref{tab:Seasonally-rms}, which presents that the calmest seasons are the spring and the summer. %, but only in the first case where the $ 90\textsuperscript{th} $ percentiles are radically reduced ???.

\begin{figure}
	\centering
	\includegraphics[width=0.48\textwidth]{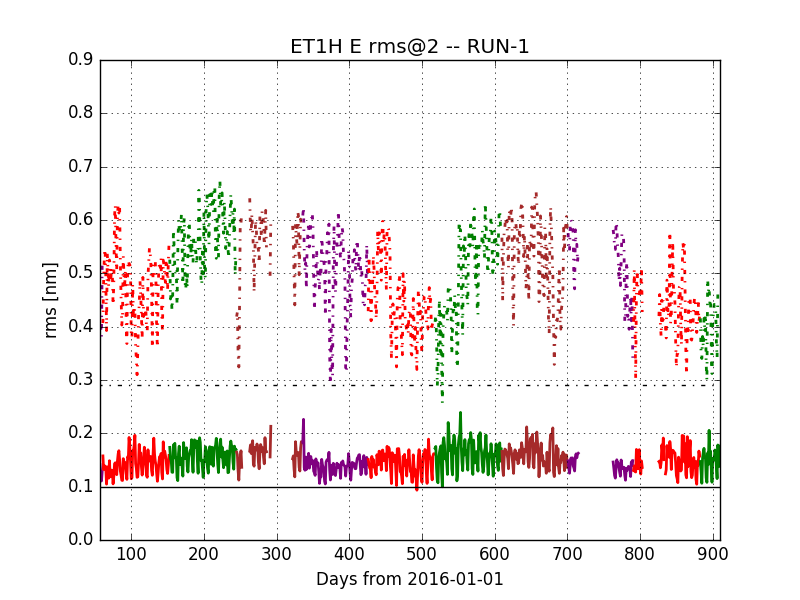}
	\includegraphics[width=0.48\textwidth]{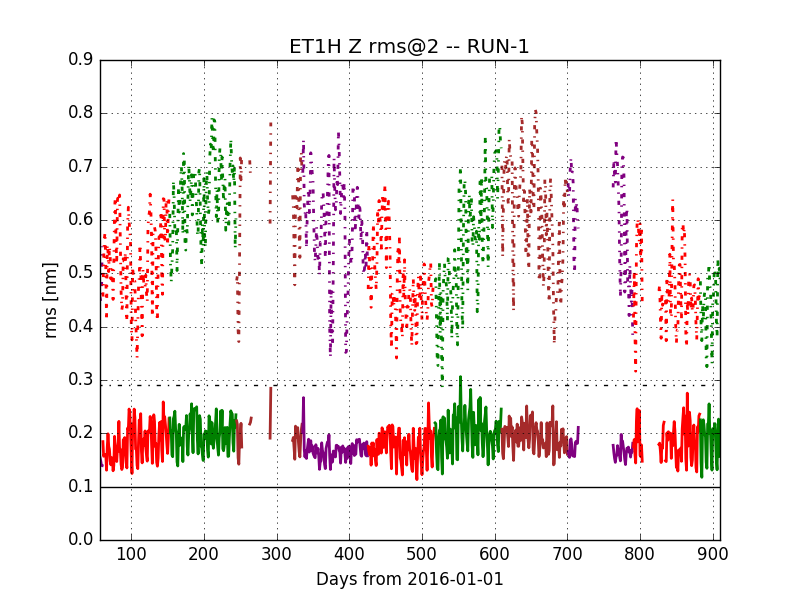}
	\caption{\label{fig:seasonal-rms} The timeline of the daily $rms$ values for the whole Run-1 in a horizontal and in the vertical direction. In the figures the mode related $rms_{2Hz}$ is the solid line and the $rms_{1-10Hz}$ from median is the dashed one. %In the last two Figures the $rms_{2-10Hz}$ is the solid line and $rms_{2Hz}$ is the dashed one, both are median.
	The colours red, green, brown and purple are for the spring, summer, autumn and winter periods, respectively. The solid and dashed black lines are the referential Black Forest values, $rms_{2Hz}= 0.1 nm$ and $rms_{1-10Hz}= 0.29 nm$.}
\end{figure}

\begin{table}
	\begin{tabular}{|c|c|c|c|c|}
		\hline
		$ rms_{2Hz} [nm] $& spring & summer & autumn & winter \tabularnewline
		\hline
		E & 0.125 & 0.134 & 0.139 & 0.132\tabularnewline
		\hline
		N & 0.122 & 0.130 & 0.131 & 0.130\tabularnewline
		\hline
		Z & 0.148 & 0.164 & 0.169 & 0.160\tabularnewline
		\hline\hline
		$ rms_{1-10Hz} [nm] $& spring & summer & autumn & winter \tabularnewline
		\hline
		E & 0.461 & 0.522 & 0.546 & 0.514 \tabularnewline
		\hline
		N & 0.487 & 0.557 & 0.581 & 0.554 \tabularnewline
		\hline
		Z & 0.505 & 0.592 & 0.625 & 0.599\tabularnewline
		\hline
	\end{tabular}
	\caption{\label{tab:Seasonally-rms} Seasonal variation of the $ rms $ values of the seasons according to the data shown in  Fig. \ref{fig:Seasonally-ASD}. The first part is the $rms_{2Hz}$ of the mode and the second one is the $rms_{1-10Hz}$ of the median.}
\end{table}
Finally, the seasonal changes are represented with a timeline plot of the  the daily $ rms $ values in Fig. \ref{fig:seasonal-rms}. As we can see, there is an annual trend in the curves, the late spring and the early summer are the most silent periods in the $ 1-10$ Hz frequency range. Note, this seasonal difference is not apparent in the $ rms_{2Hz} $ timeline, which is at the bottom panel of Fig. \ref{fig:seasonal-rms}. This seasonal variation of the lowest frequency noise is in agreement with the observable also in Fig. \ref{fig:Seasonally-ASD}.

\subsubsection{Comparing the deep and shallow \label{subsec:2-weeks-results}}

\begin{figure}
	\centering
	\includegraphics[width=0.48\textwidth]{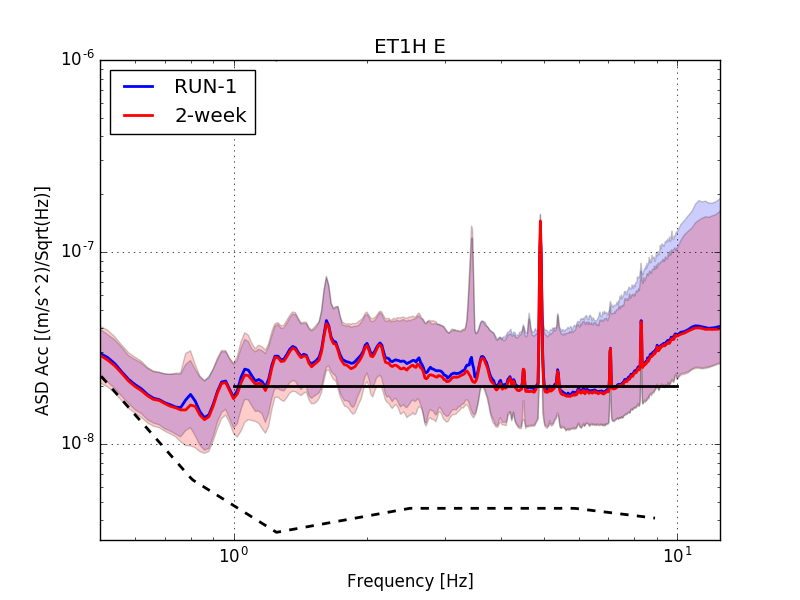}
	\includegraphics[width=0.48\textwidth]{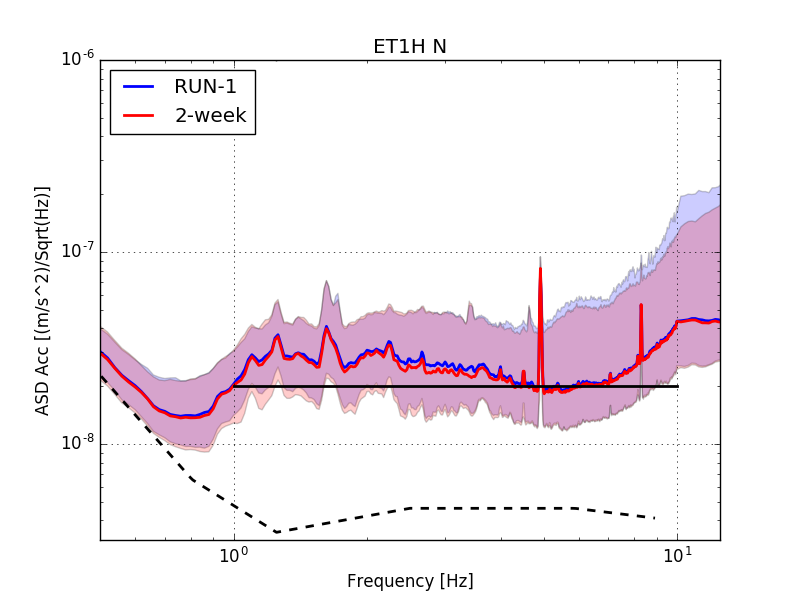}

	\includegraphics[width=0.48\textwidth]{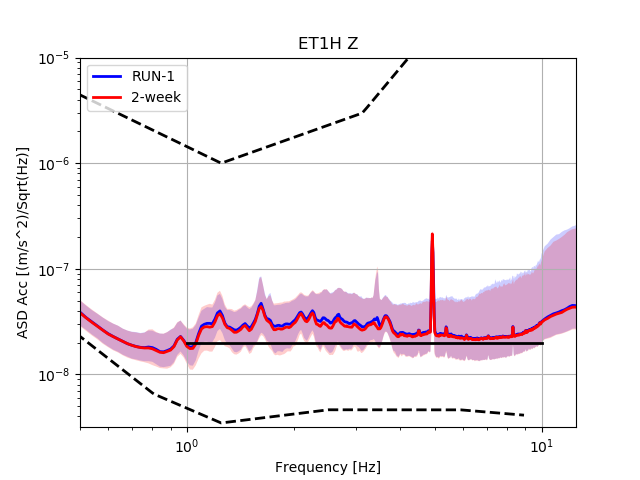}
	\caption{\label{fig:Whole-2-week}The acceleration ASD of the Run-1 (blue) and the particular two-week periods (red) of the ET1H station in each directions. The borderlines of the coloured areas are the 10\textsuperscript{th} and 90\textsuperscript{th} percentiles. The Black Forest line is solid black and the dashed black lines are NLNM curves.}
\end{figure}

In this section the GU02 (-404 m) results from a two-week observation run (1-15 June 2017) are compared to the parallel close to surface MGGL measurements of ET1H (-88 m). Because of the shorter time interval, the percentiles were computed directly from the $ 300$ s averages without INA. The modes were calculated from $ 1800$ s averages as in the previous section. In Fig. \ref{fig:Whole-2-week}  the comparison of the two-week and the total Run-1 ET1H data are plotted.  One can see, this two-week period is representative, there is no significant differences of the plots, in spite of the INA in case of Run-1.

%The acceleration ASD of the two-week data taking period is shown in Fig. \ref{fig:2-week-whole} with the $ rms $ values in Tab. \ref{tab:2-week-whole-rms}.
%\begin{figure}
%	\centering
%	\includegraphics[width=0.45\textwidth]{pics/ET1H_HHE_mode_median_plot.png}
%	\includegraphics[width=0.45\textwidth]{pics/GU02_HHE_mode_median_plot.png}
%	\includegraphics[width=0.45\textwidth]{pics/ET1H_HHZ_mode_median_plot.png}
%	\includegraphics[width=0.45\textwidth]{pics/GU02_HHZ_mode_median_plot.png}
%	\caption{\label{fig:2-week-whole}The acceleration ASD spectrum of the ET1H and GU02 stations for the two-week. The borderlines of the colored areas, that is the 10\textsuperscript{th} and 90\textsuperscript{th} percentiles were calculated from the $ 300\,s $ averages. The modes and medians are obtained from the half-hour averages. The Black Forest line is solid black and the dashed black lines are NLNM curves.}
%\end{figure}

\begin{figure}
	\centering
	\includegraphics[width=0.45\textwidth]{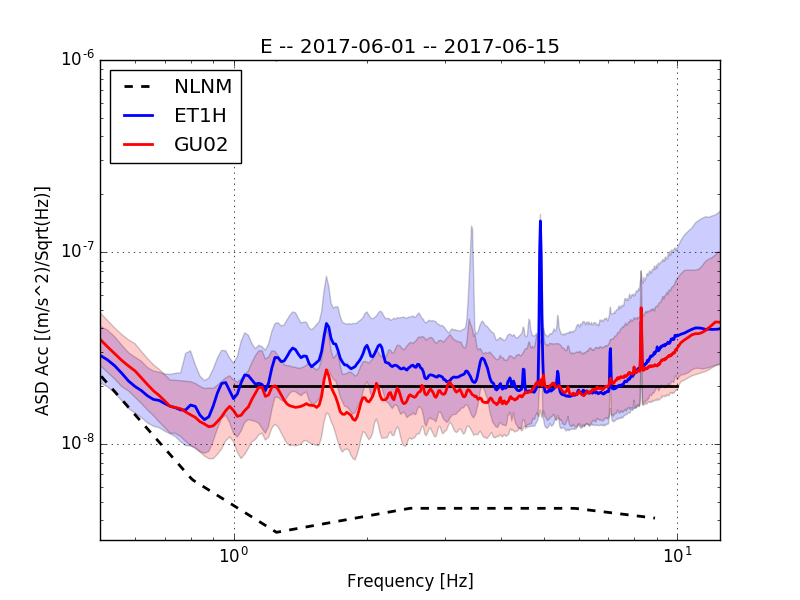}
	\includegraphics[width=0.45\textwidth]{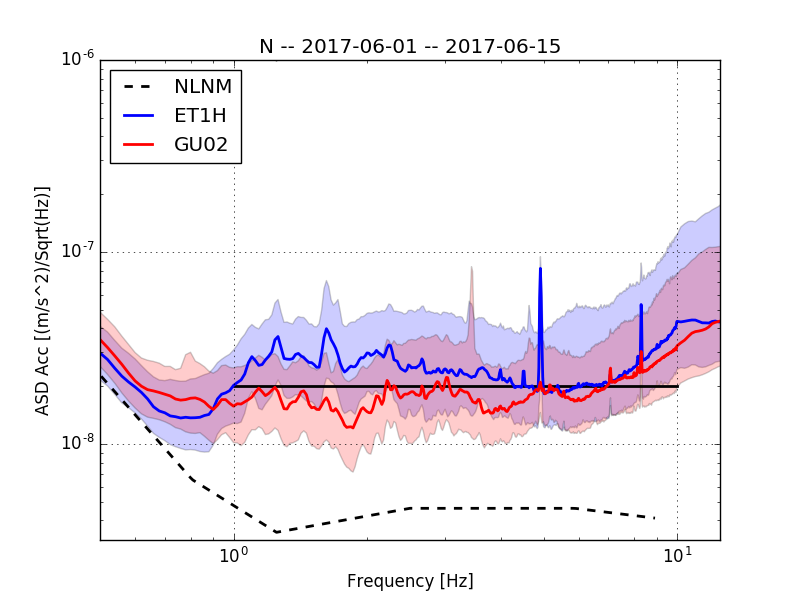}
	\includegraphics[width=0.45\textwidth]{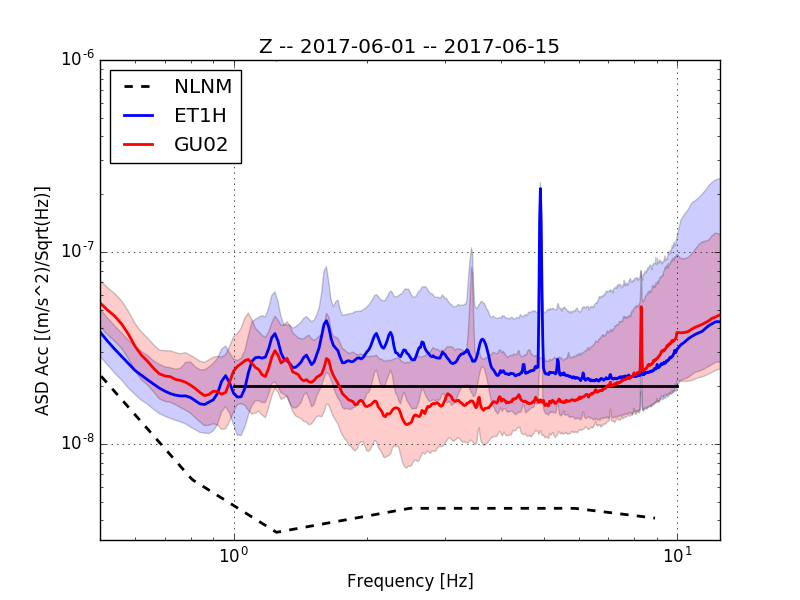}
	\caption{\label{fig:2-week-whole-together-ASD} The percentiles of the ET1H $-88$ m and (red) the GU02, $-404$ m (blue) stations are shown together in the three directions. The Black Forest line is solid black and the dashed black lines are NLNM curves.}
\end{figure}

\begin{figure}
	\centering
	\includegraphics[width=0.45\textwidth]{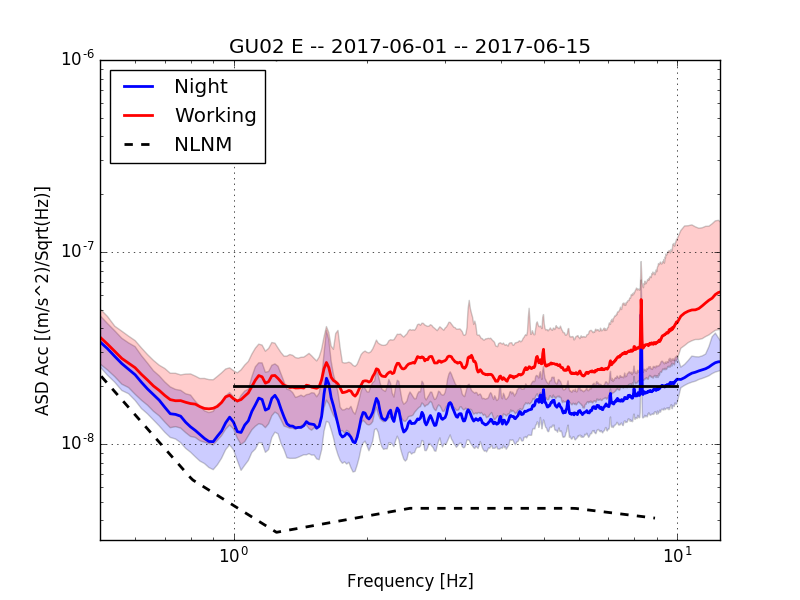}
	\includegraphics[width=0.45\textwidth]{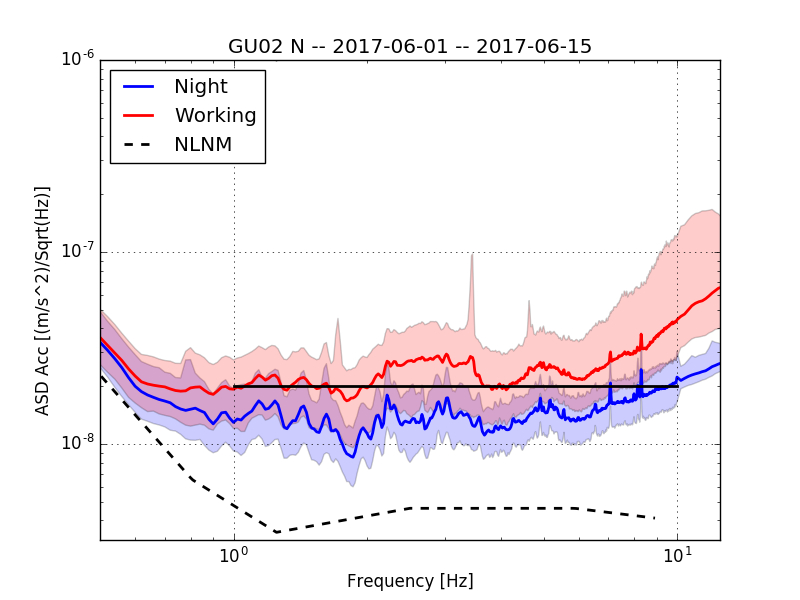}
	\includegraphics[width=0.45\textwidth]{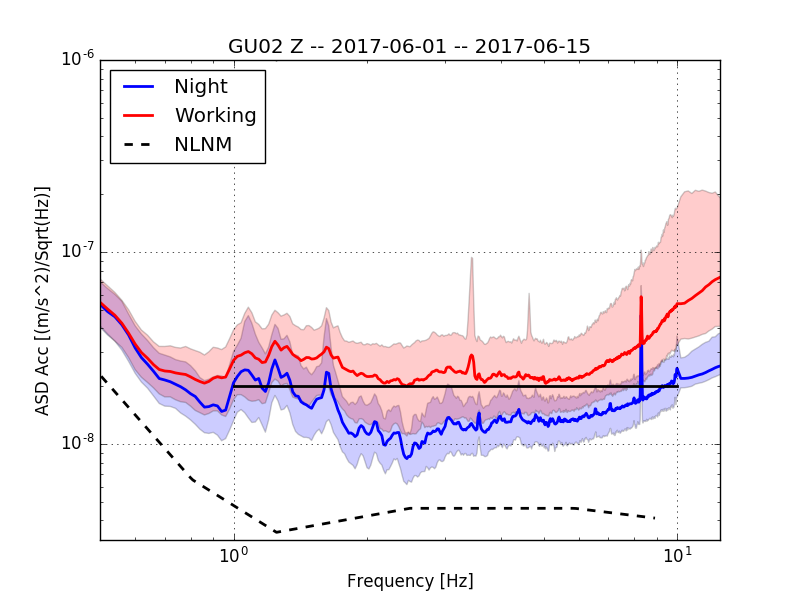}
	\caption{\label{fig:2het-night-work}Comparison of the ASD of the night and work periods of the GU02 station. The Black Forest line is solid black and the dashed black lines are NLNM curves.}
\end{figure}

\begin{figure}
	\centering
	\includegraphics[width=0.45\textwidth]{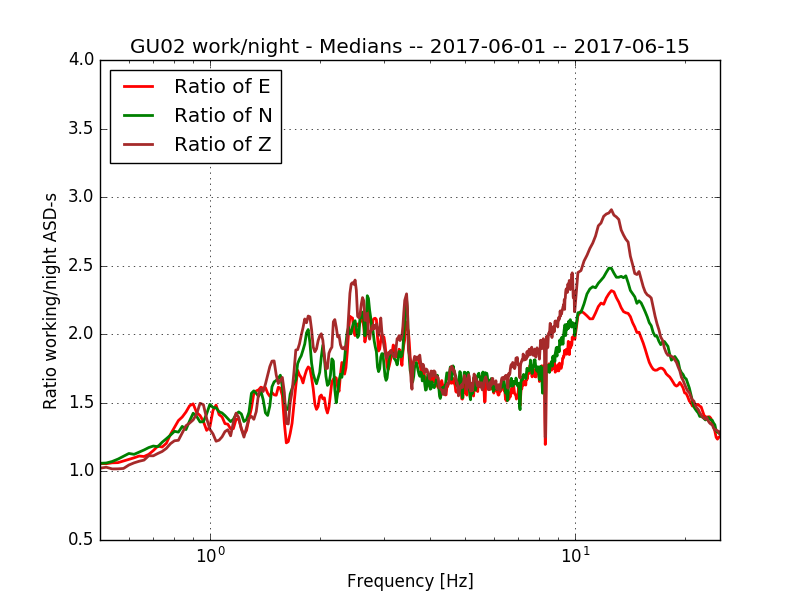}
	\caption{\label{fig:2het-working-night-ration} The ratio the noise median in the  night and in the working periods for the GU02 station.}
\end{figure}

\begin{table}
	\begin{tabular}{c|c|c|c}
		\hline
		ET1H&$ rms_{2Hz} $ & $ rms_{2-10Hz} $&$ rms_{1-10Hz} $\tabularnewline
		\hline
		E&0.126&0.140&0.385\tabularnewline
		\hline
		N&0.122&0.143&0.417\tabularnewline
		\hline
		Z&0.143&0.169&0.427\tabularnewline
		\hline
	\end{tabular}
	\begin{tabular}{c|c|c|c}
		\hline
		GU02&$ rms_{2Hz} $ & $ rms_{2-10Hz} $&$ rms_{1-10Hz} $\tabularnewline
		\hline
		E&0.0829&0.0961&0.259\tabularnewline
		\hline
		N&0.0690&0.0951&0.259\tabularnewline
		\hline
		Z&0.0690&0.0830&0.363\tabularnewline
		\hline
	\end{tabular}
	\caption{\label{tab:2-week-whole-rms}The $ rms $ values in $ nm $ for the spectral densities in Fig. \ref{fig:2-week-whole-together-ASD}. The $rms_{2Hz}$ is calculated from  the mode, the other from the median.}
\end{table}
We have plotted together the ET1H and GU02 data in Fig. \ref{fig:2-week-whole-together-ASD}. There it is apparent that the main attenuation is in the intervall of 1-4 Hz for the horizontal and  1-7 Hz for the vertical direction. This interval is crucial for the low frequency noise budget of the proposed ET, furthermore, this frequency range dominate all the $ rms $ values. The corresponding $ rms $ values for GU02 are in Tab. \ref{tab:2-week-whole-rms}.
In Fig. \ref{fig:2het-night-work} the night and working periods are shown for the GU02 station with the corresponding $ rms $ values in Tab. \ref{tab:2-nigh-working-rms}. The ratios of night and working periods are also shown in Fig. \ref{fig:2het-working-night-ration}.

\begin{table}
	\begin{tabular}{c|c|c|c}
		\hline
		Night period &$ rms_{2Hz} $ & $ rms_{2-10Hz} $&$ rms_{1-10Hz} $\tabularnewline
		\hline
		E&0.0748&0.0752&0.217\tabularnewline
		\hline
		N&0.0746&0.0732&0.213\tabularnewline
		\hline
		Z&0.0625&0.0619&0.304\tabularnewline
		\hline
	\end{tabular}

	\begin{tabular}{c|c|c|c}
		\hline
		Working period &$ rms_{2Hz} $ & $ rms_{2-10Hz} $&$ rms_{1-10Hz} $\tabularnewline
		\hline
		E&0.143&0.135&0.318\tabularnewline
		\hline
		N&0.150&0.135&0.326\tabularnewline
		\hline
		Z&0.135&0.121&0.431\tabularnewline
		\hline
	\end{tabular}
	\caption{\label{tab:2-nigh-working-rms}The calculated $ rms $ values in $ nm $ for the night and working periods of the GU02 station, according to Fig. \ref{fig:2het-working-night-ration}.}
\end{table}

\subsection{Summary\label{MGGL-seismo-summary}}
We analysed the long term Run-1 data of the ET1H station and compared the data from a representative two-week period of the $-88$ m deep ET1H station with the $-404$ m deep GU02 station. For higher than $2$ Hz frequencies, there are no significant annual changes, but for $ 1-2$ Hz the spring-summer time shows annual minimum. Comparing the deeper and shallower noise data we have observed that the decrease of seismic noise spectral amplitudes in the 1-8 Hz frequency range is approximately $ 60\% $. This range is crucial for the low frequency performance of Einstein Telescope.

We emphasize, that almost in $ 90\% $ of the observation period detected the noise level below the Black Forest line at night (see Fig. \ref{fig:2het-night-work}). The related average horizontal  $ rms_{2-10Hz} = 0.0742$ nm and the $ rms_{1-10Hz} = 0.213$ nm. The first value is equal to the mode related $ rms_{2Hz}$ value, which is $0.0745$ nm.

%----------------------------------------------------------------
\section{Study of seismic noise with the Warsaw seismometer}

As it was described in Section \ref{sec:seismo}, the custom made seismometer of the Warsaw University (WARS) is located near to the ET1H seismometer, therefore they should measure the same noise, although at low frequencies there can be differences due to  scattering of seismic waves in the mine. The y-axis of the sensor is pointing at the direction of 25 degrees NW.

\subsection{The collected and analysed data:}

The Run-1 data of the WARS seismometer statts from $24$ May 2016 till $2$ July 2018. In this period 654 days were used for analysis with the rest of the days being unavailable. Days not included in the study will be referred to as “missing-days” henceforth. The data for the study were recorded in each hour. Out of these 654 days, 30 days do not have complete 24 hours and will be referred to as “gap-days”  henceforth. After eliminating the missing durationsfrom the gap-days, total of 15 290 hours of complete data were available for analysis.

\subsection{Analysis of the WARS data:}

The acceleration power spectral density (PSD) \cite{heinzel2002spectrum}  was calculated by Welch’s method \cite{1161901}. The data for each hour were divided into 2048  length segments with each segment  windowed by Hann  window function \cite{heinzel2002spectrum,blackman1958measurement,kanasewich1981time} using half the length segment as overlap% satisfying window function constant overlap-add conditions
\cite{heinzel2002spectrum}. No zero padding was used. The acceleration amplitude spectral density (ASD) for each hour was calculated by taking square root of the acceleration PSD.
The data for each hour were available for each of the three axis for all the 15290 hours so a total of 3$\times$15290  ASDs were generated.

\subsection{Monthly analysis of the WARS data:}

\begin{figure}
\includegraphics[width=\columnwidth]{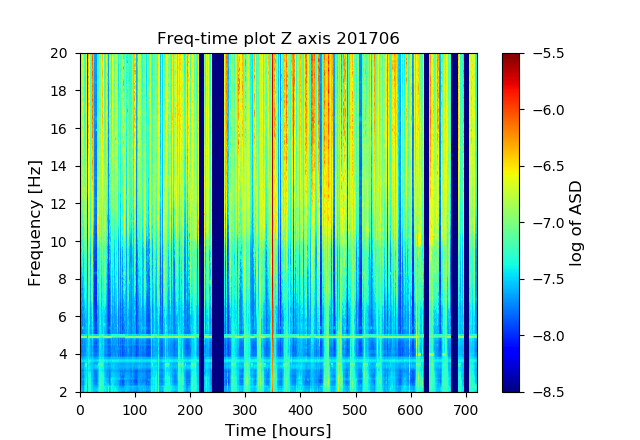}
\caption{The time frequency plot of the vertical spectrum for June 2017. The daily variation of the noise is clearly seen. %Moreover, there are several short periods where additional disturbances appear.
The data gaps are in dark blue.  Additionally weekends are clearly less noisy - which is seen in the low frequency band. The histogram unit is m$ s^{-2}$$ Hz^{-1/2}$. }
\label{TF062017}
\end{figure}

The ASD generated for each hour were  used to study the variation of ASDs with time over the period of each  month i.e spectrograms were generated for every month for the period of the study. This was done for all three axis to study the variation in detail. We  present an example time-frequency plot of the ASD for one month for the vertical axis. We have chosen to show the month of June 2017 and we present it in Fig.~\ref{TF062017}. One can clearly see some data gaps. There are persistent lines at ~3.5 Hz and at ~5 Hz. The daily variation of the noise is prominent, and one can also see that at weekends the noise is lower.

\subsection{Complete hourly analysis:}
\begin{figure}
\centering
\includegraphics[width=.49\columnwidth]{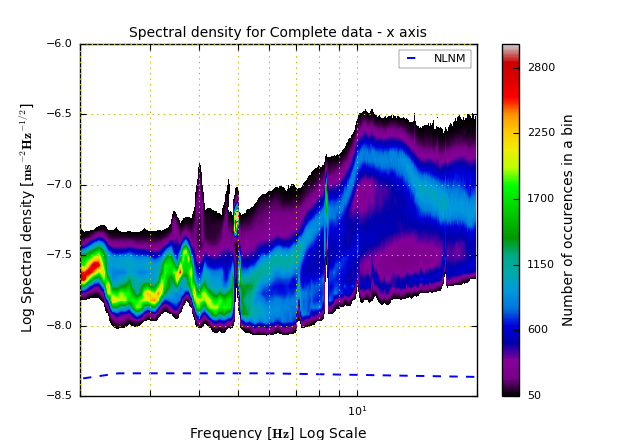}
\includegraphics[width=.49\columnwidth]{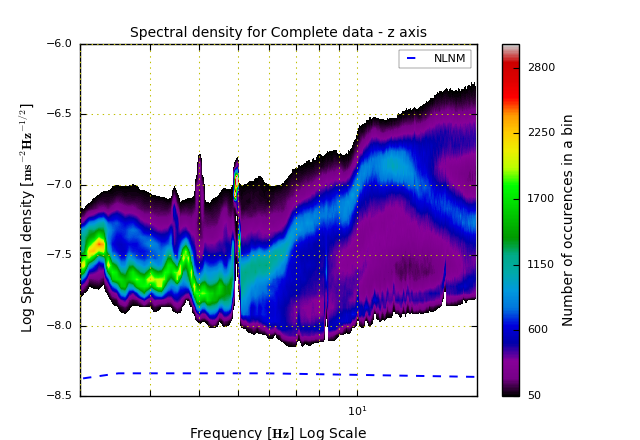}
\caption{The density plots of the spectra in the vertical and horizontal directions. The blue dashed line is the NLNM.}
\label{ShadeAll}
\end{figure}

We have also studied the variation of the ASD over the entire period of data analysis. We present  the logarithm values of hourly acceleration ASD and  binned it on a plane spanned by the frequency and the logarithm of acceleration ASD. We then plotted it  using the color shading for presenting how often each value occurs in the data. The log ASD axis was divided into twenty bins per decade. We show the results for the two axes in Fig.~\ref{ShadeAll}.  The variation of the ASD spans about 0.7 dex at low frequencies below 5Hz and goes to almost 1.5 dex at about 20Hz.
%There are several types of spectra probably corresponding  to different times of the day.

\subsection{Complete daily analysis:}

\begin{figure}
\centering
\includegraphics[width=.49\columnwidth]{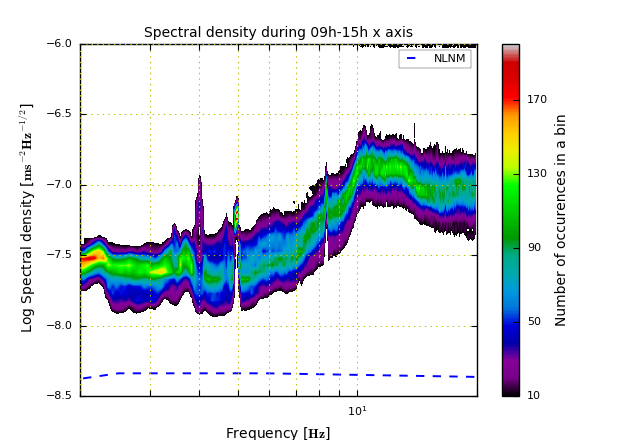}
\includegraphics[width=.49\columnwidth]{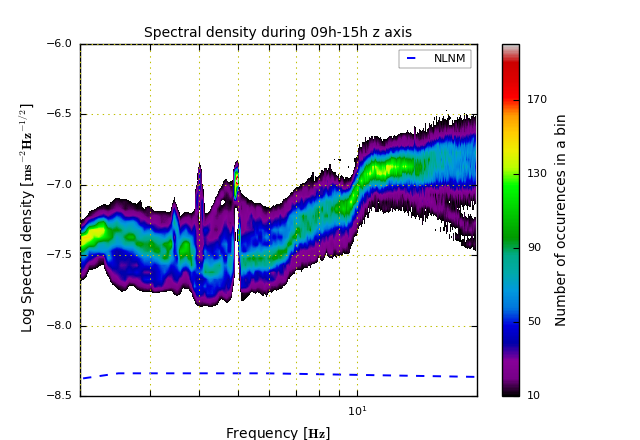}
\caption{The daily variation of the seismic noise for the work-shifts in the mine: corresponding to the time between 09:00 and 15:00 hours UTC. The blue dashed line shows the  NLNM curve.}
\label{periods0915}
\end{figure}

\begin{figure}
\centering
\includegraphics[width=.49\columnwidth]{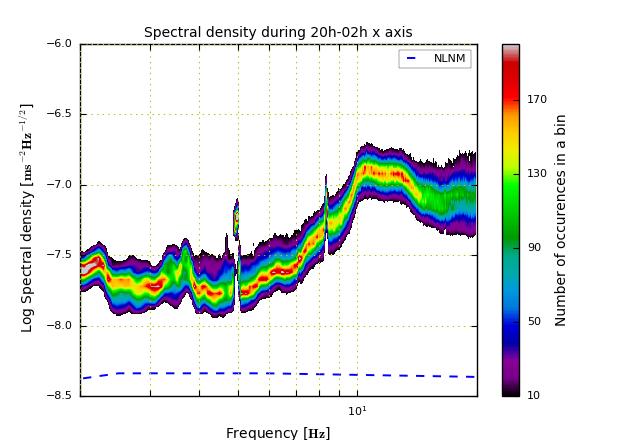}
\includegraphics[width=.49\columnwidth]{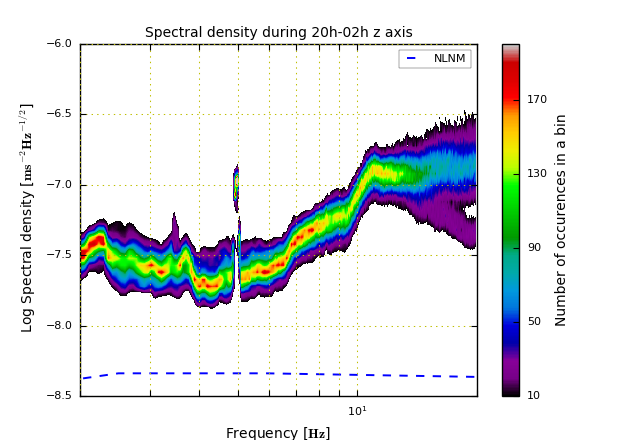}
\caption{The daily variation of the seismic noise for the off work-shifts in the mine: corresponding to the time between 20:00 and 02:00 hours UTC. The blue dashed line is the NLNM.}\label{periods2002}
\end{figure}

In order to investigate several types of spectra during different working times of the day in more detail we generated the acceleration ASD for periods corresponding to the working shifts in the mine, 09:00-15:00 UTC and 20:00-02:00 UTC.  The acceleration ASDs for these periods were then binned and plotted for each axis for the whole data set. We present these in the panels of Fig. \ref{periods0915} for the working,  and in panels of Fig. \ref{periods2002} for the night periods, respectively, like in the case of ET1H station. These plots  show less variation than the plots for the entire data set. The lack of characteristic bimodality at higher frequencies indicates lack of noisy periods at night. The working period shows additional lines, that are not visible at the night. The night spectra have very little variability less than 0.3 dex over the entire frequency band.

%Surprisingly the spectra in the first period (0-6hours) seem to be quite noisy,
%especially above 10 Hz. However, the noise is also at the level that is close or below our instrumental limit for a significant fraction of the time.
%The spectra in the day time seem to be slightly lower in this band, and less variable. All spectra  show a bimodality - two distinct bands in Figures~\ref{periods12} and \ref{periods34}.
%This suggests that there is an intense human activity at the site and at times some machines or pumps are turned on.
%However, during the quite episodes  the level of the seismic noise is extremely low.

\subsection{Site parameterization}

\begin{figure}
\centering
\includegraphics[width=\columnwidth]{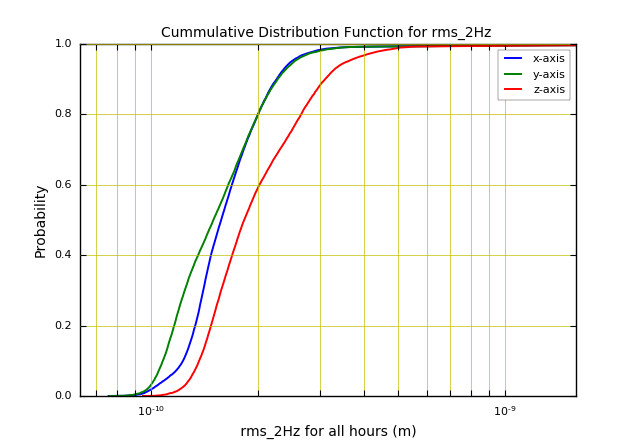}\\
\caption{The cumulative distributions of the hourly $rms_{2Hz}$. We present the results for the three directions. }
\label{cdf_rms2hz}
\end{figure}

%A cumulative characteristic of the site was given by equation (5) in \cite{BekEta15a} using $rms_{2Hz}$ where rms displacement is the square root of the displacement PSD integrated from a cutoff frequency, chosen as 2Hz, up to the  Nyquist frequency $f_{N}$. Note that the $rms_{2Hz}$ defined in this way is depends on the sampling frequency in the detector.
The values of $rms_{2Hz}$ were calculated for each hour of available data and the cumulative distribution function was plotted for the three axes of the detector in Fig. \ref{cdf_rms2hz}. The median $rms_{2Hz}$ $\approx 0.18$ nm in the vertical direction and $\approx 0.16$ nm in the horizontal directions. The two horizontal $rms_{2Hz}$ distributions are different below the median. It is unclear whether this is an instrumental effect or if this corresponds to the  properties of the seismicity in MGGL.

\subsection{Conclusion:}
In conclusion our measurements show that the seismic noise level is very low at the M\'atra site. In spite of the spectral differences the median $rms_{2Hz}$ values are in agreement with the ET1H data of Table \ref{tab:Whole-rms}, measured by a different device. %It remains to check what are the sources of the anthropogenic noise and if they can be eliminated should the site be chosen for construction of the Einstein Telescope.
%
%\newpage

\section{An electromagnetic study on the signal attenuation in the ELF range}
%Lemperger I.$^{1}$, Wesztergom V.$^{1}$, Huba G.$^{2}$, Mlynarczyk J.$^{3}$, Nov\'ak A.$^{1}$, Kis \'A$^{1}$., Szalai S.$^{1}$, \\
%$^1$HAS, Research Centre for Astronomy and Earth Sciences, \\ Geodetic and Geophysical Institute, \\H-9400, Sopron, Csatkai E. u. 6-8. \\
%$^2$HAS, Wigner Research Centre for Physics, Institute of Particle and Nuclear Physics, \\  1121 Budapest, Konkoly Thege Miklós út 29-33. \\
%$^3$Department of Electronics, AGH University of Science and Technology, Krakow, Poland

Since the gravitational wave interferometer's mirrors of the proposed ET is going to be stabilized by means of magnetic fields, man-made and natural electromagnetic (EM) signals may also result in contaminated time windows of the gravitational wave observation. Magnetic noise from local sources can be identified based on correlation analysis utilising a network of extremely low frequency (ELF) field  observatories. However, global EM components in the frequency range of interest may result in undesirable noise load \cite{2013PhRvD..87l3009T}. The global thunderstorm activity exciting the Earth-ionosphere cavity continuously keeps a background signal at certain resonance frequencies called Schumann resonances \cite{Schu52a}. Installation of the gravitational wave detector in a subsurface location may lead to lower EM noise even in the ELF range, $3-20Hz$.

An EM investigation has been carried out in close vicinity of the MGGL which aims to provide an estimation of the attenuation of EM signal with the depth in the M\'atra andesite rock. ELF range geomagnetic observation stations has been installed at the backfilled end of a 140m deep cave and in the close vicinity of the surface projection of the subsurface measurement site. For technical reasons, the subsurface and surface recordings have been run with no overlap in time. Therefore an indirect processing method has been applied utilising the recordings of the very low-noise ELF EM observation site of Hylaty station as reference observation. The amplitude rate between the subsurface and the surface station has been estimated based on the experimental transfer functions between the subsurface-Hylaty and the surface-Hylaty relation.

\subsection{The data acquisition system and the observation}
The study is based on two mobile observations and an observatory data: time series from a subsurface site, a surface station and from a reference station. The subsurface measurement location could have been accessed from the town M\'atraszentimre by an elevator crossing levels down to a tunnel at 140m depth. The observation site has been hundreds of meters away from the vertical mine of the elevator and from all active electronic facilities at that level. Notice that active loads have been in operation during the measurement in other levels of the mine, around 50m and 20m distance vertically. The subsurface data acquisition system consists of a Lemi-423 wide band magnetotelluric station and two Lemi-120 induction coils. The coils have not been fitted to a common plane to increase the separation distance to reduce the inductive coupling effect between the coils. The orientation of the coils have been set to the orientation of the tunnel: NNW-SSE and ENE-WSW, with NNW 22$^\circ$ to North. The timing of the subsurface station has been initialised at the surface by means of GPS synchronisation and had been transported in the proposed site in operating status with no sensor connected. During the whole measurement session (Period I), the inside clock of the station has not been synchronized due to the lack of GPS or NTP access. The station was powered by a fully charged 84Ah battery. The data logger has been installed 40m away from the sensors along the tunnel. The connector of the coils and the whole instrumentation have been waterproof sealed, since high humidity and leaking water could lead to shortcut of the connector part. The sampling frequency has been set to 2kHz.

The surface measurement has been carried out after the subsurface measurement (Period II) and by means of the same Lemi station which has been set up at the subsurface site before. The coils have been buried 50cm deep in a common horizontal plane in perpendicular orientation. The separation of the sensors was more than 10 meters and the Lemi-423 station has been buried 40m away. The Lemi instruments and the wires have been buried preventing harmful activity of forest animals. The orientation of the coils at the surface stations has been set to NNE-SSW and WNW-ESE adapting to the local environment conditions. The sampling frequency has been set to 2kHz at this mobile station, too.

The Hylaty geophysical station was proved to be the optimal ELF observation facility as reference station based on data quality and geographic aspects too, \cite{KulEta14a}. The Hylaty station provides very low noise ELF geomagnetic data at around 888Hz sampling rate which is optimal for the proposed study as reference.

\subsection{Data processing}
\subsubsection{Data preparation}
The primary goal of the study is to provide an estimation of EM attenuation of the overlying andesite block in the lower ELF range, especially at the first Schumann resonance frequency. In the aspect of the measurement configuration it is identical to the relation of the absolute value of the empiric amplitude transfer functions of the subsurface and the surface stations at certain frequency. Since the surface and subsurface observation time periods were not overlapping the direct comparison of the amplitude spectra was not possible. The concept of the processing is based on utilization of a high-quality, low-noise ELF reference station data covering all the studied time intervals. The indirect method is basically as follows: determination of Tc$_{deep/ref}$ power transfer coefficient in Period I. than the Tc$_{surface/ref}$ in Period II and finally the estimation of the attenuation is accounted to the relation of the two coefficient. The amplitude attenuation has been derived as the square root of the power transfer coefficient of subsurface-surface site relation. The observations took 4 and 7 days in Period I. and II., respectively.

\begin{figure}
  \centering
  \includegraphics[width=.4\columnwidth]{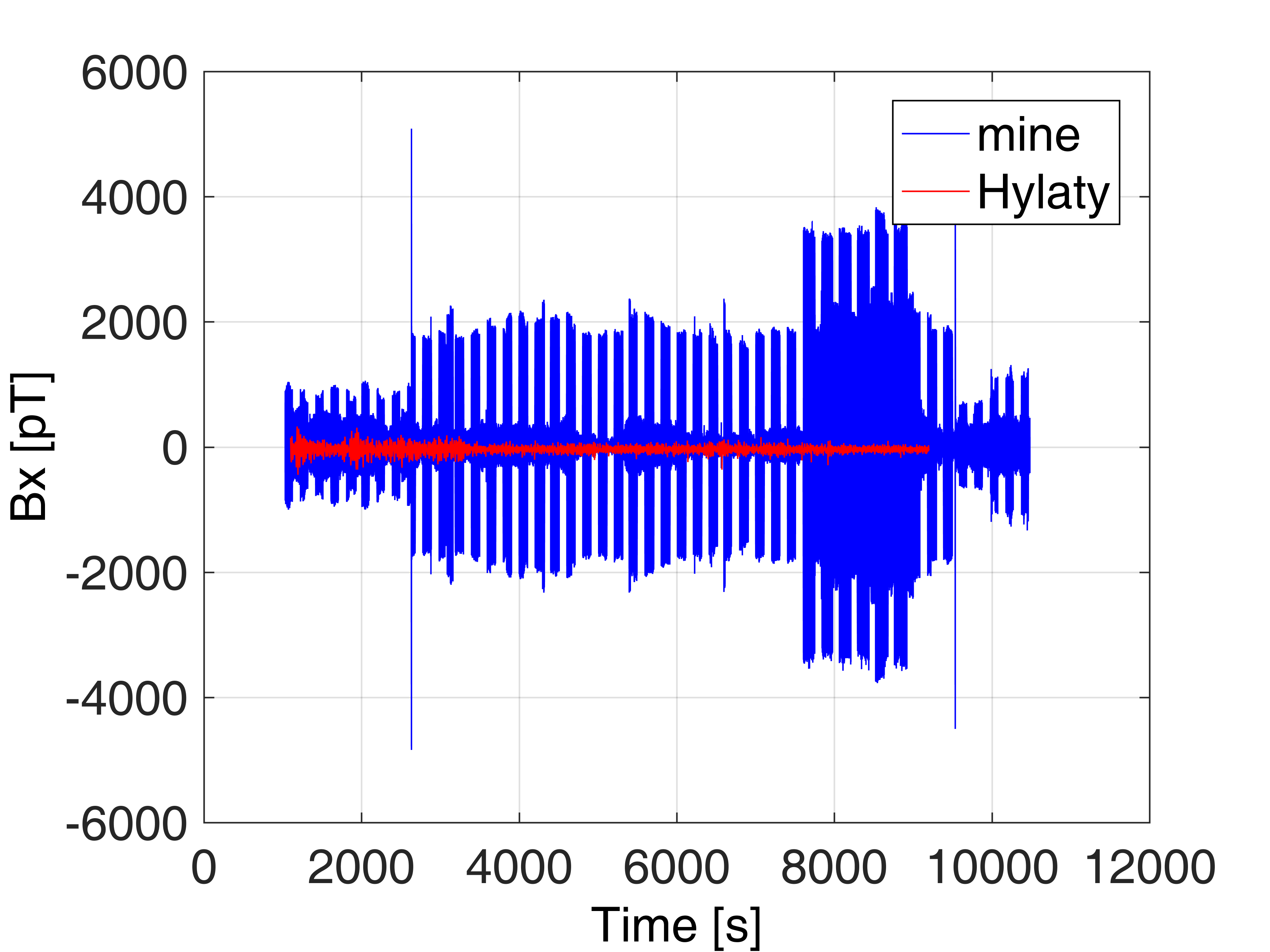}  \label{01l_ts}
  \includegraphics[width=.4\columnwidth]{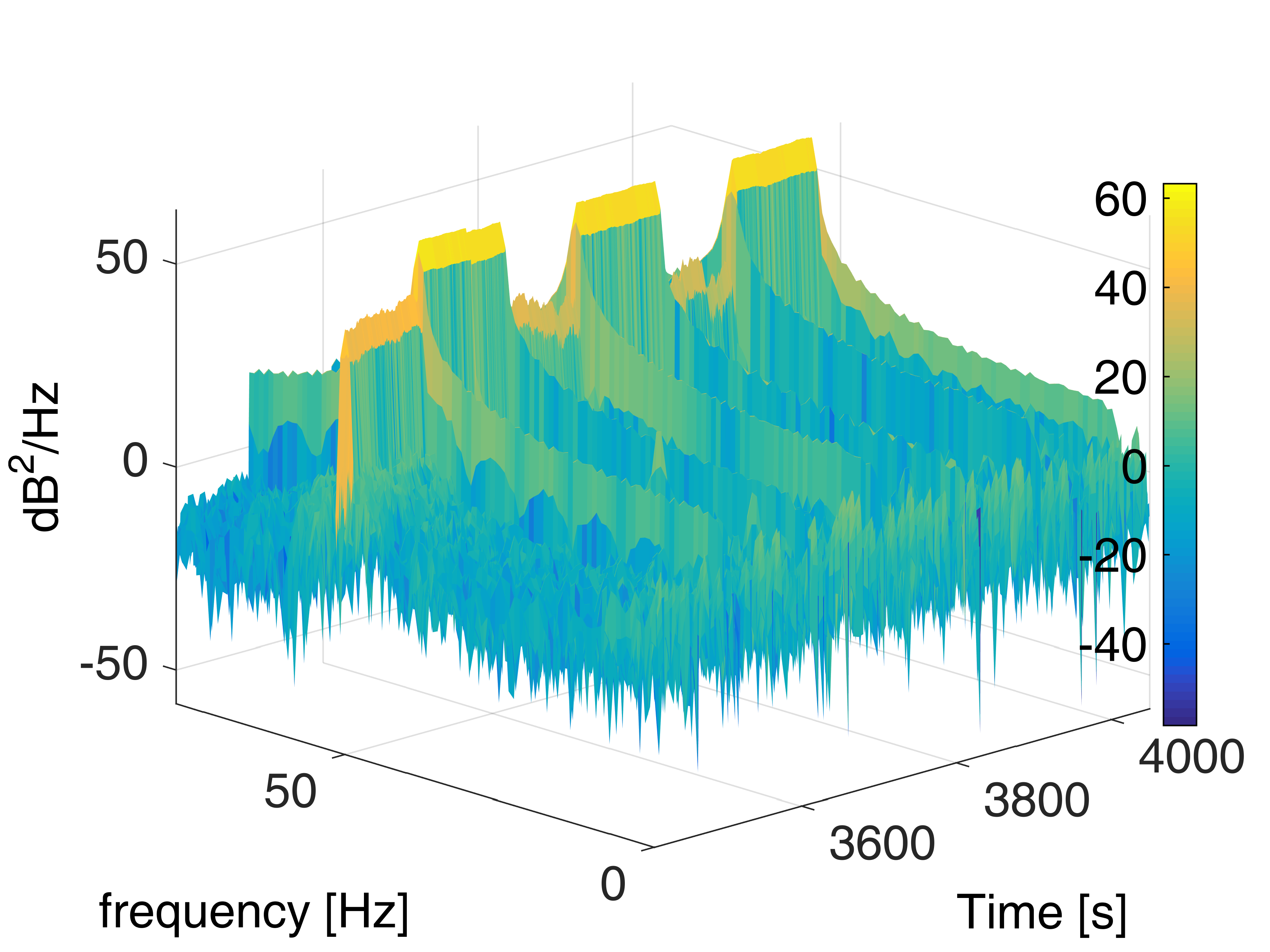} \label{01r_dynspec}
\caption{A ten minute long time series of subsurface registration (left). Blue lines are for N-S, red ones are for the corresponding section of the $x$ channel of the Hylaty reference station. A shorter section of the corresponding dynamic spectral power density of the magnetic variation at the subsurface observation site is plotted on the \textit{(right)}.}
\label{01_noiseload}
\end{figure}

After resampling the three datasets to a common frequency of 800Hz  both mobile data have been rotated in the coordinate system of the reference site. The magnetic components of the subsurface and the surface observation have been transformed by 22$^\circ$ and 55$^\circ$, respectively. Since the infrastructure of the mine has been operating during the whole observation campaign, the time and frequency domain identification of low magnetic noise intervals has formed the basis of the forthcoming processing. The variation of the \textit{x} component in time and time-frequency domain is plotted in Fig. \ref{01_noiseload}. An intermittent broadband load with maximum at 50Hz has been demonstrated for the whole observation period and besides of that a peak appears at around 15-17Hz, too for several minutes by no regular timing. These large power load periods are related to the operation phases of the elevator of the mine shaft and the latter one is originated by the horizontal transportation and the lighting on active levels of the mine. Both have been present in the whole observation period. The low-noise segments have been identified based on the dynamic magnetic spectra covering Period I. Finally the remaining subsurface data consists of thirty one 80-100 second long time series of relatively quiet time segments, selected for further processing. The low noise windows are concentrated in the early afternoon hours of 24 March, 2018, 12:00-16:00 UTC. For the reason of having similar relative position between the observation site and the source area the same temporal pattern of time segments have been selected in Period II from the surface magnetic variation time series on 13 April, 2018. From the reference data the same temporal pattern of time sections have been selected for the estimation of the transfer functions in Period I and Period II, too.

\begin{figure}
  \centering
  \includegraphics[width=.4\columnwidth]{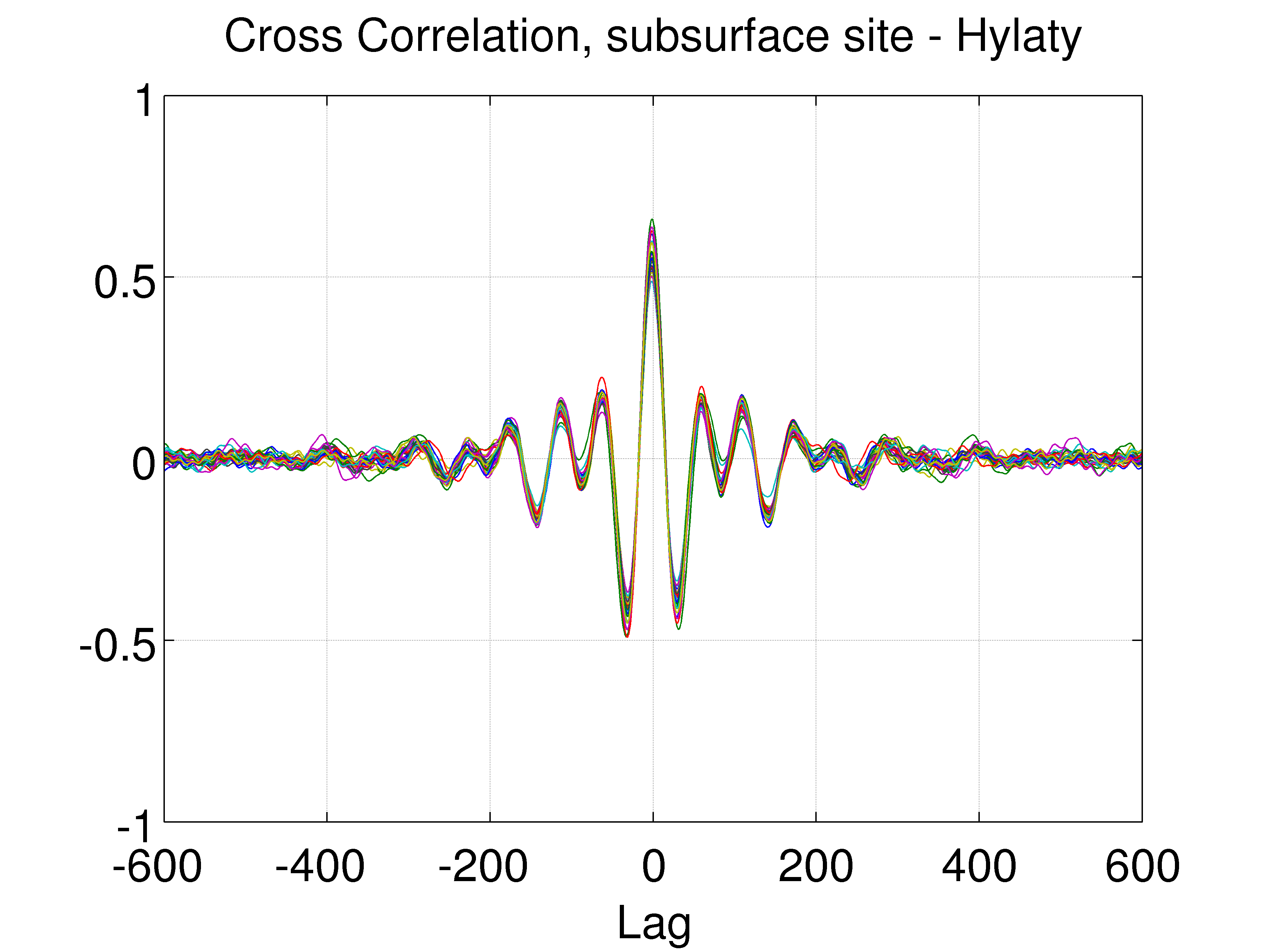} \label{02_crosscorrmH}
  \includegraphics[width=.4\columnwidth]{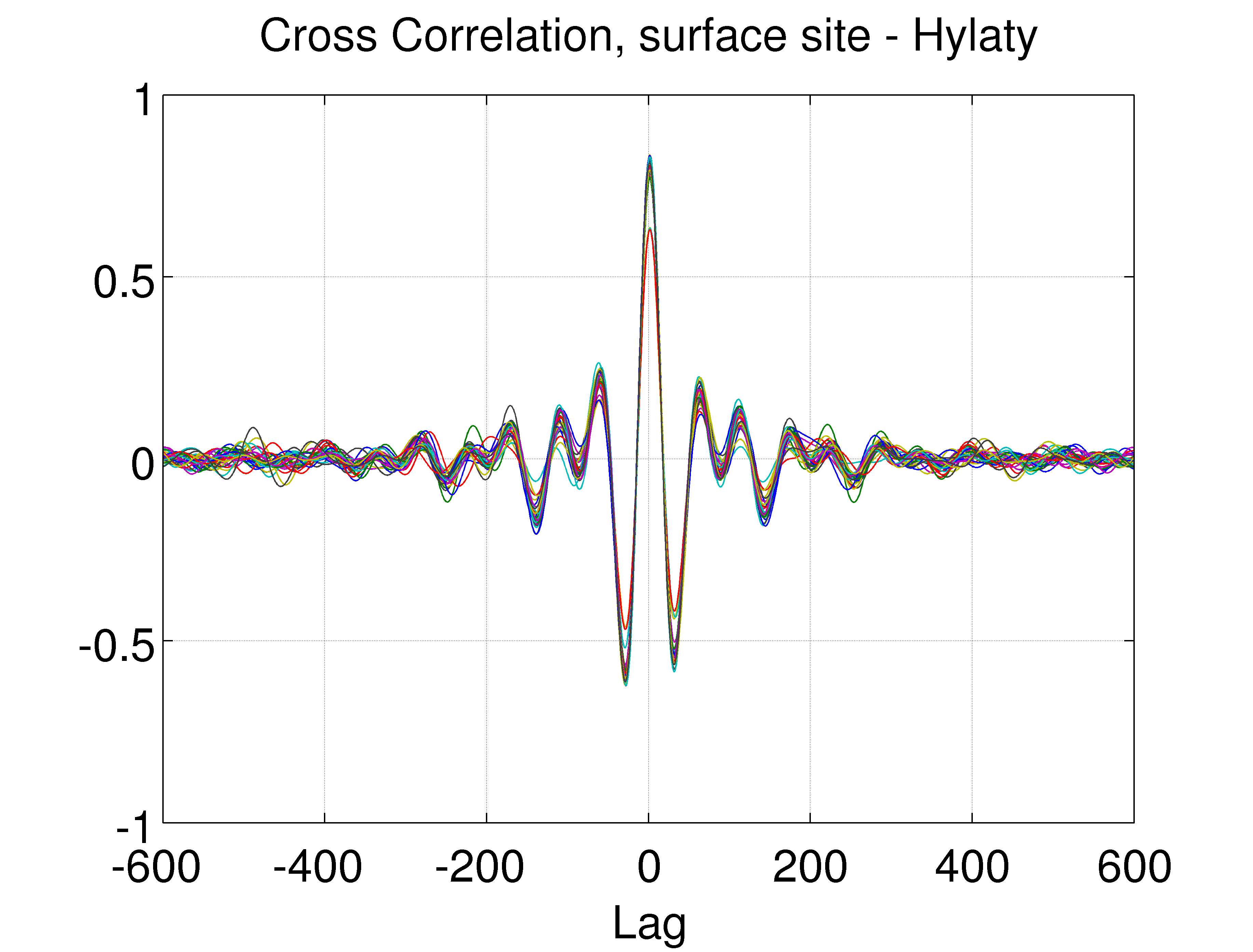}  \label{02_crosscorrsH}
\caption{Cross correlation between the bandpass filtered total magnetic variations observed in the subsurface and at the Hylaty reference site \textit{(left)}. Each color corresponds to one selected quite time window of the 31. The high peak at zero demonstrates that the subsurface station observed the same natural signal as the Hylaty station in each time sections. In the Period II. cross correlation between the bandpass filtered surface and the reference data is on the \textit{(right)}. The plot confirms the correlation between surface observation and the reference observation in the frequency band of 5.6-25Hz for the same pattern of daily distribution of time segments in Period II.}
\label{02_crosscorr}
\end{figure}

Since the inside clock of the data logger has a characteristic delay per day, post synchronization of the subsurface data have been applied first. The method is based on the comparison of the same natural signal recorded on both stations: cross correlation function of the subsurface and the Hylaty data have been estimated on the 31 pair of selected time series sections. Preparation of magnetic data sections by the application of a 5.6-25Hz bandpass Butterworth filter was necessary to attenuate the strong effect of the 50Hz peak of the power grid on the cross correlation function. The time delay according to the cross correlation was 4 samples which means 5ms for both channels. It has been corrected assuming linear drift of the inside clock time.  The total magnetic time series has been correlated in both relation of the subsurface-Hylaty and surface-Hylaty observations. The cross correlation of the bandpass filtered total variations are overlay plotted for each section pairs in Fig. \ref{02_crosscorr} after post synchronization. The cross correlated time sections contain minimum 64.000 samples each at 800Hz sampling rate.
After the same processing, the cross correlation between surface registration and reference data has also been demonstrated, see Fig. \ref{02_crosscorr}.

\begin{figure}
  \centering
  \includegraphics[width=.4\columnwidth]{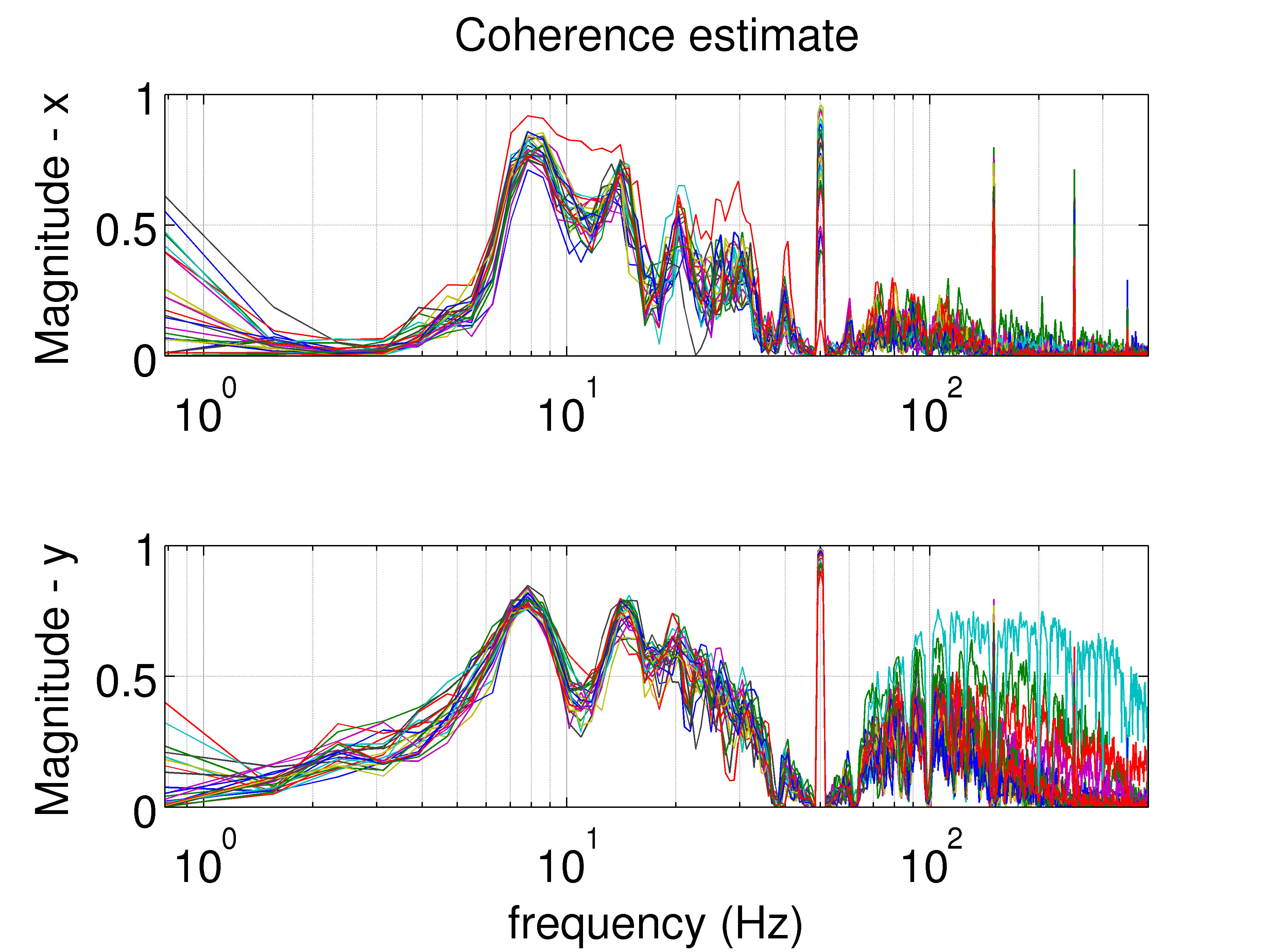}  \label{02_coherencemH}
  \includegraphics[width=.4\columnwidth]{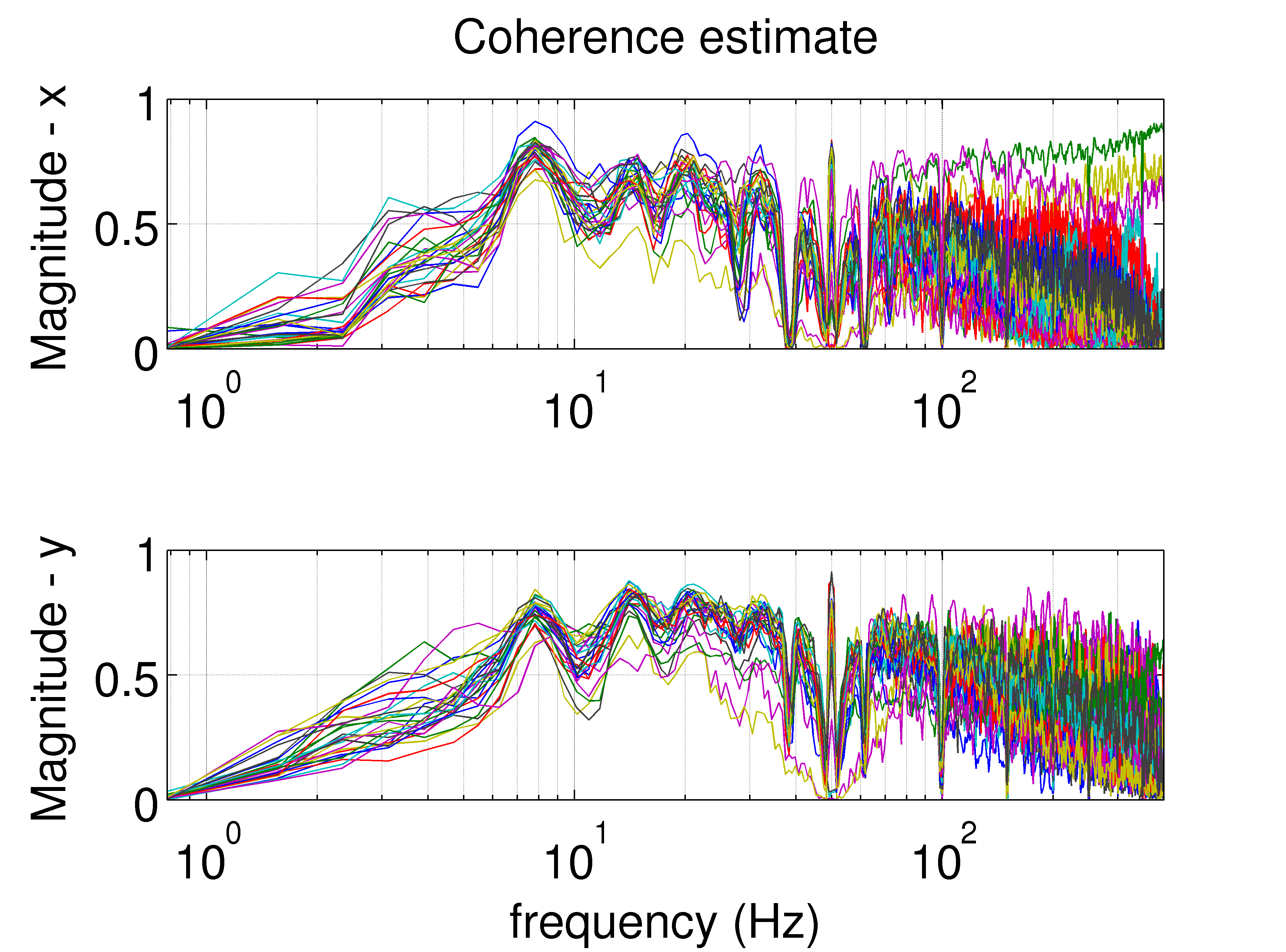}  \label{02_coherencesH}
\caption{ The coherence of magnetic variation in relation of both subsurface-Hylaty \textit{(left)} and surface-Hylaty site \textit{(right)} indicates definite correlation at the first three Schumann frequencies.}
\label{03_coherence}
\end{figure}

The sharp and high peak at zero demonstrates that the subsurface station observed the same natural signal as the reference station. In frequency domain the correlated components have also been confirmed by computing the spectral coherence between the magnetic variations in relation of subsurface-Hylaty and surface-Hylaty datasets, see Fig. \ref{03_coherence}. The spectral coherence plots indicate definite correlation at the first three Schumann resonance frequencies in relation of both subsurface-Hylaty and surface-Hylaty site.

Finally, four dataset has been prepared for the estimation of the transfer functions of subsurface-Hylaty and surface-Hylaty relation and finally to the determination of the surface-subsurface transformation:

\begin{enumerate}
\item $B_{I. mine}$: 31 selected time segments in Period I, subsurface magnetic variation;
\item $B_{I. Hylaty}$: the same 31 time segments in Period I, reference site magnetic variation;
\item $B_{II. surface}$: 31 selected time segments in Period II with the same daily distribution, surface site magnetic variation;
\item $B_{II. Hylaty}$: the same 31 time segments in Period II, reference site magnetic variation;
\end{enumerate}

\subsection{Results}
\begin{figure}
\begin{centering}
\includegraphics[width=.8\columnwidth]{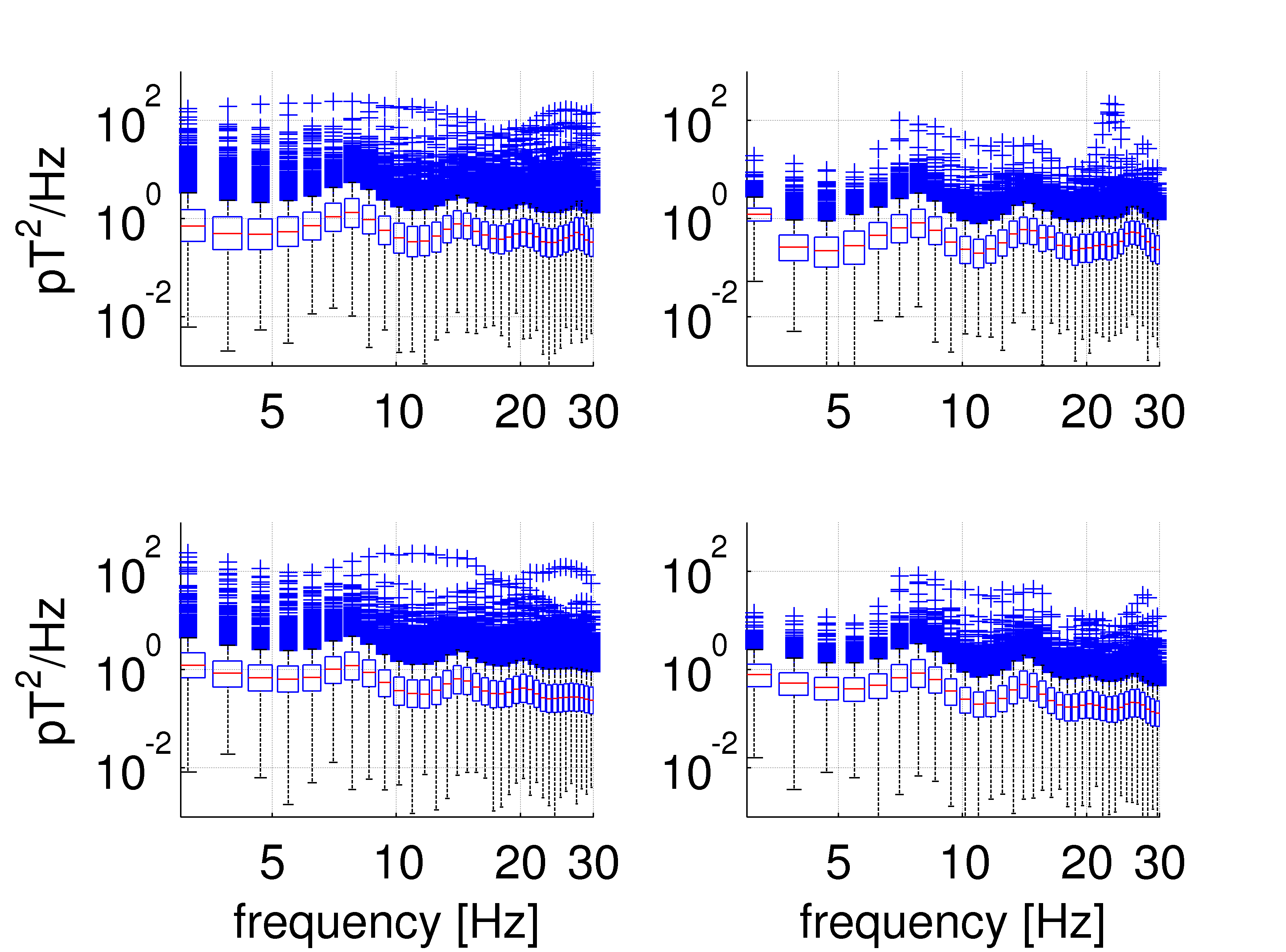}
\par\end{centering}
\protect\caption{The median, the interquartile range and the outliers of power spectra are plotted in relation of the four observations, \textit{(subsurface data: upper, left; Hylaty data - Period I: lower, left; surface data: upper, right; Hylaty data - Period II: lower, right)}. The red line in the middle of the boxes represents the median, the blue box is for the interquartile range, the black bar is for the range of the most extreme data samples not considered outliers and the blue crosses represent the outliers.}
\label{IQR}
\end{figure}

The introduced processing steps resulted in four dataset, which are accurately synchronised, transformed to the same coordinate system and resampled to a common sampling rate of 800Hz. The data has been prepared for time domain and spectral analysis.
Additionally it has been confirmed that the subsurface observation and the Hylaty reference station has observed the same natural signal which variation is considered to be homogeneous in the scale of their relative distance, ~300km. It has been demonstrated in the surface-Hylaty observation, too.  The analysis of the distribution of the spectral power values at individual frequencies demonstrated non-normal distribution of the spectral power. The mean, the median and the interquartile range (IQR) of the spectral power have been plotted in Fig. \ref{IQR}.

\begin{figure}
\begin{centering}
\includegraphics[width=.8\columnwidth]{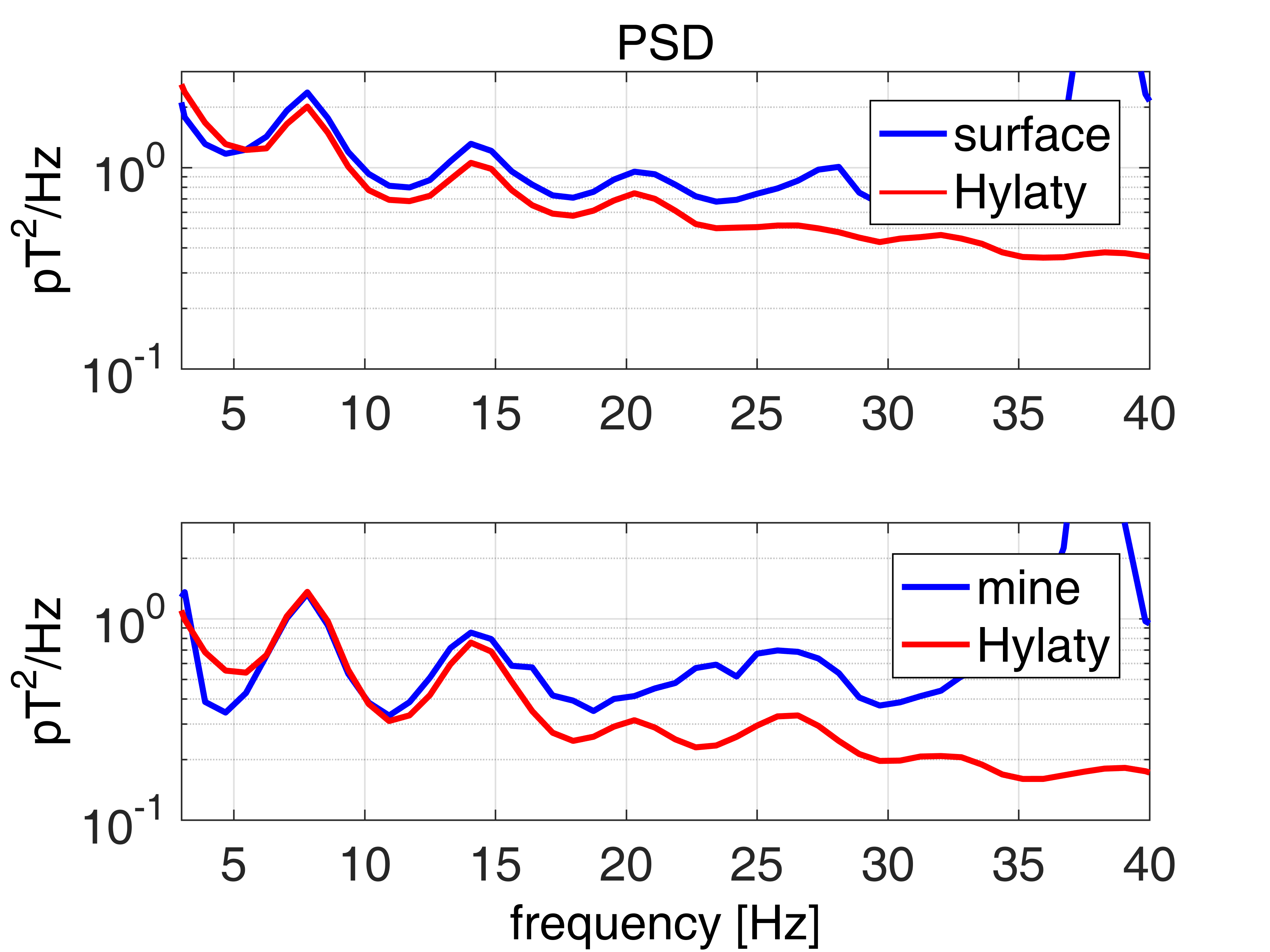}
\par\end{centering}
\protect\caption{Median of power spectral density of the EM noise of the four dataset: observation at the subsurface site in Period I., observation at the reference site in Period I., observation at the surface site at Period II. and at the reference site in Period II., respectively.}
\label{PSD}
\end{figure}

In the first approach no outlier removal algorithm has been utilised, but the simplest robust statistics, the median of the spectral power has been determined at each frequency values. For the estimation of the signal attenuation with the depth, the median of the power spectral density (PSD) has been computed related to the four datasets, $B_{I. mine}$, $B_{I. Hylaty}$, $B_{II. surface}$ and $B_{II. Hylaty}$. The four PSDs have been plotted in Fig. \ref{PSD}.

All four median PSD functions have been determined from the Fourier transform of 1024 sample long consecutive time windows, with 512 sample overlapping between adjoining segments, using Hamming window. For an estimation of the power attenuation with the depth, the power of the fundamental Schumann spectral component has been determined in case of all four observations. The following subsection introduces a method which has been applied to have accurate estimation of the fundamental Schumann resonance spectral power peaks related to the four observations.

\subsubsection{Polynomial baseline approximation}
\begin{figure}
  \centering
  \includegraphics[width=.4\columnwidth]{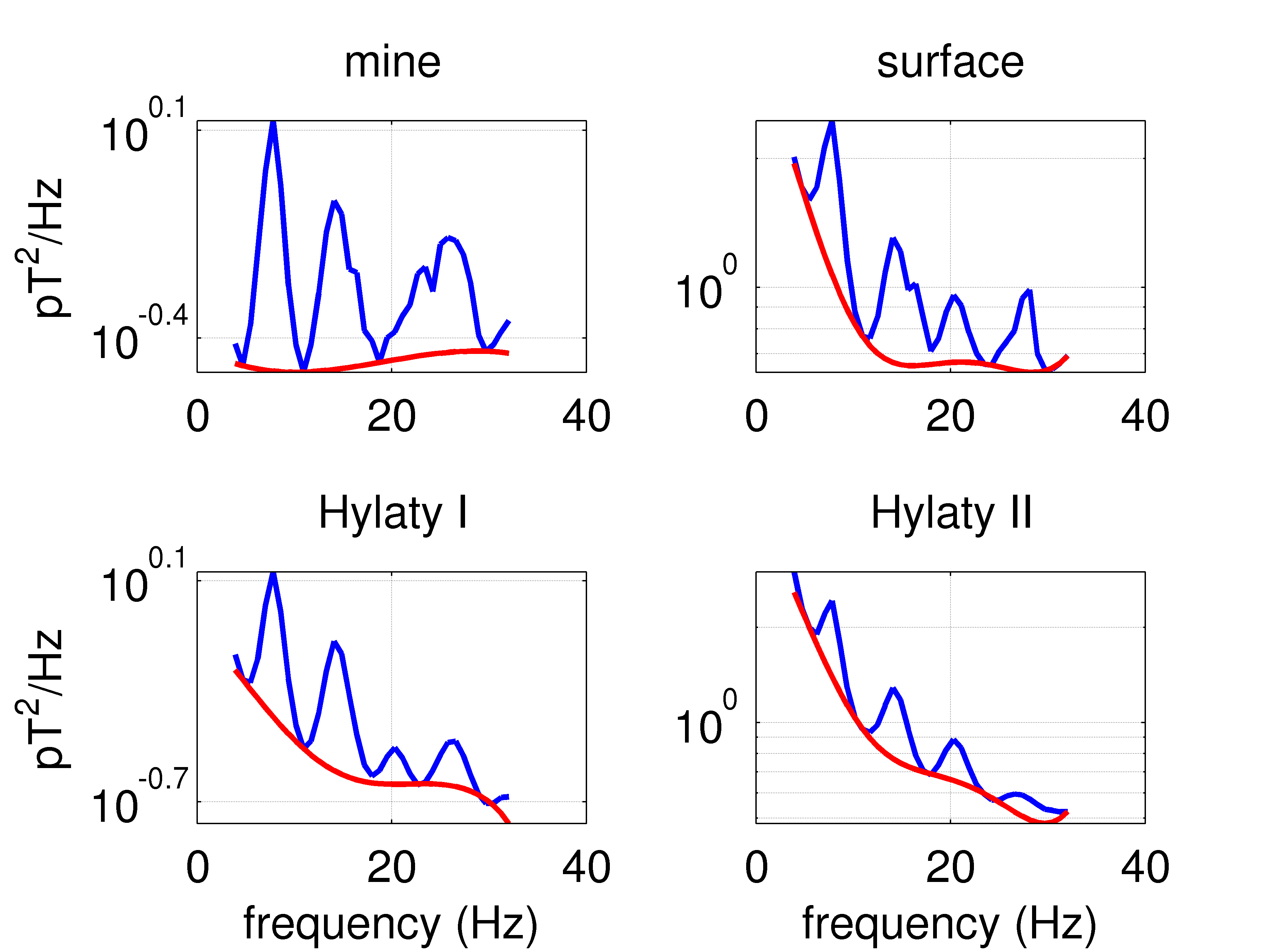}  \label{withbl}
  \includegraphics[width=.4\columnwidth]{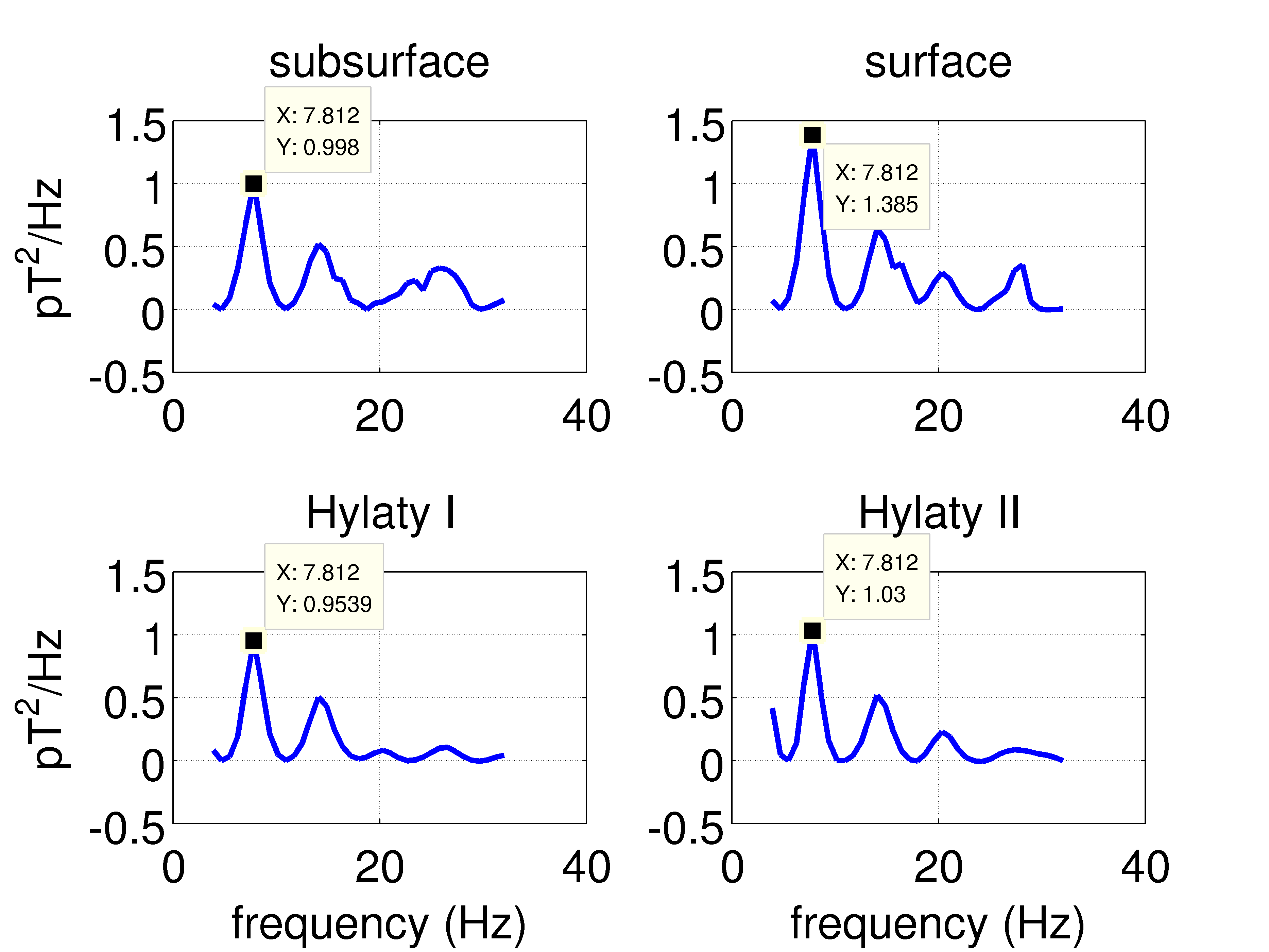}  \label{blremoved}
\caption{Polynomial baseline fitting on median PSD of the four dataset \textit{(left)} and median PSDs after baseline removal \textit{(right)}.}
\label{bl}
\end{figure}

The method to provide an estimation of the ELF range attenuation is based on utilisation of an advanced background baseline removal technique. In this method the baseline is estimated by fitting a low-order polynomial, but instead of obtaining parameters by minimising the sum of squares of the residuals a non-quadratic cost function is adapted \cite{Mazet2005}. Third order polynomials have been considered to fit the median PSDs in a limited frequency range, see left four subplots of Fig. \ref{bl}. All polynomials have been fitted by means of asymmetric truncated quadratic cost function.

The fitted baseline polynomials have been substracted from the power spectra, plotted on the right four subplots in Fig. \ref{bl}. The power ratio obtained at 7.81 Hz in the subsurface-Hylaty relation is 1.126, for the surface-Hylaty relation 1.219 which results a power ratio of 0.924. The square root of the power ratio results the estimated amplitude attenuation: 96.1\%. The rate of signal amplitude between the surface and at depth \textit{h} can generally be computed by writing the exponential decay of the magnetic signal amplitude as follows:

\begin{equation}
\frac{A(h)}{A(0)} = \exp(-h/\delta)
\end{equation}

By substitution of the empiric amplitude rate values and the 140m depth of the subsurface station the characteristic depth of the amplitude attenuation, the so called the \(\delta\) skin-depth can be estimated. It results \(\delta= 3520 \) m. The power decrease of the signal with the depth is a consequence of the nonzero electrical conductance of the underlying/overhead andesite rock. Providing an estimation of the bulk resistivity of the andesite block is possible by a simplified approach of homogenious half space approximation. This estimation is considered valid if the homogenious halfspace assumption is realistic. A simple approximation formula between the skin-depth and the bulk electrical resistivity is as follows \cite{Voz91b}:
\begin{equation}
\delta = 500\,\sqrt{\rho \times T}.
\end{equation}
Extracting the bulk resistivity by substitution of estimated skin-depth value of the Schumann fundamental resonance frequency 387 \(\Omega\)m is obtained. The estimated resistivity value is in the wide characteristic range of electrical resistivity of the andesite rock in the literature (170-45.000\(\Omega\)m)\cite{SliTel42a}. %, but may fall a bit below what is expected in the case of the Matra.

\subsubsection{Time domain investigation}
\begin{figure}
\begin{centering}
\includegraphics[width=.8\columnwidth]{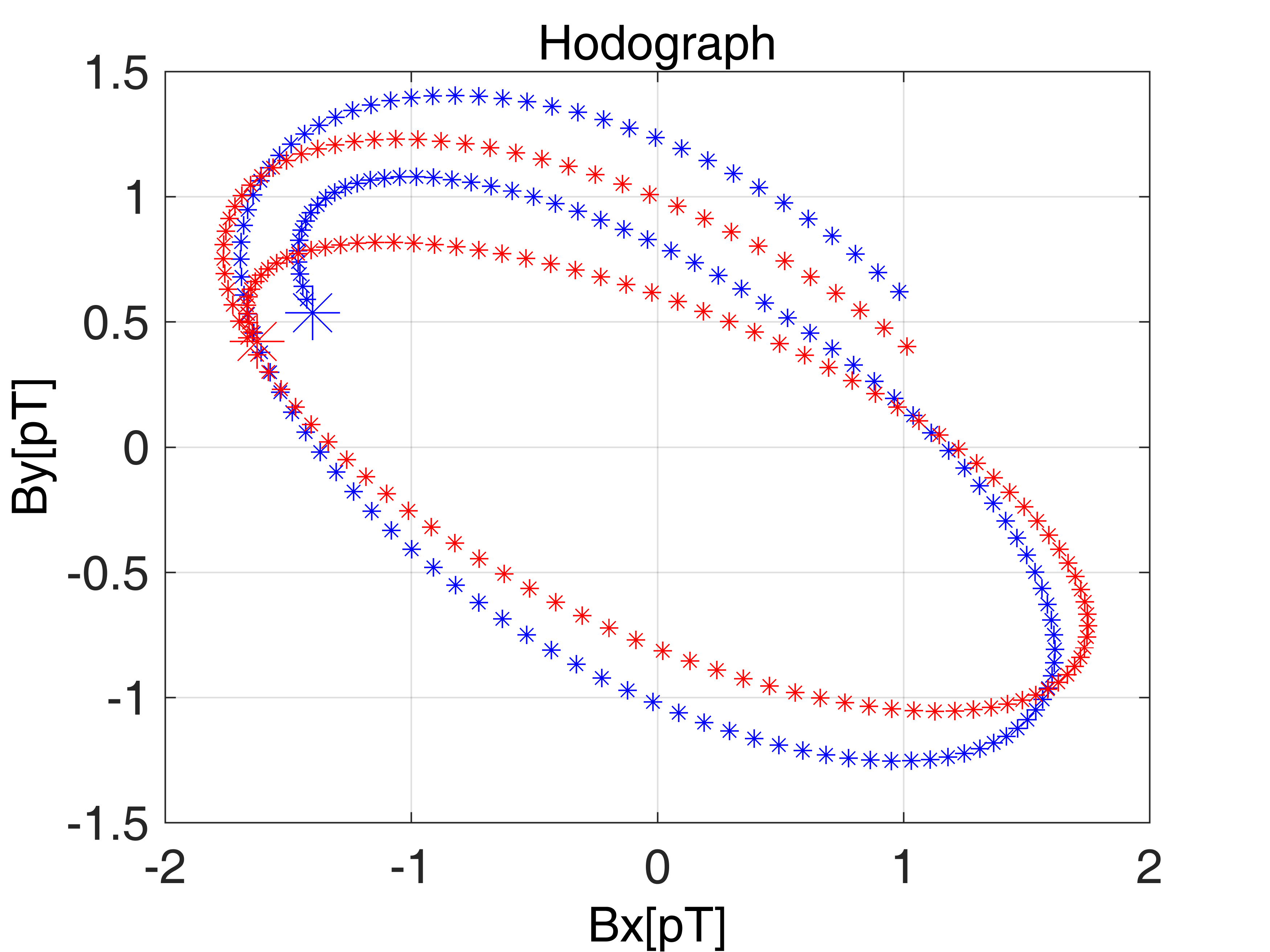}
\par\end{centering}
\protect\caption{Hodograph of a randomly chosen segment of subsurface \textit{(blue)} and the corresponding reference observation window \textit{(red)} stars respectively.}
\label{Hodograph}
\end{figure}

A recent electromagnetic study of ELF range magnetic signal attenuation could have been explained by the assumption of distortion effect of iron artifacts of the mine infrastructure. For this reason the time domain inspection of the bandpass filtered subsurface data has been carried out. Bandpass filter is tuned below common power grid frequency and below upper Schumann frequencies to have comparable pattern of the hodographs. Based on this, 5.6, 10Hz cut off frequencies has been set. About 100 sample long time windows have been picked from the subsurface data and the corresponding reference time series. The magnetic variation of a time segments has been plotted and compared in Fig. \ref{Hodograph}.

The hodograph demonstrates no strong deviation of the subsurface signal is present confirming that iron parts of the mine infrastructure did not have recognizable effect on the subsurface observation and the estimated attenuation rate at the subsurface measurement site. The noisy environment, the amount of data utilised and the indirect comparation of the signal transformation all increase the deviation of the results and the variance of the power spectra. Considering all these circumstances the obtained attenuation value confirmes that in less noisy intervals, parallel measurement of surface and subsurface relation together can significantly improve the quality of the data and may provide more accurate estimation of the amplitude attenuation in the ELF range. Application of outlier removal may result further improvement on the quality of the estimation, too. The main goal of the study was to provide an estimation of the magnetic signal attenuation with depth. The results confirm that even an indirect estimation method can be promising. The estimation of the attenuation coefficient of the overhead andesite block might be more accurate by means of the data observed in an upcoming EM silent time window and even the more with the possibility of parallel recordings at the surface and subsurface sites. These two circumstances should improve the accuracy of the estimation significantly.

%\pagebreak

\section{Infrasound measurements}
%Please indicate the authors of the section together with their affiliation, like it is  in the introduction.
%Moln\'ar J. and Fenyvesi E.

\subsection{Introduction}

Sensitivity of second generation gavitational-wave detectors at low frequencies are limited fundamentally by the thermal noise of the suspension last stage and of the test masses, the seismic noise, and the gravitational gradient noise (so-called Newtonian noise (NN) ) resulted by changes in density of medium near the detector. Seismicity is one source of this type of noise, but changes in density of air caused by pressure waves are also contribute to NN \cite{NewtonianNoise}.

Inspired by recent theoretical investigations of NN from the infrasound below the Earth surface \cite{NewtonianNoise}, we have continued our infrasound measurements in MGGL. For these measurements, new infrasound microphone (ISM) was developed by the Institute for Nuclear Research of Hungarian Academy of Sciences (Atomki) \cite{BarEta17a}. The first instrument was installed at MGGL in 2016 with the purpose on one hand to test its performance on the long-term, and on the other hand to characterize the underground site's ambient infrasound noise.

In this study, Bowman's low- and median noise models were used as a reference for our site characterization. According to Ref. \cite{Bowman}, the main sources of infrasound noise on the surface of Earth are microbaroms and the wind. These sources are not present underneath, but other sources related to the miner's work and machines produce infrasound. Our results show that ambient infrasound presumably can be supressed when going under the ground, but the local noise sources and the geometry of the underground site influences the possible extent of the noise supression at different frequencies.

\subsection{The instrument}
ISM consist of two main parts: an industrially manufactured capacitive sensor and C-V converter, developed by Atomki.  The capacitive sensor is a differential pressure sensor with two inlets. One inlet is connected to the environment,  while the other is connected to a reference volume. A diaphragm is placed in between the two pressure inlets. The diaphragm is deflected when the difference between the environment and the background pressure changes. The deflection is converted into an electrical signal by the sensor, then the signal is amplified to the  $-10-10$ V range. There is a leak with a diameter of $250$ microns, which connects the background volume to the environment. It makes possible to slowly equalize the background pressure and the environmental pressure, and hence determines the lowest frequency of the measuring range. %The instrument is equipped with a security valve, which is closed during measurement, but has to be opened when ISM is transported by plane, because large pressure changes could be destroy the membrane of the sensor.
The main characteristics of two versions of ISM are given in Table \ref{id_characteristics}.

\begin{table}
\centering
	\begin{tabular}{{ | l | c | c | }}
		\hline
		& \textbf{ISM1}     & \textbf{ISM2}    \\ \hline
		Input voltage (V)    & 6-10       & 7.5-12  \\ \hline
		Output volt. (V)     & 0-5        & -10-10  \\ \hline
		Sensitivity (V/Pa)   & 0.2        & 1       \\ \hline
		Pressure range (Pa)  & -12.5-12.5 & -10-10  \\ \hline
		Frequency range (Hz) & 0.01-10    & 0.01-10 \\ \hline
		Temperature sensor   & yes        & yes     \\ \hline
		Humidity sensor      & yes        & no    \\ \hline
	\end{tabular}
	\caption{\label{id_characteristics} The characteristics of the two infrasound sensors, ISM1 and ISM2, developed in ATOMKI.}
\end{table}

The drawback of the structure of the instrument is that fast changes in temperature $dT/dt>2 \ ^{\circ}$C/h introduce artifacts in spectrum below 1 Hz. When the frequency of the temperature changes is in the measuring range of the detectors, then the temperature change initiated pressure in the background volume cannot be distinguished from the pressure change due to environmental infrasound. Therefore a high-resolution temperature sensor is built into the detector.

\subsection{Data acquisition system and data processing}
The microphones have analog output. In order to sample, digitalize and store measurement data, a low-cost data acquisition system is added to the sensors, based on a Raspberry Pi 3 model B \cite{BarEta17a}. %An Adafuit ADS1115 ADC module converts analog signals to digital data which is stored on an SD card. The timing is obtained either with a built in DS3231M real-time clock, or from NTP servers in case of suitable internet connection. % When internet connection is accessible, timestamps for collected data can be obtained from NTP servers. In order to gain timestamps with high accuracy when proper internet connection is not accessible, the RPI3 was fitted with a DS3231M real-time clock. In the MGGL the collected data was forwarded from DAQ’s SD card to a PC through UTP cable. This PC has and optical connection with the MGGL server out of the mine. Through this server, data can be downloaded via internet.

The spectral properties of the infrasound background are characterised by the Pressure Amplitude Spectral Density ($PASD$) and the analysis of the data is similar to the seismological noise evaluation in Section \ref{sec:seismo}. The collected data was downsampled to $100$ Hz, then divided into $163.84$ s long segments. For each segment, PASD was computed. For each day, the $5\textsuperscript{th}$, $50\textsuperscript{th}$ and $95\textsuperscript{th}$ percentile of corresponding PASD data was determined.

\subsection{Comparative measurements with MB3 microbarometer}
In order to compare performance of ISM1 with the performance of MB3 microbarometers \cite{MB3}, comparative measurements were performed on 2018-04-26 at Infrasound station no. 3 of Piszkéstető observatory in M\'atra mountain range, Hungary. This station is part of the Hungarian National Infrasound Network, which is a permanent infrasound network operated by the Kövesligethy Radó Seismological Observatory of the Hungarian Academy of Sciences \cite{Infra_network}.

At Piszkéstető, the MB3 microbarometers are placed in a reservoir, which is partly covered by soil. MB3 is equipped with a wind reducing system of long pipes attached to the microbarometer. During our measurements, the pipes were detached, and the MB3 and ISM1 were placed next to each other in order to detect similar pressure as much as it is possible. The measurement lasted for three hours.

\begin{figure}[h]
	\centering
	\includegraphics[width=0.75\textwidth]{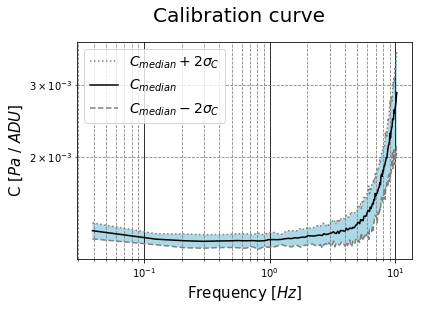}
	\caption{Calibration curve used to convert PASD given in $ADC units/\sqrt{Hz}$ of ISM1 to PASD in $Pa/\sqrt{Hz}$.}
	\label{calibration_curve}
\end{figure}

Our aim was to calibrate our instrument, that is to determine  the real PASD values in Pa/$\sqrt{Hz}$ from the PASD$_{ISM1}$ given in $ADCunits/\sqrt{Hz}$ computed from measurement data. During the data processing part, raw signals both of MB3 and ISM1 were cut into $128$ seconds length segments and PASD$_{MB3}$ and PASD$_{ISM1}$ of each segment was computed. Then values corresponding to a given frequency of PASD$_{MB3}$ and PASD$_{ISM1}$ pairs from the same time interval were collected. The number of pairs corresponding to a given frequency is equal to the number of segments used for calibration. For a given frequency, the ratios of PASD$_{MB3}$ values and PASD$_{ISM1}$ values were computed (C=PASD$_{MB3}$/PASD$_{ISM1}$). The distribution of these ratios corresponding to a given frequency can be approximated with normal distribution. Standard deviation ($\sigma_{C}$) of the distribution was estimated by using median absolute deviation (\cite{MAD}). Two standard deviations were chosen to express the uncertainty of the ratio in Fig. \ref{calibration_curve}.

\subsection{Ambient infrasound noise at MGGL}
\begin{figure}[h]
	\centering
	\includegraphics[width=0.75\textwidth]{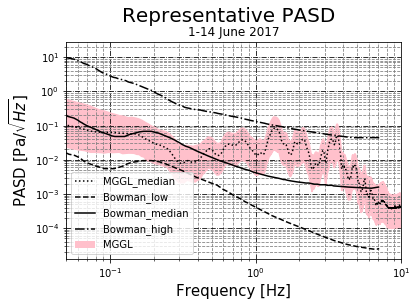}
	\caption{The pressure amplitude spectral density (PASD) of the ISM1 infrasound detector in the period 2017-06-01 -- 2017-06-14. The dashed and solid black lines are the Bowman low noise and median curves.}
	\label{representative_PASD}
\end{figure}

This study shows the results of a two-week long period of measurements that started at 2017-06-01. This is the period of the deep seismic noise measurement campaign, when the ET1H seismometer in MGGL at $-88$ m worked parallel with GU02 seismometer at $-404$ m in the mine. The colored area shows the interval between the 5\textsuperscript{th} and 95\textsuperscript{th} percentiles  in Fig. \ref{representative_PASD}. One can see, that the noise at MGGL roughly falls between Bowman's low-noise and high-noise curves. The particular shape of the PASD bar on Fig. \ref{representative_PASD} may be resulted by the geometry of the tunnels in the mine and the other circumstances related to the ongoing reclamation (e.g. sound from machines). The size of laboratory itself allow resonances with higher frequencies, out of the infrasound range. Median noise of MGGL falls below Bowman's median-noise curve only below $0.4$ Hz and above $5$ Hz, and newer falls below the low-noise curve. These results are showing that machinery and other activities at an underground site have to be minimized in order to take advantages of installing gravitational detectors under the ground.

\subsection{Relationship between seismic and infrasound noise in the MGGL}
\begin{figure}
\centering
    \includegraphics[width=.48\textwidth]{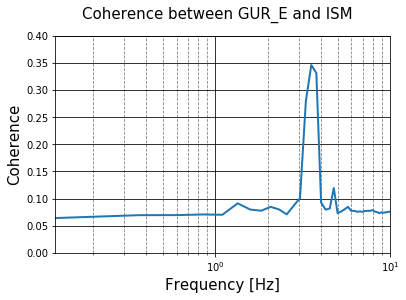}
    \includegraphics[width=.48\textwidth]{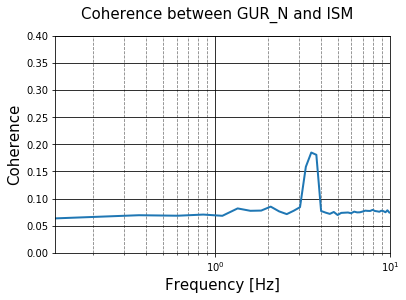}
    \includegraphics[width=.48\textwidth]{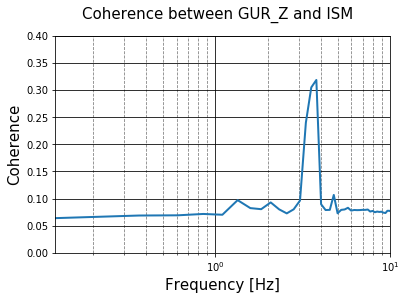}
  \caption{Average coherence between the three components (E: East, N: North, Z: vertical) of the seismic signal and the infrasound signal.}
  \label{coherence}
\end{figure}

In order to examine the relation between the signal of the seismometer and the signal of ISM1, we have computed the magnitude squared coherence estimate for each 128 s long data segment pairs from the deep measurement campaign of MGGL by using Welch’s method \cite{coherence}. In the case of the ET1H seismometer's signal, we treated the tree directions, East, North and Z (vertical) separately. This way, we got three datasets for the coherence: $C_{v_{E}p}$, $C_{v_{N}p}$ and $C_{v_{Z}p}$. We determined the average of the coherence values of each frequency of the three datasets. A peak between 3 and 4 Hz can be observed on each plot on Fig. \ref{coherence}, which indicates a common source has an effect both on the seismometer and ISM1. There were no similar coherences with the farther, GU02 seismometer. The drainage system or the ventilation system may be responsible for the peak. We will prove or refute this hypothesis with the forthcoming experiments in MGGL.

%\pagebreak

%---------------------------------------------------------------------------
\section{Measurements of the inhomogeneity and the rock-density at large-scale with cosmic muon tracking}
%%%%%%%%%%%%%%%% AUTHORS

%	\author{Barnaf\"oldi G.G.$^1$, Hamar, G.$^{1}$, Sur\'anyi G.$^{2}$, Ol\'ah, L.$^{1,2}$, Varga D.$^{1}$,
%
%
%	\address{$^1$HAS, Wigner Research Centre for Physics, Institute of Particle and Nuclear %Physics, 1121 Budapest, Konkoly Thege Miklós út 29-33.
%	$^2$G eological, Geophysical and Space Science Research Group of the HAS, 1/A %P\'azm\'any P. s., H-1117, Budapest, Hungary \\
%	$^3$Earthquake Research Institute, The University of Tokyo, 1-1-1 Yayoi, Bunkyo-ku, Tokyo, Japan 113-0032 \\
%}

%Please indicate the authors of the section together with their affiliation, like it is  in the introduction.
% Varga D. and Ol\'ah L. $^{1}$
% $^1$HAS, Wigner Research Centre for Physics, Institute of Particle and Nuclear Physics, \\
%1121 Budapest, Konkoly Thege Miklós út 29-33.
%%%%%%%%%%%%%%%%%%%%%%%%%%%%%%%%%%%%%%%%%%%%%%%%%%%%%%%%%%%%%%%%%%%%%%%%%%%%%%%%%%%

A novel cosmic muon tomography technology were used to explore the precise, large-scale density structure of the M\'atra mountain above the MGGL within the Gy\"ongy\"osoroszi mine. The idea is to measure the angular distribution of the  highly penetrating muon-component flux of the cosmic-rays, which are created in the upper atmosphere and reach the surface of the Earth. This technique, has been applied successfully by our group for various cases such as cave researches, cosmic background measurements for underground laboratories, civil engineering, and volcano muography~\cite{mt_nima:2012, mt_geo:2012, mt_ul:2013, mt_ecrs:2015, mt_npa:2016,mt_nature:2018}.

\begin{figure}[!h]
	\centering
        \includegraphics[width=4cm]{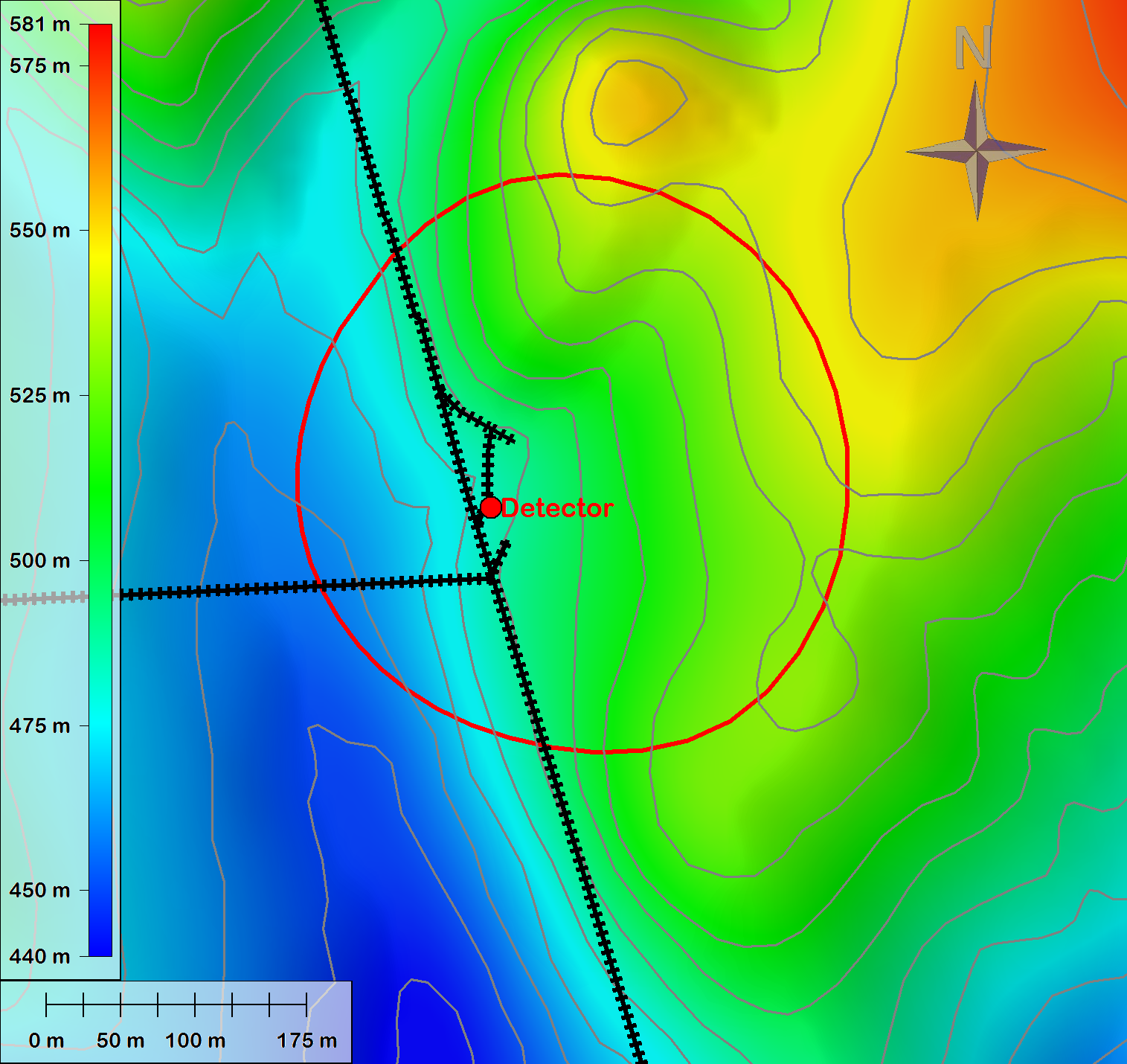}
        \includegraphics[width=4.5cm]{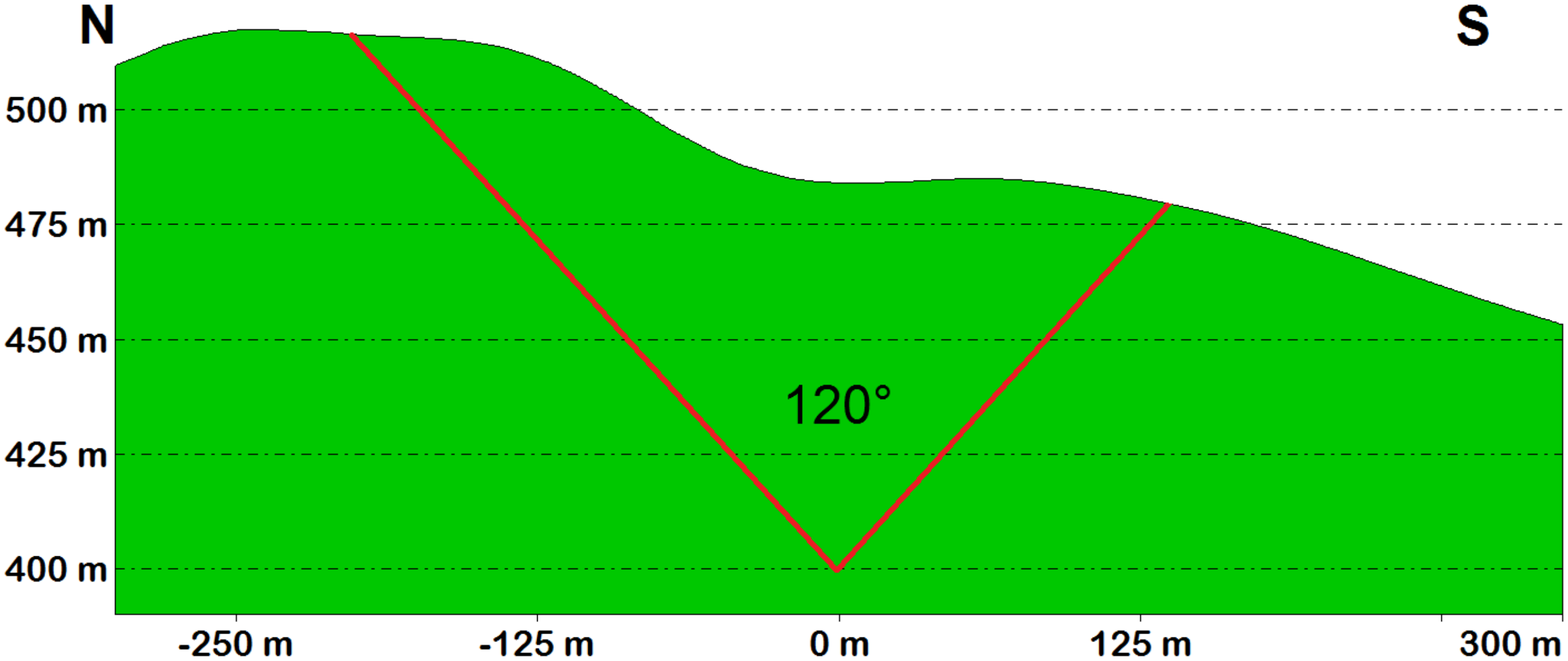}
        \includegraphics[width=4.5cm]{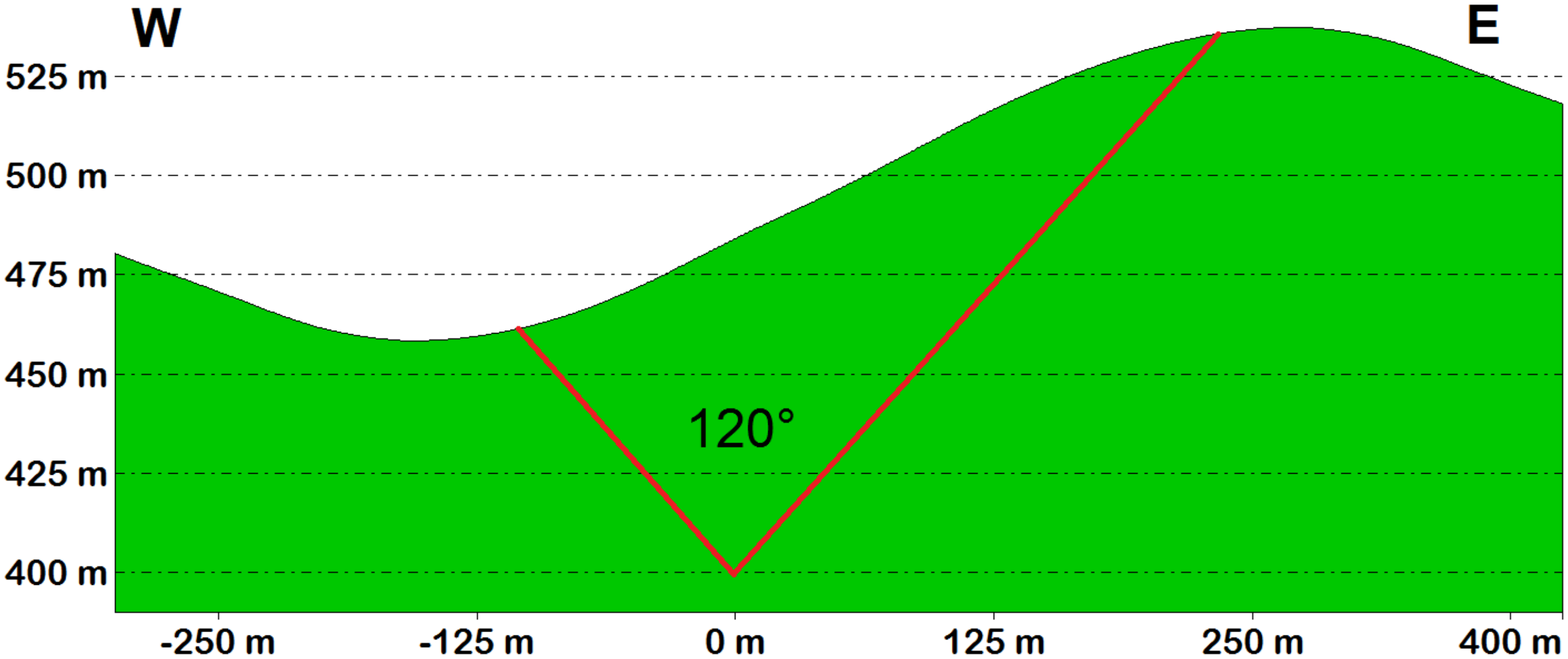}
	\caption{Left: the position of the MGGL Laboratory in the M\'atra Mountain and the acceptance of the mun-flux detector with $60^{\circ}$ zenith angle. Color-scale contours and dark contures show the elevation in m.a.s.l. Right: West-East and North-South cross-sections of the M\'atra Mountain at the MGGL and the acceptance of the mun-flux detector with $60^{\circ}$ zenith angle.}
	\label{fig:surface-map}
\end{figure}

To reach the full, 2$\pi$ coverage, muon flux measurements were continued, using a portable tracking system, called the Muontomograph. This has been developed by the Wigner Research Centre for Physics and described in details in Ref. \cite{BarEta17a}.  Based on the Run-0 experiences, we made modifications on the Muontomograph in order to get better performance in the MGGL laboratory for the Run-1. The position of the MGGL and the acceptance directions of the detector related to the surface topology are shown in Fig. \ref{fig:surface-map}. We adjusted the trigger levels for higher efficiency and a noisy chamber were also removed, thus the number of chambers were reduced to five.

%%%%%%%%%%%%%%%%%%%%%%%%%%%%%%%%
\subsection{Cosmic muon flux measurements in MGGL}

During Run-1 the cosmic muon flux measurements were also continued in the Gy\"ongy\"osoroszi mine. The Muontomograph was located at the Northwest corner of the laboratory room at a depth of 88 m. We used the same position in the MGGL, where the detector were placed earlier in the Run-0 period. The duration of Run-1 was all together 280.0 days in four compass courses. Jointly with the 124.7 days of Run-0 measurements we have 404.7 days of data taking so far.

Since we have covered the zenith- and two 90$^{\circ}$-tilted directions in Run-0, the four Run-1 measurements were all tilted to 45$^{\circ}$ from the zenith. With this choice we could cover the upper hemisphere fully, without any missing direction. The first measurement was directed to 20.5$^{\circ}$ for 77 days, then we continued to 110.5$^{\circ}$, 200.5$^{\circ}$, and 290.5$^{\circ}$ directions, respectively for 62 days, 77 days, and 64 days. Further details of the measurements are summarized in Table~\ref{tab:mt-meas}. Data were monitored and downloaded continuously from the detector via Ethernet connection during the data taking period in MGGL for online checking of data quality.
%%%%%%%%%%%%%%%%%%%%%%%%%
\begin{table}[!h]
\caption{\label{tab:mt-meas} The summary table of the Run-0 and Run-1 measurements in MGGL: the detector principal direction in azimuth to the magnetic North and the tilt in zenith, the duration of the measurements, number of events, and the number of measured muon tracks.}
\begin{center}
\begin{tabular}{lcccccc}
\hline
 Runs & Azimuth & Zenith & Duration (days) & Events & Num. of tracks\\
\hline
Run-0-M2&  zenith & $ 0^{\circ} $ & 48.3 & 4.7 M & 111,700 \\
Run-0-M3&  ENE$(65.5^{\circ}) $ & $ 90^{\circ} $ & 41.9 & 2.4 M & 18,124 \\
Run-0-M4&  NNW$(335.5^{\circ}) $ & $ 90^{\circ} $ & 34.5 & 3.1 M & 12,356  \\
\hline
Run-1-M61&  NNE$(20.5^{\circ}) $  & $ 45^{\circ} $ & 77.0 & 423.3 k & 49,827 	\\
Run-1-M62&  ESE$(110.5^{\circ}) $ & $ 45^{\circ} $ & 62.0 & 366.9 k & 35,534	\\
Run-1-M63&  SSW$(200.5^{\circ}) $ & $ 45^{\circ} $ & 77.0 & 367.4 k & 42,414	\\
Run-1-M64&  WNW$(290.5^{\circ}) $ & $ 45^{\circ} $ & 64.0 &  316.0 k & 70,099 \\
\hline
\hline
\end{tabular}
\end{center}
\end{table}
%%%%%%%%%%%%%%%%%%%%%%%%
Comparing Run-0 and Run-1 date in Table~\ref{tab:mt-meas}, one can see that, although the event number dropped by an order of magnitude, the fraction of the track number increased relatively to the number of events. This means, that the efficiency become better since noise and fake events were suppressed well in Run-1.

\subsection{Results of the rock inhomogeneity measurements}

\begin{figure}[!h]
	\centering
	\includegraphics[width=7cm]{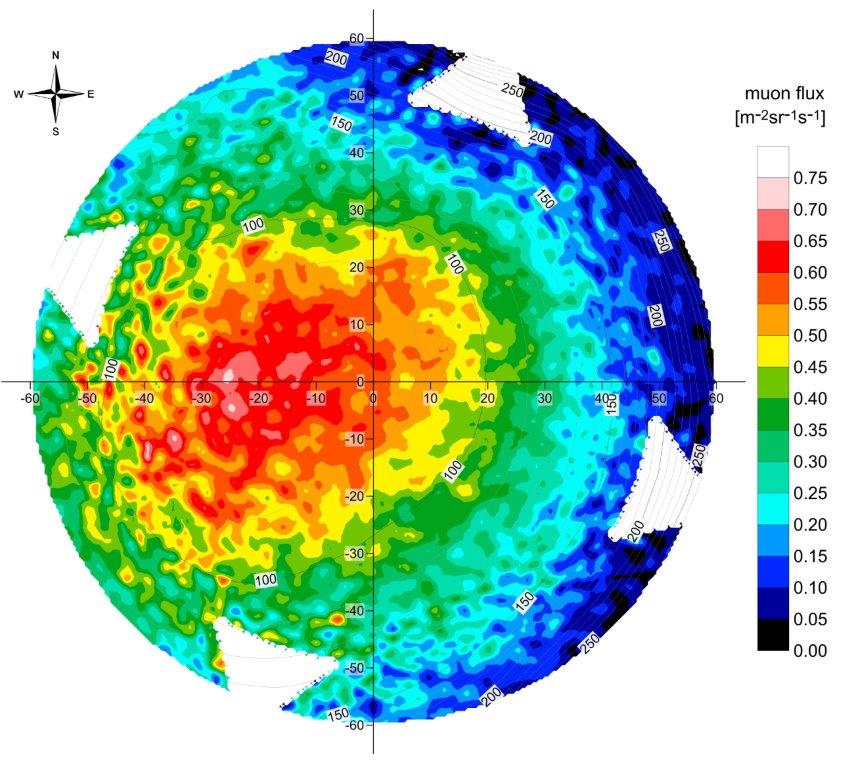}
        \includegraphics[width=7cm]{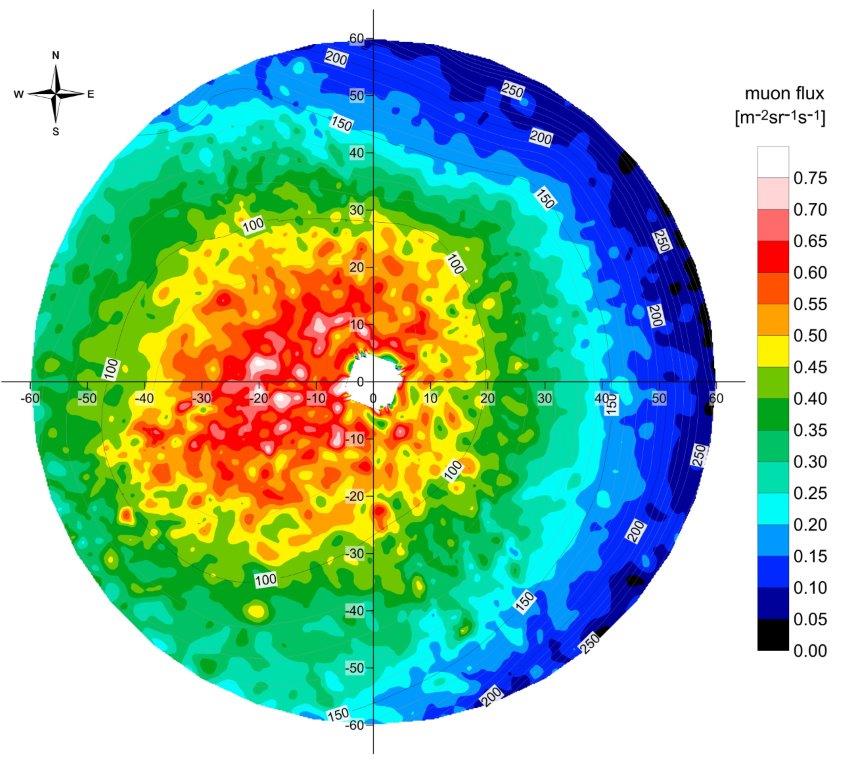}
	\caption{The cosmic muon flux map measured in MGGL is plotted as a function of azimuth and zenith angles from the detector position. Color-scale contours show the muon flux, dark contour lines show the detector-to-surface distance in meters. White areas show the out-of-acceptance directions. {\sl Left} and {\sl right panels} present the data respectively from the Run-0 and Run-1 measurements.}
	\label{fig:MTLokpolTopo}
\end{figure}

The muon flux distribution can be obtained after applying a track reconstruction algorithm and merging the maps according to the partial overlap and geometry normalization~\cite{mt_npa:2016}. The Run-0 and Run-1 flux maps are shown in the {\sl left} and {\sl right panels} of Fig.~\ref{fig:MTLokpolTopo} respectively. The muon flux distribution is plotted by color-scale contours as a function of azimuth and zenith angles.

The cosmic muon (track) rate was 0.005-0.02 Hz for both runs, which is sufficient to provide flux measurement with $5-50 \%$ statistical errors (depending on zenith angle). The measured flux has a maximum value of 0.7 m\textsuperscript{-2}sr\textsuperscript{-1}s\textsuperscript{-1} at $15^{\circ}$ zenith angle to the West. In  Fig.~\ref{fig:MTLokpolTopo} white color 'triangles' at $20^{\circ}$ zenith angle ({\sl panel left}) and 'dial' at zenith ({\sl panel right}) represents the directions, which are out of the acceptance of the current run.

The detector-to-surface distance is indicated with dark contour lines in the Fig.~\ref{fig:MTLokpolTopo}. For Run-0 in \cite{BarEta17a}, we used Shuttle Radar Topography Mission (SRTM) satellite data for the elevation map with an estimated relative error of 10\% (10 m at the zenith).
However by comparing the SRTM model with other available elevation data the accuracy of the SRTM model seemed much poorer than it was previously expected, likely because of the different height of the vegetation. Thus the elevation contours obtained from the National Hungarian Grid (EOV) have been used instead. Applying this, new detector-to-surface data, give better correlation for both Run-0 and Run-1.

\begin{figure}[!h]
	\centering
        \includegraphics[width=12cm]{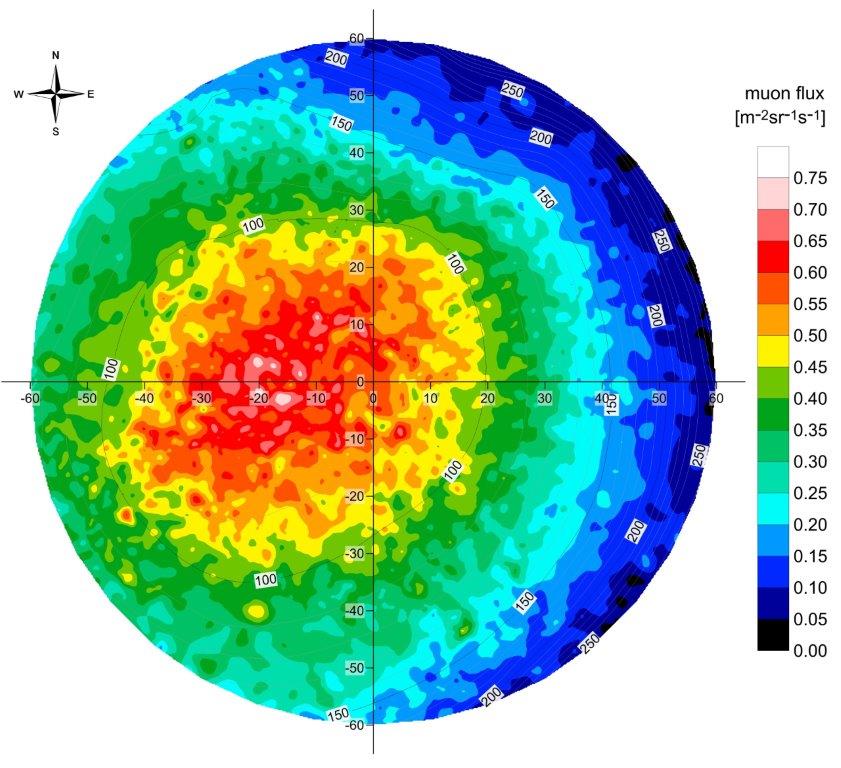}
	\caption{The cosmic muon flux map of Run-0 and Run-1 measured in MGGL plotted as a function of azimuth and zenith angles from the detector position. Color-scale contours show the muon flux, dark contour lines show the detector-to-surface distance in meters.}
	\label{fig:MTLokpolTopo_All}
\end{figure}

Merging all the data of Run-0 and Run-1 the muon flux distribution can be obtained, which is plotted in Fig.~\ref{fig:MTLokpolTopo_All} up to zenith angle 60$^{\circ}$. Correlation between muon flux map and the detector-to-distance contour lines is excellent. We found fluctuation is in the order of $\pm 0.05$ m\textsuperscript{-2}sr\textsuperscript{-1}s\textsuperscript{-1} as it was obtained in Run-0 as well. One can observe that the surface distance and the muon flux reasonably match each other in Fig~\ref{fig:MTLokpolTopo_All}.
Combining this with the acceptance plotted on the surface map plot on Fig.~\ref{fig:surface-map} the maximal flux matches the shortest distance-to-surface directions on the hill's West slopes, while to the East, the muon flux drops at the large zenith angles with the longest traveling lengths.

\begin{figure}[!h]
	\centering
        \includegraphics[width=12cm]{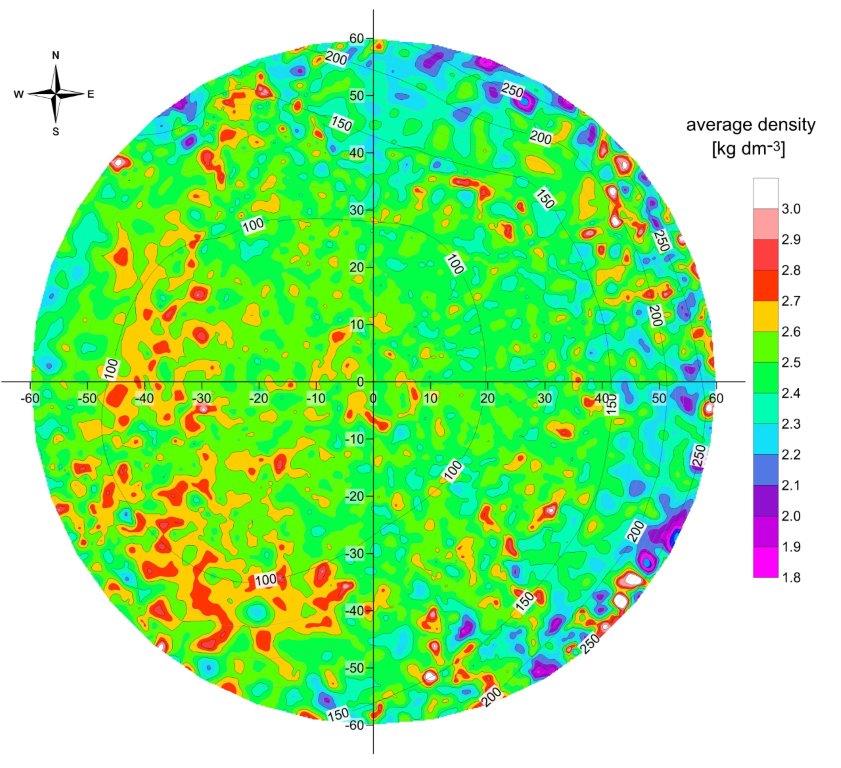}
	\caption{The angular distribution of the density based on Run-0 and Run-1 measured from the MGGL is plotted as a function of azimuth and zenith angles from the detector position. Color-scale contours show the rock densities, dark contour lines show the detector-to-surface distance in meters.}
	\label{fig:DensTopo}
\end{figure}

The collected $\sim 340,000$  tracks of the muon flux measurement provided high-enough statistics to estimate the rock thickness, based on the muon absorption model~\cite{mt_Tang:2006uu}. This model has been used successfully in other, similar measurements up to few 100 meters depth. In Fig.~\ref{fig:DensTopo} the angular distribution of the rock density is presented. The average of the rock density is $2.6 \pm 0.1$ kg\,dm\textsuperscript{-3}, which is the density of the andesite-based host rock. The data do not show any large-scale density inhomogeneities or cavities above the MGGL. However, minor, $\sim 0.2$ kg\,dm\textsuperscript{-3}, positive density deviation ridge is observed along the $40^{\circ}-50^{\circ}$ zenith angle from the southwest to northwest. This may correspond to several percent metallic ore in the unexplored part of the Gy\"ongy\"osoroszi mine. A negative $\sim 0.5$ kg\,dm\textsuperscript{-3} density anomaly valley was also observed at the lowest $50^{\circ}-60^{\circ}$ zenith angle from the south to north. This is the direction of the known tunnels and caverns of the mine, although here the uncertainty of the measurement is also significant at large zenith angles.

\subsection{Conclusions} Based on the muon-flux measurements, the improved detector-to-surface date, and the muon absorption model, we can state, that the rock density map correlates well with the unexplored and mined areas above the MGGL. This unique measurement support to place the Einstein Telescope at Gy\"ongy\"osoroszi mine, since the lack of large-scale density inhomogenities would reduce the higher-order tensor-like gravitational corrections during the measurements.
%
%\begin{figure}[!h]
%	\centering
%        \includegraphics[width=12cm]{pics/qwe_6.png}
%	\caption{The position of the MGGL Laboratory in the M\'atra Mountain and the acceptance of the mun-flux detector with $60^{\circ}$ zenith angle. Color-scale contours and dark contures show the elevation in m.a.s.l.}	\label{fig:surface-map} \end{figure}

%+\pagebreak

\section{Summary \label{sec:Conclusion1}}

The Mátra Gravitational and Geophysical Laboratory was established to investigate on a long term basis the conditions and requirements of next generation gravitational wave detectors in case of underground construction and operation. In more general, the aim was to measure various geological and rheological properties of the base rock in addition to test experimentally novel theoretical approaches on the noise penetration and suppression. Also the Mátra mountain range is studied as a possible site of the planned Einstein Telescope.

Our first investigations were published in Ref.~\cite{BarEta17a} including the technical details of the measurements. In this paper we summarize the first two years of measurements of the laboratory by different methods: geophysical environment, electromagnetic attenuation, infrasound noise, cosmic muon tomography of the sorrounding rock mass, and long term seismic noise. The timeline of the data taking periods of the various measurements are shown in Fig. \ref{fig:datatakeingper}. Our topical results are:

\begin{figure}
\centering
\includegraphics[width=0.8\textwidth]{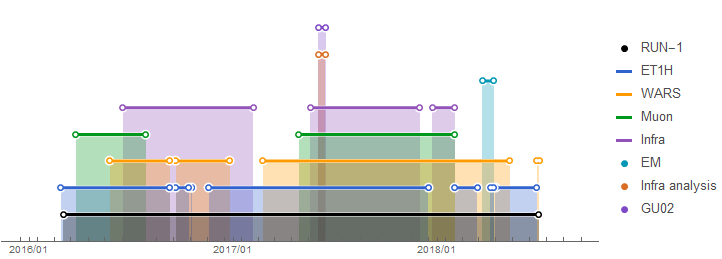}
\caption{\label{fig:datatakeingper} The data taking periods of the various measurements in MGGL during RUN-1.}
\end{figure}

\begin{description}
\item[Geophysical environment] A short survey of the geological and seismological conditions in the Mátra mountains reveals a homogeneous composition of hard andesite rock with low seismic activity. We performed rock mechanic experiments and shown, that a typical grey andesite from the vicinity of the laboratory is not ideal elastic. We characterised and modeled the deviation from ideal elasticity and determined the corresponding rheological parameters (see Tables~\ref{tab_reopar} and~\ref{tab_bpar}). {\sl The independent measurements of dynamic elastic moduli verify our novel rheological model, which properties are important when calculating the Newtonian noise from rock deformations}.

\item[Electromagnetic attenuation] With EM noise measurements we estimated the electromagnetic attenuation by the andesite rock mass in the lower ELF range, especially at the frequency of the first Schumann resonance component.  Comparing the data of the external surface reference station, we obtained the skin depth $3520$ m. For the bulk resistivity at the Schumann resonance frequency  $387\Omega$m was obtained.  {\sl This value -- supporting the validity of this measurement -- fits well to the literature value of vulcanic, andesite rocks ($170$-$45000$ $\Omega$m).}% The measured attenuation found to be 4\% lower for this underground vulcanic environment, supporting our proposal for the Einstein Telescope's underground location at the Mátra.

\item[Infrasound noise] Using the custom-made infrasound detector designed by the MTA Atomki, we measured the pressure amplitude spectral distribution in the MGGL (see Fig.~\ref{representative_PASD}). This is relatively noisy in the frequency range $1-7$ Hz. {\sl The recorded infrasound noise was found to be in accordance with the seismic noise at 3-4 Hz in the Laboratory.}
% presenting the validity of the non-linear noise propagation model for our rheological approach.

\item[Cosmic muon tomography] During the 404 days long measurement with the Muontomograph merging Run-0 and Run-1 data, led us to map the rock density and its inhomogeneities above MGGL at large scale (see Fig.~\ref{fig:DensTopo}).{\sl  We verified this novel measurement technique by obtaining the typical andesite rock density ($2.6 \pm 0.1$ kg\,dm\textsuperscript{-3}). According to this investigations, we have found, there were no large scale density inhomogeneities ($\le 0.2$ kg\,dm\textsuperscript{-3}) measured in the rock mass.}

\item[Seismic noise] The long term seismic noise was registered by two seismometers inside the MGGL at depth $-88$ m almost for the whole two-years period. In parallel, as a cross check, a third seismometer measured the seismic noise for two-weeks duration. This data record were farther from the MGGL in the mine in $-404$ m depth. {\sl Seismic noise measurement were found to be consistent.}
\end{description}

In the MGGL we had the opportunity to combine data taken at the same position, and in addition to this we could test the insights gained from noise analysis. According to our experience the proper evaluation of long term data requires some refinements of the methods applied in the ET Design Report~\cite{ETdes11r}. Our suggestions has been published in Ref.~\cite{SomEta18m} and in the recent study we evaluated the data accordingly. Therefore we calculated and show also the median of the spectra and the integrated noise measures $rms_{2-10\textrm{Hz}}$ and $rms_{1-10\textrm{Hz}}$ from the median, here. The analysis of the data from the Guralp ET1H and GU02 stations is based on these characteristics and the evaluation of the data from the WARS station also considered the problems of the more traditional methods.

\begin{figure}
\centering
\includegraphics[width=0.8\textwidth]{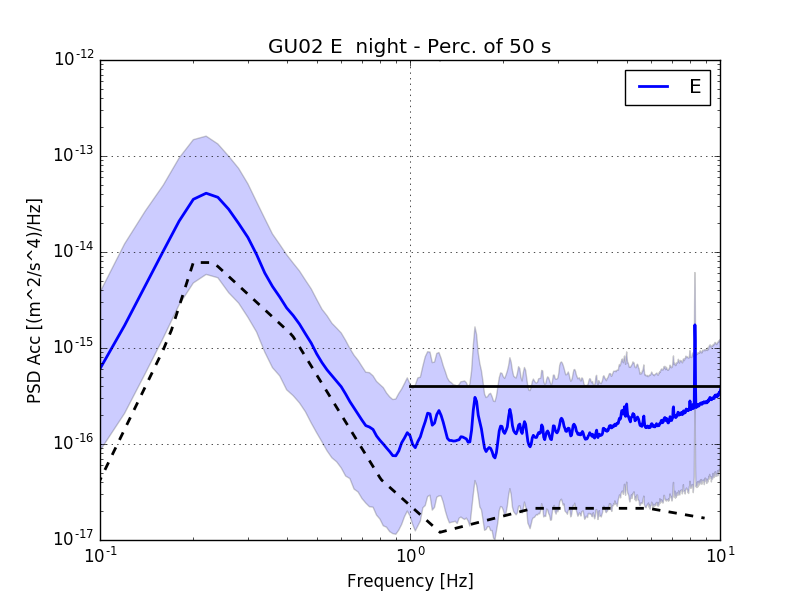}
\caption{\label{fig:2-week-PSD-HHE-night} Acceleration ASD values of the representative two-week period of GU02 seismometer (-404 m), from the 50 s long night percentiles in one of the horizontal directions. The median is solid blue and the borderlines of the blue area are the 10\textsuperscript{th} and the 90\textsuperscript{th} percentiles. The dashed line is the NLNM curve of Peterson, the Black Forest line is solid black.}
\end{figure}

We found, that the long term data does not show yearly changes in the median, but the intensity of the noisy periods reflects the changes in the activity inside the mine (see Fig.~\ref{fig:Year-night-ASD} and Tab.~\ref{tab:Yearly-night-rms-media}). The seasonal changes show minimal noise level in the late spring early summer period (see Fig. \ref{fig:Seasonally-ASD}). The comparison of the noise level at the shallower and the deeper locations revealed that the two-weeks measurement is unexpectedly representative for the whole  Run-0 and Run-1 period (see Fig. \ref{fig:Whole-2-week}). We compared also working and night periods in MGGL in order to estimate an achievable minimum noise level (see  Fig. \ref{fig:Whole-work-night-ratio}). Considering our whole analysis the median spectrum of the night noise at -404m in the mine in Fig. \ref{fig:2-week-PSD-HHE-night} is convincingly representative in this respect. Here the percentiles are calculated from the spectra of the 50s long periods of data. One can see that in 90\% of these periods the noise level is below the Black Forest line.

In general the noise levels our long term measurements in the shallower MGGL and also the two-week data from the deeper location are remarkably similar to the noise levels of the previous study \cite{ETdes11r,Bek13t,BekEta15a}, the mode $rms_{2Hz}$ are almost the same. As we have already mentioned, the median rms values are preferable, because they are more stable, less sensitive to the uncertainties is the data and in the evaluation. The ET1H and WARS median $rms_{2Hz}$ are close to each other. However, from the point of view of operational requirements, a higher percentile limit is a better estimation of the sensitivity of the detector for continuous operation. Therefore, the 90\textsuperscript{th} percentile rms values, given in Tab. \ref{tab: rms-90-perc}, estimate the noise level for a continuous operation of the Einstein Telescope.

\begin{table}
	\centering
	\begin{tabular}{|c|c|c|c|}
		\hline
		90\textsuperscript{th} & GU02 & ET1H 2ws & ET1H 2ys\tabularnewline
		\hline
		\hline
		$rms_{2Hz} [nm]$ & 0.171 & 0.257 & 0.261\tabularnewline
		\hline
		$rms_{2-10Hz} [nm]$ & 0.167  & 0.243 & 0.238\tabularnewline
		\hline
		$rms_{1-10Hz} [nm]$ & 0.401  & 0.600 & 0.583\tabularnewline
		\hline
	\end{tabular}
	\caption{\label{tab: rms-90-perc} The $rms$ values of the 90\textsuperscript{th} percentile in the E direction with different frequency ranges for the MGGL seismometer ET1H and for the deeper location GU02, too. The first two columns are calculated from the $90\,th$ percentile of the $300\,s$ data of the two-week interval, the last one from the average of daily $90\,th$ percentiles.}
\end{table}
%90\textsuperscript{th} & GU02 & ET1H 2ws & ET1H 2ys\tabularnewline
%\hline
%\hline
%$rms_{2Hz} [nm]$ & 0.181 & 0.270 & 0.281\tabularnewline
%\hline
%$rms_{2-10Hz} [nm]$ & 0.177  & 0.258 & 0.266\tabularnewline
%\hline
%$rms_{1-10Hz} [nm]$ & 0.441  & 0.658 & 0.853\tabularnewline
According to the recent study a reliable and comparable characterisation of long term seismic noise data requires the evaluation of various noise measures with a uniform methodology. The separation of reducible and irreducible cultural  noise, the seasonal changes and the attenuation by depth are important aspects for site description. Also it is very important to publish the raw seismological data for open evaluation and comparison. For the ET1H station of MGGL and for the GU02 station these are available at \cite{MGGL_RUN1dat}.

%%%%%%%%%%%%%%%%%%%%%%%%%%%%%%%%%%%%%%%%%%%%%%%%%%%%%%%%%%%%%%%%%%%%%%%%%%%%%%%%%%%%%%%%%%%%%%%%%%%%%%%%%%%%%%%%%%%%%
\section{Conclusions \label{sec:FinalConclusion}}

During the Run-0 and Run-1 periods of the MGGL in 2016-2018 we performed long-term monitoring the Mátra mountain range as a possible underground site of the planned Einstein Telescope. We used various standard methods in parallel to novel approaches of investigating the geophysical environment, electromagnetic attenuation, infrasound noise, cosmic muon tomography of the surrounding rock mass, and long term seismic noise. The collected data could enable us to cross check and and compare standard measurements and techniques applied in earlier investigations with the new ones. Alongside this, the geological and rheological properties of the base rock were summarized in this paper. In addition to the analysis of the noise background relevant for a next generation, underground-based gravitational wave detector, especially in the low frequency regime, at 1-10 Hz. We strongly believe that applying our results for the site selection will significantly improve the signal to nose ratio of %planned underground gravitational wave detectors of
the multi-messenger  astrophysics era. Our conclusion was that for the background noise analysis, it is necessary to perform long term data taking and apply the state-of-the-art techniques presented here.

%%%%%%%%%%%%%%%%%%%%%%%%%%%%%%%%%%%%%%%%%%%%%%%%%%%%%%%%%%%%%%%%%%%%%%%%%%%%%%%%%%%%%%%%%%%%%%%%%%%%%%%%%%%%%%%%%%%%%

\section{Acknowledgement}

The contribution and support of Nitrokemia Zrt. in particular \'A. V\'aradi and V. Rofrits  is acknowledged. We also thank the construction work of Geofaber Zrt.

This research was supported by the MTA EUHUNKP grant. Authors PV and RK thank the Hungarian National Research, Development and Innovation Office NKFIH grants K124366, K104260, K116197, K120660, NK106119, and K116375 for funding. TB, MC, MS, DR and TS were supported by the NCN Grant UMO-2013/01/ASPERA/ST9/00001. The muon flux measurements have been supported by the Momentum ("Lend\"ulet") grant of the Hungarian Academy of Sciences under contract LP2013-60. The support of the PHAROS (CA16214) and G2net (CA17137) COST Actions is also acknowledged. We acknowledge the support of the grant TEAM/2016-3/19 from the Foundation for Polish Science. The support of the European Regional Development Fund and Hungary in the frame of the project GINOP-2.2.1-15-2016-00012 is acknowledged, too. We thank the Wigner GPU Laboratory for their help, too.

\bibliographystyle{unsrt}

\end{document}